\documentclass[12pt]{article}

\pdfoutput=1

\usepackage{tikz-feynman}
\tikzfeynmanset{warn luatex=false}

\usepackage{cancel}
 
\usepackage{amsmath,amssymb,amsfonts,amscd,mathrsfs}
\usepackage{xcolor}
\definecolor{darkblue}{rgb}{0.1,0.1,.7}

\usepackage{hyperref} 
\hypersetup{colorlinks, linkcolor=darkblue, citecolor=darkblue, urlcolor=darkblue, linktocpage}

\usepackage[square, comma, compress,numbers]{natbib}

\usepackage[]{graphicx}
\usepackage[space]{grffile}
\usepackage{geometry}
\usepackage[margin=10pt,font=small,labelfont=bf]{caption}
\usepackage{ifthen}
\usepackage{tikz}
\usepackage{subcaption}
\usepackage{booktabs,multirow}
\usepackage{hhline}
\usepackage{tablefootnote}
\usepackage{braket}

\usepackage{mathtools,etoolbox}
\DeclarePairedDelimiterX{\abs}[1]{\lvert}{\rvert}{\ifblank{#1}{{}\cdot{}}{#1}}

\definecolor{myorange}{RGB}{199,146,32}

\usepackage{ragged2e}
\usepackage{array}
\newcolumntype{L}[1]{>{\raggedright\let\newline\\\arraybackslash\hspace{0pt}}m{#1}}
\newcolumntype{C}[1]{>{\centering\let\newline\\\arraybackslash\hspace{0pt}}m{#1}}
\newcolumntype{R}[1]{>{\raggedleft\let\newline\\\arraybackslash\hspace{0pt}}m{#1}}

\usepackage{dsfont} 

\usepackage{booktabs}

\usepackage{titlesec}
\titleformat*{\section}{\large\bfseries}
\titleformat*{\subsection}{\normalsize\bfseries}
\titleformat*{\subsubsection}{\normalsize\it}
\titleformat*{\paragraph}{\normalsize\bfseries}
\titleformat*{\subparagraph}{\normalsize\bfseries}

\newcommand{\reef}[1]{(\ref{#1})}

\def\eps{\epsilon}

\newcommand{\beq}{\begin{equation}} 
\newcommand{\eeq}{\end{equation}}

\def\bZ {\mathbb{Z}}

\def\bZ {\mathbb{Z}}

\def\geq{\geqslant}
\def\leq{\leqslant}
\newcommand{\diffop}[2]{\ifthenelse{\equal{#2}{1}}{\frac{\mrm{d}}{\mrm{d} #1}}{\frac{\mrm{d}^#2}{\mrm{d} #1^#2}}}

\newcommand{\bk}{{\mathbf k}}

\newcommand{\mrm}[1]{{\mathrm #1}}

\usepackage[normalem]{ulem}

 \def\cE{{\cal E}}
\newcommand{\roig}{\color{red}}
\newcommand{\gray}{\color{orange}}
\newcommand{\be}{\begin{equation}}
\newcommand{\ee}{\end{equation}}
\newcommand{\bea}{\begin{eqnarray}}
\newcommand{\eea}{\end{eqnarray}}

 \def\om{\omega}
  
  \def\dblu{\color [rgb]{0,.69,.30}}
    \def\blu{\color [rgb]{0,.09,.60}}

\usepackage{accents}
\newlength{\dhatheight}

    \newcommand{\Va}{
     \begin{minipage}[h]{0.015\linewidth}
\begin{tikzpicture}
\begin{feynman}[small]
 \node [crossed dot] (i1) at (.3,0);
\diagram*{
   (i1)
};
  \end{feynman}
\end{tikzpicture}
  \end{minipage} 
  }    
               
    \newcommand{\Vb}{
     \begin{minipage}[h]{0.04\linewidth}
\begin{tikzpicture}
\begin{feynman}[small]
\vertex (i1) at (0,0);
\vertex (i2) at (.6,0); 
\diagram*{
   (i1) --[thick] (i2)
};
  \end{feynman}
\end{tikzpicture}
  \end{minipage} 
  }
                                  
    \newcommand{\Vc}{
     \begin{minipage}[h]{0.04\linewidth}
\begin{tikzpicture}
\begin{feynman}[small]
\vertex (i1) at (0,0);
\vertex (i2) at (.6,0); 
\node [crossed dot] (i3) at (.3,0);
\diagram*{
   (i1) --[thick] (i3)--[thick] (i2)
};
  \end{feynman}
\end{tikzpicture}
  \end{minipage} 
  }
                    
    \newcommand{\Vd}{
     \begin{minipage}[h]{0.04\linewidth}
\begin{tikzpicture}
\begin{feynman}[small]
\vertex (i1) at (0,0);
\vertex (i2) at (.6,0); 
\node [crossed dot] at (.3,.2);
\diagram*{
   (i1) --[thick] (i2)
};
  \end{feynman}
\end{tikzpicture}
  \end{minipage} 
  } 
    \newcommand{\Vdin}{
     \begin{minipage}[h]{0.04\linewidth}
\begin{tikzpicture}
\begin{feynman}[small]
\vertex (i1) at (0,0);
\vertex (i2) at (.6,0); 
\node [crossed dot, rotate=45] at (.3,.2);
\diagram*{
   (i1) --[thick] (i2)
};
  \end{feynman}
\end{tikzpicture}
  \end{minipage} 
  }
          
    \newcommand{\VVa}{
     \begin{minipage}[h]{0.045\linewidth}
\begin{tikzpicture}
\begin{feynman}[small]
 \node [dot] (i1) at (0,0);
 \node [dot] (i2) at (.6,0);  
\diagram*{
   (i1) -- [ quarter right, looseness=.6,  thick] (i2)  -- [ quarter right, looseness=.6 ,  thick] (i1) ,
     (i1) -- [ half right, looseness=1.,  thick] (i2)  -- [ half right, looseness=1. ,  thick] (i1)
};
  \end{feynman}
\end{tikzpicture}
  \end{minipage} 
  }

    \newcommand{\VVb}{  
  \begin{minipage}[h]{0.07\linewidth}
\begin{tikzpicture}
\begin{feynman}[small]
 \vertex (i0) at (-.3,0);
 \node [dot] (i1) at (0,0);
 \node [dot] (i2) at (.6,0);  
 \vertex (i3) at (.9,0);
\diagram*{
   (i0) -- [thick]   (i3) ,
     (i1) -- [ half right, looseness=1.,  thick] (i2)  -- [ half right, looseness=1. ,  thick] (i1)
};
  \end{feynman}
\end{tikzpicture}
  \end{minipage}
  }
  
      \newcommand{\VVc}{ \begin{minipage}[h]{0.07\linewidth}
\begin{tikzpicture}
\begin{feynman}[small]
 \vertex (im2) at (.9,0);
  \vertex (im1) at (.8,0);
 \vertex (i0) at (.7,-.1);
 \node [dot] (i1) at (0,0);
 \node [dot] (i2) at (.6,0);  
 \vertex (i3) at (-.1,.1);
 \vertex (i4) at (-.2,0);
  \vertex (i5) at (-.3,0);
\diagram*{
(im2)--[thick](im1)--[quarter right, thick]   (i0) -- [half left, looseness=1.3, thick] (i1) --[thick]   (i2) 
   --  [thick, half right, looseness= 1.3] (i3)     -- [thick, quarter left , looseness = 1] (i4) --[thick] (i5)  ,
     (i1) -- [ half right, looseness=1.,  thick] (i2)  -- [ half right, looseness=1. ,  thick] (i1)
};
  \end{feynman}
\end{tikzpicture}
  \end{minipage}
  }
 
     \newcommand{\VVd}{
     \begin{minipage}[h]{0.065\linewidth}
\begin{tikzpicture}
\begin{feynman}[small]
 \node [dot] (i1) at (0,0);
 \node [dot] (i2) at (.6,0);  
 \vertex (i0) at (-.3,-.3);
 \vertex (i3) at (.9,-.3);
\diagram*{
   (i1) -- [ quarter right, looseness=.6,  thick] (i2)  -- [ quarter right, looseness=.6 ,  thick] (i1) ,
     (i1) -- [ half right, looseness=1.,  thick] (i2)  -- [ half right, looseness=1. ,  thick] (i1) , 
     (i0) -- [thick] (i3)
};
  \end{feynman}
\end{tikzpicture}
  \end{minipage} 
  }
          \newcommand{\VVe}{ \begin{minipage}[h]{0.07\linewidth}
\begin{tikzpicture}
\begin{feynman}[small]
 \vertex (im2) at (.9,0);
  \vertex (im1) at (.8,0);
 \vertex (i0) at (.7,-.1);
 \node [dot] (i1) at (0,0);
 \node [crossed dot] (i2) at (.55,0);  
 \vertex (i3) at (-.1,.1);
 \vertex (i4) at (-.2,0);
  \vertex (i5) at (-.3,0);
\diagram*{
(im2)--[thick](im1)--[quarter right, thick]   (i0) -- [half left, looseness=1.3, thick] (i1)    ,
 (i5) -- [thick]   (i1) -- [ quarter right, looseness=1,  thick] (i2)  -- [  quarter  right, looseness=1 ,  thick] (i1)
};
  \end{feynman}
\end{tikzpicture}
  \end{minipage}
  }    
    \newcommand{\VVVa}{
       \begin{minipage}[h]{0.065\linewidth}
\begin{tikzpicture}
\begin{feynman}[small]
   \node [dot] (i2) at (0,0);
 \node [dot] (i3) at (.5,0);
 \node [dot] (i4) at (1.,0);  
\diagram*{
 (i4)    -- [half right, thick, looseness=.7] (i2)  -- [half right, thick, looseness=.7] (i4),
           (i2) -- [ quarter right, looseness=.9 ,  thick] (i3) -- [ quarter right, looseness=.9 ,  thick] (i2)  , 
           (i3) -- [ quarter right, looseness=.9 ,  thick] (i4) -- [ quarter right, looseness=.9 ,  thick] (i3)  , 
};
  \end{feynman}
\end{tikzpicture}
  \end{minipage}
    }
    
    \newcommand{\VVVb}{
       \begin{minipage}[h]{0.085\linewidth}
\begin{tikzpicture}
\begin{feynman}[small]
 \vertex (i1) at (-.3,-.35);
   \node [dot] (i2) at (0,0);
 \node [dot] (i3) at (.5,0);
 \node [dot] (i4) at (1.,0);  
 \vertex (i5) at (1.3, -.35);
\diagram*{
(i1)  -- [thick] (i5),
 (i4)    -- [half right, thick, looseness=.7] (i2)  -- [half right, thick, looseness=.7] (i4),
           (i2) -- [ quarter right, looseness=.9 ,  thick] (i3) -- [ quarter right, looseness=.9 ,  thick] (i2)  , 
           (i3) -- [ quarter right, looseness=.9 ,  thick] (i4) -- [ quarter right, looseness=.9 ,  thick] (i3)  , 
};
  \end{feynman}
\end{tikzpicture}
  \end{minipage}
    }
    \newcommand{\VVVc}{
       \begin{minipage}[h]{0.095\linewidth}
\begin{tikzpicture}
\begin{feynman}[small]
 \vertex (i1) at (-.3,0);
   \node [dot] (i2) at (0,0);
 \node [dot] (i3) at (.5,0);
 \node [dot] (i4) at (1.,0);  
 \vertex (i5) at (1.3, 0);
\diagram*{
(i1) -- [thick] (i2),
(i4) -- [thick] (i5),
 (i4)    -- [half right, thick, looseness=.7] (i2),
           (i2) -- [ quarter right, looseness=.9 ,  thick] (i3) -- [ quarter right, looseness=.9 ,  thick] (i2)  , 
           (i3) -- [ quarter right, looseness=.9 ,  thick] (i4) -- [ quarter right, looseness=.9 ,  thick] (i3)  , 
};
  \end{feynman}
\end{tikzpicture}
  \end{minipage}
    }
        \newcommand{\VVVd}{
       \begin{minipage}[h]{0.095\linewidth}
\begin{tikzpicture}
\begin{feynman}[small]
 \vertex (i1) at (-.3,-.05);
   \node [dot] (i2) at (0,0);
 \node [dot] (i3) at (.5,0);
 \node [dot] (i4) at (1.,0);  
\vertex (ii1) at (.6,-.21-.03);  
 \vertex  (ii2) at (.3,-.25-.03);  
\vertex (ii3) at (.63,-.21-.04);  
 \vertex  (ii4) at (.33,-.25-.06);  
 \vertex (i5) at (1.3, -.05);
\diagram*{
 (i4)    -- [half right, thick, looseness=.7] (i2),
           (i2) -- [ quarter right, looseness=.9 ,  thick] (i3) -- [ quarter right, looseness=.9 ,  thick] (i2)  , 
           (i3) -- [ quarter right, looseness=.9 ,  thick] (i4) -- [ quarter right, looseness=.9 ,  thick] (i3)  , 
    (i2) --[ quarter right , thick, looseness=.9] (i5) ,
      (ii1) -- [white, thick] (ii2),
      (ii3) -- [white, thick] (ii4),
  (i1)    -- [quarter right, thick, looseness=.9] (i4),
};
  \end{feynman}
\end{tikzpicture}
  \end{minipage}
    }

        \newcommand{\VVVe}{
       \begin{minipage}[h]{0.095\linewidth}
\begin{tikzpicture}
\begin{feynman}[small]
 \vertex (i1) at (-.3,-.05);
   \node [dot] (i2) at (0,0);
 \node [dot] (i3) at (.5,0);
 \node [dot] (i4) at (1.,0);  
\vertex  (ii1) at (.97,-.058);  
\vertex  (ii2) at (.5,-.38);  
 \vertex  (ii3) at (.95,-.065);  
\vertex  (ii4) at (.5,-.32);  
 \vertex (i5) at (1.3, 0);
\diagram*{
   (i2) -- [thick]   (i4) ,
     (i2) -- [ quarter right, looseness=1.,  thick] (i3)  , 
     (i3) --[ quarter right , thick, looseness=1.1] (i5) ,
     (i2) --[ quarter left , thick, looseness=.7] (i4) ,
      (ii1) -- [white, thick] (ii2),
 (ii3) -- [white, thick] (ii4),
 (i4)    -- [half right, thick, looseness=.9] (i2),
  (i1)    -- [quarter right, thick, looseness=.9] (i4),
};
  \end{feynman}
\end{tikzpicture}
  \end{minipage}
    }    
        \newcommand{\VVVf}{
       \begin{minipage}[h]{0.095\linewidth}
\begin{tikzpicture}
\begin{feynman}[small]
 \vertex (i1) at (-.3,-.05);
   \node [dot] (i2) at (0,0);
 \node [dot] (i3) at (.5,0);
 \node [dot] (i4) at (1.,0);  
 \vertex (i5) at (1.3, 0);
\diagram*{
   (i2) -- [thick]   (i5) ,
     (i2) -- [ quarter right, looseness=1.,  thick] (i3)  , 
     (i1) --[ quarter right , thick, looseness=1.1] (i3) ,
     (i2) --[ quarter left , thick, looseness=.7] (i4) ,
 (i4)    -- [half right, thick, looseness=.9] (i2),
};
  \end{feynman}
\end{tikzpicture}
  \end{minipage}
    }
    
    \newcommand{\VVVg}{
      \begin{minipage}[h]{0.095\linewidth}
\begin{tikzpicture}
\begin{feynman}[small]
 \vertex (i1) at (-.3,0);
   \node [dot] (i2) at (0,0);
 \node [dot] (i3) at (.5,0);
 \node [dot] (i4) at (1.,0);  
 \vertex (i5) at (1.3, -.05);
\diagram*{
   (i1) -- [thick]   (i4) ,
     (i3) -- [ half right, looseness=1.,  thick] (i4)  -- [ half right, looseness=1. ,  thick] (i3) , 
     (i2) --[ quarter right , thick, looseness=.9] (i5) ,
 (i4)    -- [half right, thick, looseness=.9] (i2),
};
  \end{feynman}
\end{tikzpicture}
  \end{minipage}
    }
    
    \newcommand{\VVVh}{
       \begin{minipage}[h]{0.095\linewidth}
\begin{tikzpicture}
\begin{feynman}[small]
  \node [dot] (i2) at (0,0);
 \node [dot] (i3) at (.5,0);
 \node [dot] (i4) at (1.,0);  
  \vertex (i7) at (1.3,0);
  \vertex (im1) at (-.3,0);
\diagram*{
   (i2) -- [thick]   (i4) ,
     (i2) -- [ quarter left, looseness=.4,  thick] (i4)   , 
      (i2) -- [ half left, looseness=.55,  thick] (i4)   , 
     (i3) --[ quarter right , thick, looseness=.7]  (i7) ,
 (i4)    -- [half right, thick, looseness=1] (i2) ,
   (i3) --[ quarter left , thick, looseness=.7]  (im1) ,
};
  \end{feynman}
\end{tikzpicture}
  \end{minipage} 
    }
    
\numberwithin{equation}{section}
\usepackage[tocgraduated]{tocstyle}

\begin{document}

\vspace*{-.6in} \thispagestyle{empty}
\begin{flushright}
\end{flushright}
\vspace{1cm} {\Large
\begin{center}
\textbf{
Exploring  Hamiltonian Truncation  in $\bf{d=2+1}$
}
\end{center}}
\vspace{1cm}
\begin{center}

{\bf  Joan Elias Mir\'o$^{a,b}$, \, Edward Hardy$^{c}$} \\[2cm] 
{
$^a$  ICTP,    Strada Costiera 11, 34135, Trieste, Italy  \\
$^b$  CERN, Theoretical Physics Dep., 1211 Geneva 23, Switzerland\\
$^c$  University of Liverpool, Dept. of Mathematical Sciences, Liverpool L69 7ZL, United Kingdom \\
}

\vspace{3cm}
\end{center}

\vspace{-2cm}

\begin{abstract}

We initiate the application of Hamiltonian Truncation methods to solve  strongly coupled   QFTs in $d=2+1$. 
By  analysing perturbation theory with a Hamiltonian Truncation regulator, we pinpoint the challenges of such an approach and propose a way that these can be addressed. This enables us to formulate  Hamiltonian Truncation theory for $\phi^4$  in $d=2+1$, and to study its spectrum at weak and strong coupling. The results obtained agree well with the predictions of a weak/strong self-duality possessed by the theory. 
The $\phi^4$  interaction is a strongly relevant UV divergent perturbation, and represents a case study of a  more general scenario.  Thus, the approach developed should be applicable to many other QFTs of 
interest.

 \end{abstract}
\vspace{.2in}
\vspace{.3in}
\hspace{0.7cm}

\setcounter{tocdepth}{2}

\newpage

{
\tableofcontents
}

\newpage

\section{Introduction}

There is currently no known universal and efficient method to derive the phenomenological implications of strongly coupled Quantum Field Theories (QFTs). 
 Thus any new strategy to understand the strongly coupled regime merits scrutiny. In this paper we analyse one such approach: the Hamiltonian Truncation (HT) method.

The basic idea behind HT is quite transparent, consisting of a generalisation of the Rayleigh-Ritz method of Quantum Mechanics to QFT: Consider a theory whose Hamiltonian can be decomposed as $H=H_0+V$, where $H_0$ denotes a solvable Hamiltonian~\footnote{A free-theory or an interacting integrable theory, be it a solvable CFT or and integrable massive QFT.} -- $H_0 |E_i\rangle = E_i |E_i\rangle$ --  and $V$ denotes a perturbation. 
Then, HT proceeds by  truncating the Hamiltonian $H$ into a large finite matrix $H_{ij}$ with $E_i \leq E_T$,  and diagonalising it numerically. 
There is a  systematic error due to the truncation energy $E_T$,  however in many instances  the spectrum of the full theory can be recovered  from results obtained at finite $E_T$ by performing precise extrapolations $E_T\rightarrow \infty $.~\footnote{We will formulate HT more precisely   in section \ref{sec:ham1}.}
 
Although conceptually simple, the strength of the HT idea is that it can be used to tackle strongly coupled QFTs.
 Indeed, it has been applied very successfully in $d=2$ spacetime dimensions.  
There are various incarnations of the method differing in the quantisation frame and  the choices of basis and $H_0$. 
One important version is the Truncated  Conformal Space Approach (TCSA) introduced in \cite{Yurov:1989yu,Yurov:1991my}, which sparked numerous further studies. In TCSA the  Hamiltonian $H_0$ is that of a solvable  two dimensional CFT.~\footnote{More precisely, the CFT is placed on the cylinder $\mathbb{R}\times S_R^1$ of radius R. Then, due to the state-operator map, $H_0$ is the dilatation operator and  the role of $E_T$ is played by the  dimension of a  heavy operator $E_T\sim \Delta_\text{max}/R$.}
More recently, a closely related version, coined Conformal Truncation, exploits  light-cone quantisation to truncate the wave-functions at infinite volume \cite{Katz:2014uoa,Katz:2013qua,Katz:2016hxp,Fitzpatrick:2018ttk,Delacretaz:2018xbn}.
Another guise  of HT    uses a massive Fock-Space basis to truncate the Hamiltonian.  Early work on this direction was done in \cite{Brooks:1983sb}. 
Since then, the method has been  developed in  \cite{Rychkov:2014eea,Rychkov:2015vap,Elias-Miro:2015bqk,Bajnok:2015bgw}  and  results with a precision that is competitive with other up to date techniques  have been obtained \cite{Elias-Miro:2017xxf}. 

The literature on HT in $d=2$ is by now extensive, with a plethora of fascinating results for strongly coupled QFTs. For example, HT has allowed  strongly coupled real time dynamics, and strongly coupled perturbations of interacting fixed points, to be studied.  We refer the reader to \cite{James:2017cpc} for a review where further references may also be found.

Despite this success in $d=2$, it has remained an unsolved challenge to apply the ideas of HT to  $d\geq  3$ strongly coupled QFTs. The main obstacle is the appearance of UV divergences that require regularisation. 
In particular, divergences appear in many QFTs that we care about for $d\geq 3$, when the QFT is formulated by perturbing a solvable theory by a relevant operator,  $H_0\rightarrow H_0+V$. The $E_T$ cutoff is a natural regulator for the divergences. But then, what are the counter-terms required to formulate HT? And, is a covariant Lorentz spectrum recovered as the $E_T$ regulator is removed? In this paper we will address these, and other, key open questions.

We will be pragmatic and approach these problems in a particular instance of HT, using the massive Fock-Space basis. We will study the  $\phi^4_3$ theory, namely we will perturb  the free massive theory by the operator $V=g\int\phi^4$ in $d=2+1$. This is a relevant operator, in the RG sense. Thus, very naively, we may expect that the spectrum will converge in a power-like manner $g/E_T$ as $E_T$ is increased. 
However, the $\phi_3^4$ theory has a linear UV divergence associated to the vacuum and a logarithmic UV divergence associated to the mass. 
Thus, it will act as a prototype case study, capturing the key features of a far more general class of theories. In particular, many of the techniques that we will develop should be useful for formulating HT in any dimension for theories in which $V$ is a strongly relevant perturbation. 

We will study  the theory at finite volume. This allows us to focus on the UV conundrums without having to deal with other problems associated to the large volume dynamics, such as the orthogonality catastrophe recently analysed in \cite{Elias-Miro:2017tup}. It would be very interesting to carry out a separate investigation of the infinite volume limit in the future.

There have been various  interesting works prior to us in $d>2$ that we would like to highlight.
In  \cite{Hogervorst:2014rta} the TCSA method was  developed    at $2< d \leq 2.5$ spacetime dimensions. In this range of $d$ the UV divergences of the $\phi^4_d$ theory are absent.
Then, ref.~\cite{Rutter:2018aog} investigated,  among other things, the allowed counter-terms at second order in perturbation theory  in TCSA   for general $d$.~\footnote{As we will show below many interesting effects start at fourth and higher order. }
In the Conformal Truncation  line of development, ref.~\cite{Katz:2016hxp} studied  $\phi^4_3$ at weak coupling, and the $(\vec{\phi}.\vec{\phi})^2$ perturbation for large vector size $N$.  Meanwhile ref.~\cite{Anand:2019lkt}  laid down formulas to  efficiently  compute the matrix elements of $\phi^4_3$ in the Conformal Truncation approach. 
Finally, other recent  work has been done in \cite{Hogervorst:2018otc}. There, the main results for $d=3$ [on a $S_3$ manifold]  are restricted to $\phi^2$ and $i\phi^3$ perturbations, where the UV divergences are either absent or only logarithmic.

First, we  review the basics  of HT and Hamiltonian perturbation theory in  section \ref{sec:ham1} and  \ref{hptdiags}, respectively.
Then, our strategy is as follows. The results of Hamiltonian Truncation at weak coupling must match  those of perturbation theory. 
Therefore, we begin our main investigation by analysing perturbation theory regulated with an $E_T$ cutoff. 
In section \ref{secphi2} we analyse the $\phi^2_3$ perturbation from three  perspectives: the exact solution, the Hamiltonian perturbation theory solution, and HT. 
Then, in section \ref{pertUV} we study the $\phi^4_3$ theory. Although in this case we do not know the exact solution, we carry out an analogous perturbation theory analysis. 
The $E_T$ cutoff  is a rather unusual regulator and  various surprises lie ahead.  
For instance, disconnected vacuum diagrams cancel in an intricate manner in Hamiltonian Perturbation theory; and we find that  such cancelation is spoiled with the ET regularisation, introducing new UV divergences.
We manage to precisely delineate all such challenges and we offer a solution in section  \ref{pertUV}.

After understanding perturbation theory, we uplift our formalism into Hamiltonian Truncation. 
In section \ref{sNI} we start our numerical explorations. 
First we implement HT and find agreement with perturbation theory.  Having matched perturbation theory and seen that extrapolations to large $E_T$ are feasible, we increase the coupling in section \ref{sCM}. There we perform a   crosscheck of the strong coupling spectrum by making use of a strong/weak self-duality that the theory possesses. 
Finally, we conclude and discuss some of the many interesting possible directions  for future work in section~\ref{concs}.

 \section{Hamiltonian Truncation }
\label{sec:ham1}

We begin our study of Hamiltonian Truncation in higher dimensions  focusing on the  $(\phi^2)_3$ and $(\phi^4)_3$ perturbations at finite volume. In this section we review the basics of  HT and set our conventions.

\subsection{Definitions}
\label{defham}

The   theory  that we study is defined by deforming the free massive theory by a strongly relevant perturbation, $(\phi^4)_3$. 
The free  theory is quantized on a flat torus space of size $L\times L$,  and the time direction is left uncompact $t\in \mathbb{ R}$. 
We impose periodic boundary conditions on the operator $\phi(t,x,y)=\phi(t,x,y+nL)=\phi(t,x+nL,y)$  for $n\in \mathbb{Z}$. Since the space directions are compact the spectrum of the free theory  is discrete and free of IR divergences.
In canonical quantization $\phi$ can be expanded in terms of creation and annihilation operators as
\beq
\phi(0,x)=\sum_{k} a_{k}e^{-i k \cdot \mathbf{x}}/\sqrt{2L^2\om_{k}}+h.c. \label{field}
\eeq
where  $k = (k_1,k_2)$, $ k_i= 2\pi n_i/L$,  $n_i\in \bZ$,  $\om_{k}=(k_1^2+k_2^2+m^2)^{1/2}$
and $a^\dagger$, $a$ satisfy the algebra of  creation/annihilation operators $
[a_{k},a_{k^\prime}^\dagger]= \delta_{k,k^\prime} \, , \quad [a_{k},a_{k^\prime}]=0$. 
The free Hamiltonian is
 $
H_0 = \sum_{\mathbf{k}} \om_{k} a_{k}^\dagger a_{k}
$ 
and it is diagonalized by the $H_0$ eigenbasis basis
 $
\ket{E_i}=\frac{a_{k_N}^{\dagger n_N}}{\sqrt{n_N!}}\cdots   \frac{a_{k_2}^{\dagger n_2}}{\sqrt{n_2!}} \frac{a_{k_1}^{\dagger n_1}}{\sqrt{n_1!}} \ket{0} \label{obv}
 $
 with eigenvalues $E_i=\sum_{s=1}^N n_s \sqrt{k_s^2+m^2}$.~\footnote{In \reef{obv}  $k_i$ is obviously a two-dimensional vector [instead of components of $k$, below \reef{field}]. From here on   any subscripted momenta is a label for a two-dimensional vector. }  
  The free vacuum is defined by $H_0\ket{0}=0$.

The interacting Hamiltonian is
\be
 H(E_T)=H_0+ V   \quad \text{with} \quad V=g_2V_2 + g_4 V_4 + C(E_T) \label{ham1}
\ee
where 
\be
V_n=  \frac{1}{n!}\int_{-L/2}^{+L/2} d^2x \,: \phi^n(x): ~,
\ee
and $:\phi^n(x):$ indicates normal ordering, i.e. annihilation operators are placed to the right of creation operators.~\footnote{Normal ordering in finite 
volume differs from the normal ordering  in infinite volume. The difference is a scheme choice and it can be accounted for by finite counterterms that are exponentially small in the large volume limit \cite{Rychkov:2014eea}. } 
The operator $C$ in \reef{ham1} is a counter-term that must take a somewhat unusual form and will be explained in the sections below.
 For future use we define  $O_{ij}\equiv \bra{E_i} O\ket{E_j}$ and $g_4\equiv g$. 


\subsection{The approach}

The Hamiltonian defined in section \ref{defham} acts in the Hilbert space ${\cal H}$ spanned by the free states $\ket{E_i}$. In the Hamiltonian Truncation approach, we study the theory by considering only states $E_i \leq E_T$,  with the counter-terms in \reef{ham1}  evaluated at $E_T$.
In this case the Hamiltonian \reef{ham1} can be viewed as the finite dimensional matrix
\be
H_{ij} = \bra{E_i} H  \ket{E_j} \, .  \label{mat1} 
\ee
The spectrum of $H$ can then be obtained by applying a numerical routine to obtain the lowest few eigenvalues of $H_{ij}$. 
Similarly to Hamiltonian Perturbation theory, the actual theory that we are interested in is recovered by removing the regulator, i.e. taking the limit $E_T\rightarrow \infty$. 
In practice this is done by considering a  series of Hamiltonians  $H(E_T^{(i)})$ with $E_T^{(0)} < E_T^{(1)}< \cdots < E_T^{(n)}$ such that $E_T^{(n)}\gg m$, and extrapolating the eigenvalues to infinite $E_T$.

\begin{figure}[t]
\begin{center}
\includegraphics[width=0.6\textwidth]{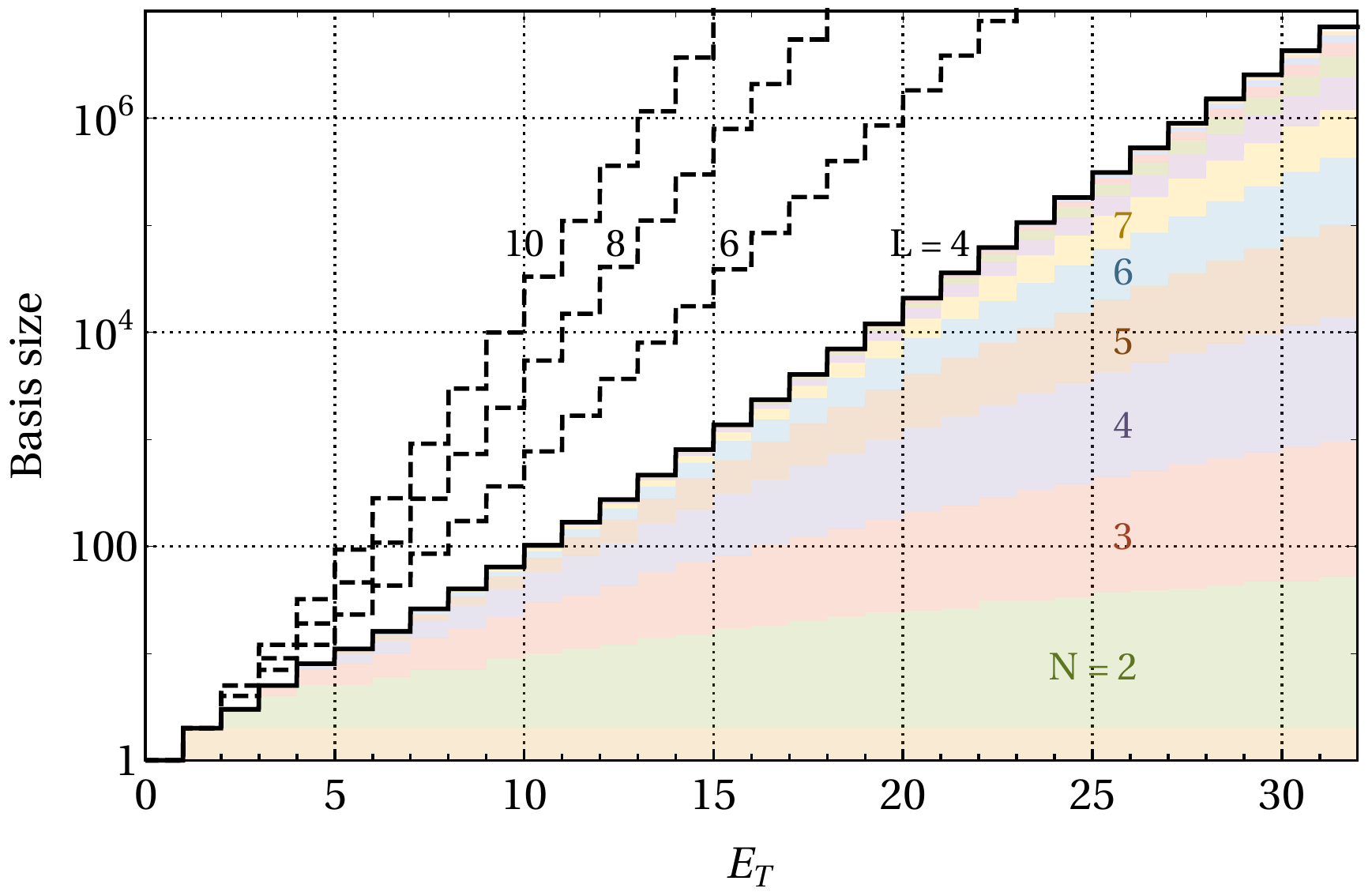}
\end{center}
\caption{The size of the Fock space basis as a function of $E_T$ for $m=1$  and different box sizes $L$. For $L=4$ the contributions from basis elements with different occupation numbers $N$ is shown.    \label{figbasis}} 
\end{figure}

The symmetry transformations of the theory \reef{ham1} can be used to simplify the task of obtaining the eigenvalues of $H$. 
 These transformations are the field parity $\mathbb{Z}_2 $   ($\phi(x)\rightarrow -\phi(x)$), the momentum, and the finite subgroup of O(2) that corresponds to the symmetry group of a flat torus [further details may be found in appendix \ref{symmetriesT2}]. The matrix \reef{mat1} can therefore be diagonalised separately in the different selection sectors. Throughout this paper we focus on states with zero total momenta,  and we diagonalise the  $\mathds{Z}_2 = \pm $ sectors separately. We have verified that the lowest lying states are in the singlet sector of the square torus transformations, as expected from perturbation theory. Therefore we restrict to this sector in what follows.

A basis of the Hilbert space up to a given cutoff  $E_T$ can be straightforwardly constructed in a computer program. Then the Hamiltonian can be calculated, although some care is required in the algorithm to enable large basis sizes to be reached. The lowest lying eigenvalues can be obtained using standard algorithms, for example based on the Lanczos method. Further details on our numerical approach can be found in appendix~\ref{codeapp}.

In Figure~\ref{figbasis} we plot the size of the singlet sector of the basis for a theory with $m=1$ as a function of the cutoff energy for $E_T$ [including both field parity sectors]. Using moderate computational resources we are able to obtain eigenvalues for bases of size up to $\sim 10^7$. We see that our numerical reach corresponds to $E_T / m \sim 32$ for a box of length $L=4/m$, which will turn out to be sufficient to perform reasonable extrapolations $E_T\rightarrow \infty$.

\section{Perturbation theory }
\label{hptdiags}

At weak coupling, the spectrum of the truncated Hamiltonian can be computed using Hamiltonian Perturbation Theory (HPT), a.k.a. Old Fashioned Perturbation Theory. 
Throughout the paper we will make extensive use of HPT. 
 Since HPT is somewhat in disuse, in this section we provide a basic review and set up the notation   -- the reader familiar with such material may safely skip this section. 
  
  In perturbation theory we compute the $i$th energy level in a truncated power series of the interaction strength $\cE_i=\sum_n \cE_i^{(n)}$.
  The general structure of the corrections $\cE_i^{(n)}$ is always the same, namely, the $n$th order correction is given by 
   \be
\cE^{(n)}_i=   \bra{E_i} \hat{V} ( [E_i-H_0]^{-1}\hat{V})^{n-1}   \ket{E_i} -\text{subtraction terms}  \, ,  \label{subte}
   \ee
   where we have defined $\hat{V}=P V P$, and $P$   is the projector $P=\mathds{1}- |E_i\rangle\langle E_i|$.
   The role of the \emph{subtraction terms} will be clarified in the sections below, and they will turn out to be crucial for our work.  
For instance, the first terms of perturbation theory for the energy levels are given by  
      \bea
   \cE_i &=& \underbrace{V_{ii}}_{\cE_i^{(1)}} + \underbrace{ V_{ik}\frac{1}{E_{ik}}V_{ki}}_{\cE_1^{(2)}} +\underbrace{ V_{ik}\frac{1}{E_{ik}}V_{kk^\prime} \frac{1}{E_{ik^\prime}}V_{k^\prime i} - V_{ii}   \,  V_{ik}\frac{1}{E_{ik}^2}V_{ki}}_{\cE_0^{(3)}} \nonumber \\
   &+& 
   \underbrace{ \frac{V_{ik}V_{kk^\prime} V_{k^\prime k^{\prime\prime}}V_{k^{\prime\prime} i}}{E_{ik} E_{ik^\prime} E_{ik^{\prime\prime}} }   -\cE_i^{(2) } \, \frac{V_{ik}V_{ki}}{E_{ik}^2} -2 V_{ii}  \, \frac{ V_{ik}V_{kk^\prime} V_{k^\prime i}    }{E_{ik}^{2}  E_{ik^\prime}} - V_{ii}^2  \,  \frac{V_{ik}V_{ki}}{E_{ik}^{3}}  }_{\cE_0^{(4)}} + \, O(V^5) \, ,   \label{basicpert}
   \eea
   where  $E_{ij}=E_i-E_j$, and a sum over intermediate states $k, k^\prime, k^{\prime\prime} \neq i$ is implicit.
   See appendix \ref{basicHPT} for a derivation of \reef{basicpert}.    
      
To calculate \reef{basicpert},  we can proceed by evaluating the  expressions 
 $
 \bra{E_i}F_n(z)\ket{E_i}= \bra{E_i} V ( [z-H_0]^{-1}V)^{n-1}   \ket{E_i} 
 $. These  can be conveniently written as 
\be
F_n(z+E_i)=  (-i)^{n-1}  \int_0^\infty dt_1\cdots dt_{n-1} \ e^{i z(t_1+\cdots t_{n-1})} \,   V\left( T_{n-1}\right) V\left( T_{n-2}\right) \cdots   V\left( 0\right)   \, , \label{dWe}
\ee
where $T_k= \sum_{i=1}^{i=k} t_i$, $V(t)= e^{iH_0 t} V e^{-i H_0 t}$.~\footnote{ Generically we will need $F(z_1,z_2,\dots , z_n)=V \prod_{i=1}^n (z_i-H_0)^{-1}V$ and its derivatives, which involves a straightforward generalisation. Note also that the integrals in \reef{dWe} converge for $\text{Im}z>0$, and for other values of $z$ analytic continuation is understood.} 
From \reef{dWe} it is easy to extract \reef{basicpert}.
 Eq.~\reef{dWe} is time ordered in the whole integration domain and thus we can apply Wick's theorem.
For concreteness let us take $V=g_k \phi^k/k!$, generalisations are straightforward. 
Applying Wick's theorem to  the product $ V\left( T_{n-1}\right) V\left( T_{n-2}\right) \cdots   V\left( 0\right) $ we get   
\be
\frac{g_k^n}{k!^n} \int_{L/2}^{L/2}  \prod_{i=0}^{n-1} d^2y_i \,   s^k_{\{n_{a, b}\}} 
\prod_{a=1}^{n-1}\prod_{b=0}^{a-1} 
D^{n_{a,b}}_{ab}
 \, :  \phi^{k-\Sigma_{s\neq n-1} n_{n-1,s}}_{Y_{n-1},T_{n-1}}  \phi^{k-\Sigma_{s\neq n-2} n_{n-2,s}}_{Y_{n-2},T_{n-2}}\cdots  \phi^{k-\Sigma_{s\neq 0} n_{0,s}}_{Y_{0},T_{0}} : \ ,   \label{wicks}
\ee
where $\phi_{x,t} \equiv \phi(x,t)$. $D_{ij}\equiv D(x_i^\mu-x_j^\mu)$ is the propagator joining the points $(\vec Y_i, T_i)$ and $(\vec Y_j, T_j)$, where $\vec Y_i= \vec y_0+ \sum_{p=1}^i \vec y_p$, given by
\be
\theta(t) D(\vec z,t) = \sum_{\vec k } \frac{1}{2L^2 \om_{\vec k}} e^{-i \om_{\vec k } t+ i \vec k \cdot \vec z} \, .  \label{props}
\ee
The tensor  $n_{a,b}=n_{b,a}$ is symmetric and takes indices in the numbered vertices $V(T_a)$. Thus,  if we have a number $n_{1,2}$ of propagators connecting the $1$ and $2$ vertices, these same number of fields is subtracted from the vertices $\phi_{Y_{1},T_{1}}^k$ and $\phi_{Y_{2},T_{2}}^k$.  
Finally the combinatoric factor $s_{\{n_{a, b}\}}$ is given by
\be
s_{\{n_{a,b}\}}^k  =  \frac{k!^n}{\prod_{j=0}^{n-1} (k-\Sigma_{s\neq j} n_{j,s})! } \frac{1}{\prod_{a=1}^{n-1}\prod_{b=0}^{a-1}  n_{a,b}!} \label{sym} \, . 
\ee
Upon plugging the field mode expansion \reef{field} and the propagators \reef{props} into equations \reef{dWe} and  \reef{wicks},  we can perform the  straightforward integrals over the  exponentials and we are left with a sum over the momentum modes. 
Since the index structure in Formula \reef{wicks} is somewhat involved, let us  give  a couple of examples.

 \subsection{Examples}
For instance for $k=4$, the $O(g^2)$  correction to the vacuum is computed as follows
\be
V_{0k}E_k^{-1}V_{k0}=  \frac{g^2L^2 s^4_4 }{4!^2}  \int_0^\infty dt_1 e^{i z t_1} \int_{-L/2}^{L/2} d^2y_1   D^4(\vec y_1,t_1) \, ,  \label{basicex}
\ee
with $s^k_p=k!^2/(k-p)!^2/p!$. 
Upon performing the straightforward integrals of the previous expression we get
\vspace{-.5cm}
\be
 \begin{minipage}[h]{0.058\linewidth}
\begin{tikzpicture}
\begin{feynman}[small]
 \node [dot] (i1) at (0,0);
 \node [dot] (i2) at (.8,0);  
  \vertex (XX) at (.4,1.15); 
 \vertex (x1) at (.4,.5); 
 \vertex (x2) at (.4,-.9) {\roig \scriptsize $-E$};
\diagram*{
          (XX) -- [white, scalar, thick] (x1),
   (i1) -- [ quarter right, looseness=.8,  thick] (i2)  -- [ quarter right, looseness=.8 ,  thick] (i1) ,
     (i1) -- [ half right, looseness=1.3,  thick] (i2)  -- [ half right, looseness=1.3 ,  thick] (i1),
               (x1) -- [red, scalar, thick] (x2)
};
  \end{feynman}
\end{tikzpicture}
  \end{minipage} =    \frac{(g L)^2}{24} \sum_{\vec{k_i}}  \frac{L^2 \delta_{\vec k_1+\vec k_2+\vec k_3+\vec k_4}}{\underbrace{-(\om_{\vec k_1}+\om_{\vec k_2}+\om_{\vec k_2}+\om_{\vec k_4})}_{  \scriptsize \roig -E}} \, , \label{1diag}
\ee
where we have introduced the  Feynman  diagrammatic notation that we review later. 
Here and in  many examples below,  it will be convenient to write the sum over the intermediate states in \reef{1diag} in terms of the relativistic phase-space. 
At $d=2+1$, the   finite volume   phase space is given by
\be
\Phi_n (P^\mu)=\sum_{k's}   \frac{1  }{\prod_j 2L^2\om_{\vec k_j}}  L^2 \delta_{\Sigma_{i}\vec{k}_i, \vec P}  (2\pi)\delta(\Sigma_i \om_{\vec k_i}- E )\, ,  \label{ps1}\ee
where $P^\mu=(E,\vec{P})$. 
Its infinite volume limit can be  computed in $d=2+1$ and it is given by
\be
\Phi_n(\sqrt{s}) \ \stackrel{L\rightarrow \infty}{  \longrightarrow} \ \int_{-\infty}^\infty \prod_{s=1}^n\frac{ d^2k_s  }{  (2\pi)^2 2\om_{\vec{k}_s}}  
\delta^{(3)}(\Sigma_{i=1}^n k^\mu_i-P^\mu) =  \frac{\pi^{2-n}4^{1-n}}{(n-2)!} \frac{(\sqrt{s}-n\sqrt{m^2})^{n-2}}{\sqrt{s}} \, , \label{vacu22}
\ee
where $s=P_\mu^2$, see for instance \cite{Groote:1998wy}. 
Thanks to the closed form expression in \reef{vacu22} we will be able to analytically evaluate many  HPT diagrams with a large number of loops.  
Now, upon inserting $1= \int_{4m}^{E_T} dE \delta(E - \Sigma_ i \om_{k_i}) $ in \reef{1diag} we get
\be
  \reef{1diag} =   - \frac{(g L)^2}{24}  \int_{4m}^{E_T} \frac{dE}{2\pi}\frac{\Phi_4(E)}{E} \, ,  \label{vall}
\ee
where we have introduced  a regulator by cutting the maximal energy of the intermediate state $E_k\leq E_T$. 
More generally, regulating the theory with an energy cutoff $E_T$ -- so that 
all Fock space states have $H_0$-energies $E_i\leq E_T$ -- corresponds to requiring that the energy of the states between the vertices in \reef{dWe}  is $E_i\leq E_T$.
This can be easily implemented in the HPT diagrams by imposing that each of the energy propagators is bounded from below by $1/E_T$.~\footnote{This constraint on the propagator applies when calculating the vacuum energy. The generalisation when the propagator is shifted by non-zero external energy is straightforward.}
The value of  \reef{vall} in the infinite volume limit is straightforwardly computed using \reef{vacu22}.

As a second example, consider a mass perturbation, i.e. $k=2$. Then \reef{dWe}-\reef{wicks}  for $n=2$ gives
\be
 -\frac{i g^2}{2!^2}\int_0^\infty dt \int_{-L/2}^{+L/2} d^2x d^2 z \sum_{p=0}^2 s^2_{p}  D_F^{p}(\vec{z},t)\, : \phi^{2-p}_{(\vec{x}+\vec{z}, t/2) }\phi^{2-p}_{(\vec{x},-t/2)}: 
\ee
For concreteness let us compute the $p=1$ Wick's contraction and take the expectation value with the one particle state at rest $|m\rangle$, i.e. 
\be
- i \frac{g_2^2  L^2 }{2}  \int_0^\infty dt  \, e^{i t (\eps-m)} \int_{-L/2}^{+L/2}   d^2 z D_{(\vec z,t)} \, \langle m|:  \phi( \vec z,t ) \phi(0,0 ): |m\rangle   \label{basci3}
\ee
     where $\eps\rightarrow i 0_+$.
          Carrying out all the integrals of the previous expression we get 
          \vspace{-.3cm}
    \be
    \reef{basci3} =  \frac{g_2^2}{4 m^2} \underbrace{ \frac{1}{m- 3m}}_{\roig m-E_1}   + \frac{g_2^2}{4m^2} \frac{1}{\underbrace{m- m}_{\roig m-E_2} + \eps}  = 
    \begin{minipage}[h]{0.11\linewidth}
\begin{tikzpicture}
\begin{feynman}[small]
\vertex (i1) at (0,0);
 \node [dot] (i2) at (.5,.2);
  \node [dot] (i3) at (1.2,0);
 \vertex  (i4) at (1.7,.2);  
   \vertex (XX) at (.85,1.15);    
 \vertex (x1) at (.85,.5); 
 \vertex (x2) at (.85,-.7) {\roig \scriptsize $m-E_1$};
\diagram*{
          (XX) -- [white, scalar, thick] (x1),
   (i1) -- [  thick] (i3) -- [  thick] (i2)  -- [  thick] (i4) ,
           (x1) -- [red, scalar, thick] (x2)
};
  \end{feynman}
\end{tikzpicture}
  \end{minipage} 
          +  \
           \begin{minipage}[h]{0.07\linewidth}
\begin{tikzpicture}
\begin{feynman}[small]
\vertex (i1) at (0,0);
 \node [dot] (i2) at (.5,0);
  \node [dot] (i3) at (1.2,0);
 \vertex  (i4) at (1.7,0);    
    \vertex (XX) at (.85,1.);    
 \vertex (x1) at (.85,.5); 
 \vertex (x2) at (.85,-.7) {\roig \scriptsize $m-E_2$}; 
\diagram*{
               (XX) -- [white, scalar, thick] (x1),
   (i1) -- [  thick] (i2) -- [  thick] (i3)  -- [  thick] (i4) ,
        (x1) -- [red, scalar, thick] (x2),
};
  \end{feynman}
\end{tikzpicture}
  \end{minipage}  \, .  \label{vvv1}
    \ee
    If this was the calculation of   $V_{1k}E_{1k}^{-1}V_{k1}$ in \reef{basicpert}, the second diagram would not contribute and needs to be discarded because the energy of the state between the two vertices is $E_k=1$.
This is a particular instance of a general rule: when computing \reef{basicpert}  through the  correlation functions \reef{wicks} we need to discard those contributions in which the state being propagated between   two consecutive vertices is equal to $E_i$. 
Note that there are two possible vertex orderings in  \reef{vvv1} giving rise to two different expressions. 

Indeed the order of the vertices matters  because  the integrand  \reef{dWe} is time-ordered in the whole integration domain, thus diagrams differing by the  vertex ordering  give rise to different expressions. 
A more dramatic instance occurs in the following corrections  to the 1-particle state
\vspace{-.2cm}
  \bea
  \begin{minipage}[h]{0.115\linewidth}
\begin{tikzpicture}
\begin{feynman}[small]
 \vertex (i1) at (-.4,0);
   \node [dot] (i2) at (0,0);
 \node [dot] (i3) at (.7,0);
 \node [dot] (i4) at (1.4,0);  
 \vertex (i5) at (1.8, 0);
 \vertex (X1) at (.3,.5);
 \vertex (X2) at (.3,-.7) {\roig \tiny $m{-}E_1$};
 \vertex (Y2) at (1.1,-.7)  {\roig \tiny $\ \ m{-}E_2$};
 \vertex (Y1) at (1.1,.5);
 \vertex (g1) at (.5,.9);
  \vertex (g2) at (1.5,.9);
\diagram*{
   (i1) -- [thick]   (i4) ,
     (i3) -- [ half right, looseness=1.,  thick] (i4)  -- [ half right, looseness=1. ,  thick] (i3) , 
     (i2) --[ quarter right , thick, looseness=.9] (i5) ,
 (i4)    -- [half right, thick, looseness=.9] (i2),
 (X1) -- [scalar, thick, red] (X2),
 (Y1) -- [scalar, thick, red] (Y2) ,
 (g1) -- [white] (g2)
};
  \end{feynman}
\end{tikzpicture}
  \end{minipage}
 &=&   \frac{g^3}{6m}    \sum  \frac{L^2\delta_{q+p,0}}{\underbrace{m-( \om_{p}+\om_{q}+m)}_{\scriptsize \roig m-E_1}}   \frac{L^2\delta_{p+k_1+k_2+k_3,0}}{\underbrace{m-(\om_{p}+\om_{k_1}+\om_{k_2}+\om_{k_3}+m)}_{\scriptsize \roig m-E_2}}   \, ,  \label{exa3} \\[-.15cm]
   \begin{minipage}[h]{0.115\linewidth}
\begin{tikzpicture}
\begin{feynman}[small]
  \node [dot] (i2) at (0,0);
 \node [dot] (i3) at (.7,0);
 \node [dot] (i4) at (1.4,0);  
  \vertex (i7) at (1.8,0);
  \vertex (im1) at (-.4,0);
 \vertex (X1) at (.3,.5);
 \vertex (X2) at (.3,-.7) {\roig \tiny $m{-}E_1$};
 \vertex (Y2) at (1.1,-.7)  {\roig \tiny $\ \ m{-}E_2$};
 \vertex (Y1) at (1.1,.5);
 \vertex (g1) at (.5,.9);
  \vertex (g2) at (1.5,.9);
\diagram*{
   (i2) -- [thick]   (i4) ,
     (i2) -- [ quarter left, looseness=.4,  thick] (i4)   , 
      (i2) -- [ half left, looseness=.55,  thick] (i4)   , 
     (i3) --[ quarter right , thick, looseness=.7]  (i7) ,
 (i4)    -- [half right, thick, looseness=1] (i2) ,
   (i3) --[ quarter left , thick, looseness=.7]  (im1) ,
 (X1) -- [scalar, thick, red] (X2),
 (Y1) -- [scalar, thick, red] (Y2),
 (g1) -- [white] (g2)
};
  \end{feynman}
\end{tikzpicture}
  \end{minipage}
  & =&  \frac{g^3}{6m}    \sum   \frac{L^2\delta_{k_1+k_2+k_3+p,0}}{\underbrace{m-(\om_{k_1}+\om_{k_2}+\om_{k_3}+\om_{p}+m)}_{\roig \scriptsize m-E_1}}  \frac{L^2\delta_{k_1+k_2+k_3+q,0}}{\underbrace{ m-(\om_{k_1}+\om_{k_2}+\om_{k_3}+\om_{q}+m)}_{\scriptsize \roig m-E_2}}   \, , \ \  \ \   \ \  \label{exa4}  
      \eea
where for simplicity we defined
$
\sum \equiv \sum_{\substack{\text{momenta}\\ \text{s.t. }E_i \leq E_T} }\frac{1}{\prod_i 2\om_{i} L^2 } 
$
i.e. all internal lines of the diagrams are  summed over including relativistic factors $1/(2\om L^2)$
and under the constraint that the state $E_i$ flowing between any two consecutive vertices is smaller than the energy cutoff,  $E_i\leq E_T$. 
Even-though the diagram \reef{exa4} is a permutation of the vertices of diagram \reef{exa3}, it is apparent that the expression in Eq.~\reef{exa3} differs from Eq.~\reef{exa4}.
The latter is finite as $E_T\rightarrow \infty$, while diagram \reef{exa3} diverges logarithmically in the limit $E_T\rightarrow \infty$.  Indeed, 
\be
 \reef{exa3} = \frac{g^3}{6}  \sum \frac{L^2\delta_{q+p,0}}{ \om_{p}+\om_{q}}    \sum  \frac{L^2\delta_{k_1+k_2+k_3+p,0}}{\om_{k_1}+\om_{k_2}+\om_{k_3}+\om_{p}}     + O(m^2/E_T)  
\ee
where the second factor
 $
   \sum    \frac{L^2\delta_{k_1+{k}_2+{k}_3,0}}{\om_{k_1}+\om_{k_2}+\om_{k_3}}  \sim \log (E_T)
 $.
Further details and examples are given in appendix~\ref{HPTapp}.

For clarity,  in each of the  preceding diagrams  we have drawn vertical lines cutting between consecutive vertices to signal the state propagating.  We will not draw   such lines in what follows. However, one should remember that the vertices of all the diagrams below are ordered.

\subsection{Rules}
The  general rules  to compute the first term in   \reef{subte} can be summarised as follows. 
First, consider all possible Feynman diagrams for the transition $i\rightarrow i$, including both connected and disconnected diagrams. Now, draw each $n$th order  Feynman diagram $n!$ times ordering the $n$ vertices in every possible way, with the external lines  kept fixed. 
Label each internal line with space momenta $\vec p$. Then the rules associated to each particular vertex-ordered diagram are
\begin{enumerate}
\item[$\circ$]  For every vertex except the last [leftmost], include a factor $L^2$ times a momentum conservation Kronecker delta. 
\item[$\circ$]  For every intermediate state $j$, i.e. a set of lines between any to consecutive vertices, include a factor
\be
[E_i-E_j+i\eps]^{-1} \label{todr}
\ee
where $E_j\equiv \sum \om$ is the total energy of the state $j$.
\item[$\circ$]  For every internal line include a factor $[(L^{2}(2\om_{\vec p})]^{-1}$ while every external line counts a factor  $[(L^{2}(2\om_{\vec p})]^{-1/2}$.
\item[$\circ$]  Multiply by $g^n/n!$ and by the symmetry factor associated to the diagram. If in doubt, resort to the general formula given in \reef{sym}.
\item[$\circ$]  Integrate the product of these factors over all the internal momenta. 
\end{enumerate}
Finally, sum all the expressions associated to the vertex ordered diagrams. The rules to compute the pieces in the  subtraction terms are analogous, only differing in some extra powers of $\left(E_i-E_j+i\eps\right)^{-1}$ which are readily traced.   In order to get \reef{basicpert} we need to remember not to include the state $E_i$ in the sum over internal states, then the $\eps$ in \reef{todr} can be safely dropped. 
At infinite volume the internal lines are changed as $1/\left(L^{2}2\om_{\vec p}\right)\rightarrow 1/\left((2\pi)^{2}2\om_{\vec p}\right)$, the Kronecker deltas $L^2 \delta_{\vec k}\rightarrow (2\pi)^2\delta^{(2)}(\vec k )$ and the sums $\sum_{\vec k}\rightarrow \int d^2k$. 
The rules just described are similar to those to compute scattering matrix elements in Old Fashioned Perturbation Theory, see e.g. \cite{Weinberg:1966jm}.

\section{$\phi^2$ test} \label{secphi2}

In this section we apply Hamiltonian Truncation techniques  to a mass perturbation  $\phi^2$ in  $2+1$ dimensions. This perturbation is exactly solvable. It is nevertheless interesting to compare the exact solution with perturbation theory and HT since these latter two approaches are of much more general.  
In our detailed comparison  with perturbation theory we will learn a lesson that will be crucial when generalising the  HT method to the $\phi^4$ perturbation. 

\subsection{Analytic solution}

The free theory is given by  the action 
\be
S_0=\int d^3x \left(\frac{1}{2}(\partial \phi)^2 -\frac{1}{2} m^2 \phi^2 +\frac{ m^2}{2} Z\right) \, , \label{free1}
\ee
where $Z=\int \frac{d^3p}{(2\pi)^2} \frac{1}{p^2+m^2}$. The only potential divergence of the theory, which consists of a one loop diagram  with a single  mass insertion, is  canceled by the $m^2Z$ term. 
Thus, the vacuum energy of the theory in  \reef{free1} is zero and the excited states are the free theory Fock states $|E_i\rangle$, with energies $E_i$. 
Next we perturb the theory  by $m^2\rightarrow m^2+g_2$,  
\be
S_{g_2} =\int d^3x\left( \frac{1}{2}(\partial \phi)^2 -\frac{1}{2}M^2 \phi^2 + \frac{M^2}{2} Z \right) \, , \label{pert1}
\ee
with $M^2=m^2+g_2$.
In the perturbed theory \reef{pert1} the vacuum energy is non-zero and measurable.
Indeed, the vacuum energy density   is given by the effective potential  
\be
{\cal V}_\text{eff}(g_2) =   \frac{2}{24\pi} \Big[(m^2)^{3/2} -  (m^2+g_2)^{3/2}  +     \frac{3}{2} g_2 m \Big]  \, ,\label{veff}
\ee
where  \reef{veff}
 only receives contributions from a single loop, i.e. the Coleman-Weinberg \cite{Coleman:1973jx} potential ${\cal V}_\text{eff}(g_2)=\frac{1}{2} \int \frac{d^3p}{(2\pi)^2} \log (p^2+M^2) - M^2 Z$.
The excited states  are Fock space states with energy
\be
L^2 \, V_\text{eff}(g_2)+ E_i \, ,  \label{exst}
\ee
where $L^2$ is the volume of the space. Eq.~\reef{exst} is valid for large volumes, so that winding corrections can be ignored.~\footnote{We do however include winding corrections in our subsequent comparison with HT calculations, since they have a small but not negligible effect for the box sizes we use.}

\subsection{Perturbative solution}

Next we compute the first terms of the perturbative expansion of  $V_\text{eff}$ and of the mass gap $\Delta=\cE_1-\cE_0$ using  time ordered perturbation theory [for the calculations of this section we set $m=1$].
We will do the calculation up to $O(g_2^4)$ where the effect that we want to discuss first arises for the vacuum energy. The conclusions that we draw in this section are insensitive to whether we work in finite or infinite volume, so we use the infinite volume phase space formula throughout.

   \subsubsection{Vacuum}
   \label{anagphi2}
   The vacuum is given by
 $
   \cE_0 = \cE_0^{(2)} + \cE_0^{(3)} + \cE_0^{(4)}+ O(g_2^5)$. 
At $O(g_2^2)$ and $O(g_2^3)$ we have  the   contributions  
   \be
     \begin{minipage}[h]{0.07\linewidth}
\begin{tikzpicture}
\begin{feynman}[small]
 \node [dot] (i1);
 \node [dot, left = of i1] (i2);
 \vertex [  above right =.7cm of i2] (x1);
 \vertex [  below right = .7cm of i2] (x2);      
\diagram*{
   (i1) -- [ half left, looseness=.8,  thick] (i2) ,  
    (i2) -- [ half left, looseness=.8 ,  thick] (i1) ,
};
  \end{feynman}
\end{tikzpicture}
  \end{minipage}
 =  \frac{g_2^2}{2}  \int_2^\infty \frac{dE}{2\pi} \frac{\Phi_2(E)}{-E}  =- \frac{g_2^2}{32\pi} \quad \text{and} \quad 
     \begin{minipage}[h]{0.1\linewidth}
\begin{tikzpicture}
\begin{feynman}[small]
 \node [dot] (i1);
 \node [dot, left = .7cm of i1] (i2);
 \node [dot, left = .7cm  of i2] (i3);
 \vertex [  above right =.7cm of i2] (x1);
 \vertex [  below right =.7cm of i2] (x2);    
 \vertex [  above right =.7cm of i3] (x3);
 \vertex [  below right =.7cm of i3] (x4);      
\diagram*{
   (i1) -- [ thick] (i2) ,  
    (i2) -- [ thick] (i3) ,
        (i3) -- [ half left, looseness=.8 ,  thick] (i1) ,
};
  \end{feynman}
\end{tikzpicture}
  \end{minipage}
 =  \int_2^\infty \frac{dE}{2\pi} \frac{\Phi_2(E)}{E\cdot E^2}  =\frac{g_2^3}{192\pi } \, . \label{I2}
   \ee
At $O(g_2^4)$ we have three connected diagrams
\be
     \begin{minipage}[h]{0.1 \linewidth}
\begin{tikzpicture}
\begin{feynman}[small]
 \node [dot] (i1);
 \node [dot, right = .5cm of i1] (i2);
 \node [dot, right = .5cm of i2] (i3); 
 \node [dot, right  = .5cm of i3] (i4); 
\diagram*{
   (i1) --[   thick] (i2) ,  
    (i2) -- [  thick] (i3) ,
    (i3) --[  thick]  (i4) ,
        (i4) -- [ half right , looseness=.8,  thick] (i1) ,
};
  \end{feynman}
\end{tikzpicture}
  \end{minipage}
 +\ 
     \begin{minipage}[h]{0.1\linewidth}
\begin{tikzpicture}
\begin{feynman}[small]
 \node [dot] (i1);
 \node [dot, right = .5cm of i1] (i2);
 \node [dot, right = .5cm of i2] (i3); 
 \node [dot, right  = .5cm of i3] (i4); 
\diagram*{
   (i1) -- [ thick] (i2) ,  
    (i3) -- [thick] (i4) ,
        (i2) -- [ half right , looseness=.8,  thick] (i4) ,
        (i1) -- [ half left , looseness=.8,  thick] (i3) ,
};
  \end{feynman}
\end{tikzpicture}
  \end{minipage}
 + \ 
     \begin{minipage}[h]{0.1\linewidth}
\begin{tikzpicture}
\begin{feynman}[small]
 \node [dot] (i1);
 \node [dot, right = .5cm of i1] (i2);
 \node [dot, right = .5cm of i2] (i3); 
 \node [dot, right  = .5cm of i3] (i4); 
\diagram*{
   (i2) -- [thick] (i3) ,  
        (i1) -- [ half left , looseness=.8,  thick] (i3) ,
                (i1) -- [ half right , looseness=.8,  thick] (i4) ,
                (i2) -- [ quarter right , looseness=.8,  thick] (i4) ,
};
  \end{feynman}
\end{tikzpicture}
  \end{minipage}
 =-\Big(1+1+\frac{1}{2}\Big)  \int_2^\infty \frac{dE}{2\pi} \frac{1}{E^2 \cdot  E^3} =- \frac{g_2^4}{512 \pi} \, ,   \label{I3}  
   \ee
and two disconnected diagrams
\be  
     \begin{minipage}[h]{0.1 \linewidth}
\begin{tikzpicture}
\begin{feynman}[small]
 \node [dot] (i1);
 \vertex [right = .5cm of i1] (i2);
 \vertex [ right = .5cm of i2] (i3); 
 \node [dot, right  = .5cm of i3] (i4); 
  \vertex [ above =.5cm of i1] (j1);
 \node [dot, right = .4cm of j1] (j2);
 \node [dot, right = .7cm of j2] (j3); 
 \vertex [ right  = .4cm of j3] (j4); 
\diagram*{
        (i4) -- [ half right , looseness=.3,  thick] (i1) ,
        (i4) -- [ half left , looseness=.3,  thick] (i1) ,
        (j2) -- [ half right , looseness=.6,  thick] (j3) ,
        (j3) -- [ half right , looseness=.6,  thick] (j2) ,
};
  \end{feynman}
\end{tikzpicture}
  \end{minipage}
\  + \ \, 
     \begin{minipage}[h]{0.1 \linewidth}
\begin{tikzpicture}
\begin{feynman}[small]
 \node [dot] (i1);
 \vertex [ right = .5cm of i1] (i2);
 \node [dot, right = .5cm of i2] (i3); 
 \vertex [ right  = .5cm of i3] (i4); 
  \vertex [ above =.5cm of i1] (j1);
 \node [dot, right = .5cm of j1] (j2);
 \vertex [ right = .5cm of j2] (j3); 
 \node [dot, right  = .5cm of j3] (j4); 
\diagram*{
        (i3) -- [ half right , looseness=.5,  thick] (i1) ,
        (i3) -- [ half left , looseness=.5,  thick] (i1) ,
        (j4) -- [ half right , looseness=.5,  thick] (j2) ,
        (j2) -- [ half right , looseness=.5,  thick] (j4) ,
};
  \end{feynman}
\end{tikzpicture}
  \end{minipage}
  = - \frac{1}{4}  \int_2^\infty \frac{dE_1}{2\pi}     \frac{dE_2}{2\pi}   \frac{\Phi_2(E_1)}{E_1+E_2} \frac{\Phi_2(E_2)}{E_2} \left(\frac{1}{E_1}+\frac{1}{E_2} \right)  = -  \frac{g_2^4}{256(4 \pi)^2}  \, ,  \label{I1} 
   \ee
all arising from the $V H_0^{-1}V H_0^{-1}V H_0^{-1}V$ piece. Meanwhile the second term of $\cE^{(4)}_0$ in \reef{basicpert} gives a fully disconnected piece  
  \be
   -\cE_0^{(2) } \, V_{0k}E_k^{-2}V_{k0}= \frac{g_2^2}{2}\int_2^\infty \frac{dE}{2\pi} \frac{\Phi_2(E)}{E}  \frac{g_2^2}{2} \int_2^\infty \frac{dE}{2\pi} \frac{\Phi_2(E)}{E^2} = \frac{g_2^4}{256(4 \pi)^2} \, .   \label{I6}
   \ee
As expected, the disconnected pieces cancel, $
\reef{I1}+\reef{I6}=0$; and adding up all the corrections we  reproduce  \reef{veff}
\be
L^2{\cal V}_\text{eff}=   - \frac{g_2^2}{32\pi} +\frac{g_2^3}{192\pi } - \frac{g_2^4}{512 \pi} + O(g_2^5)  \, . 
\ee

The point of this exercise is to note that if we  set up a sharp $E_T$ cutoff on the energy of the states propagating in between the vertices, Eq.~\reef{I1} is now given by
\be
-\frac{1}{4}  \int_2^{E_T-2} \frac{dE_2}{2\pi}\frac{\Phi_2(E_2)}{E_2^2} \int_2^{E_T-E_2} \frac{dE_1}{2\pi}  \frac{\Phi_2(E_1)}{E_1} \label{I1et}
\ee
while the integrals of  \reef{I6}  are cut off independently
\be
 \frac{1}{2}\int_2^{E_T} \frac{dE}{2\pi} \frac{\Phi_2(E)}{E}  \frac{1}{2} \int_2^{E_T} \frac{dE}{2\pi} \frac{\Phi_2(E)}{E^2}  \, .  \label{I6et}
\ee
Therefore, when the $E_T$ cutoff is introduced, the disconnected diagrams do not cancel  by terms of $O(1/E_T)$,
\be
\reef{I1et}+\reef{I6et} =  \frac{1}{512\pi^2} \frac{1}{E_T^2} \,  .  \label{toinc1}
\ee
 This effect is harmless in practice for this theory because it decouples fast enough. 
   However, these kind of effects will turn out to be very important for the $\phi^4$ theory [or generic UV divergent theories] because for $\phi^4$ theory the non-cancelation of disconnected bubbles is not suppressed by powers of $E_T$. 
   This will be discussed in detail in section~\ref{pertUV}.

   \subsubsection{First excited state}
  
To develop our intuition further, we now carry out the analogous calculation of the energy of the first excited state.  
   We start by computing the connected diagrams, and  then we compute the disconnected diagrams  contributing to $\cE_1$.

   \underline{Connected diagrams}. 
   The leading correction to the first excited state $\cE_1$ is given by $\cE_1^{(1)}=
   V_{11} = 
   \begin{minipage}[h]{0.05\linewidth}
\begin{tikzpicture}
\begin{feynman}[small]
\vertex (i1) at (0,0);
\node [dot] (i2) at (.4,0);
 \vertex  (i3) at (.8,0);
\diagram*{
   (i1) -- [  thick] (i2) -- [  thick] (i3) ,  
};
  \end{feynman}
\end{tikzpicture}
  \end{minipage} 
  =
   \frac{g_2}{2}
$. Next,  at  $O(g_2^2)$,  
      \be
       \cE_1^{(2)}|_\text{Conn.} = 
       \begin{minipage}[h]{0.11\linewidth}
\begin{tikzpicture}
\begin{feynman}[small]
\vertex (i1) at (0,0);
 \node [dot] (i2) at (.5,.2);
  \node [dot] (i3) at (1.2,0);
 \vertex  (i4) at (1.7,.2);    
\diagram*{
   (i1) -- [  thick] (i3) -- [  thick] (i2)  -- [  thick] (i4) ,  
};
  \end{feynman}
\end{tikzpicture}
  \end{minipage} 
        = -   \frac{g_2^2}{8}  \label{befcite}\, ,
      \ee
meanwhile at  $O(g_2^3)$ we find 
      \be
      \cE_1^{(3)}|_\text{Conn.} = 
      \begin{minipage}[h]{0.14\linewidth}
\begin{tikzpicture}
\begin{feynman}[small]
\vertex (i1) at (0,0);
\node [dot] (i2) at (.5,.3);
 \node [dot] (i3) at (1.2,.15);
 \node [dot] (i4) at (1.9,.0);
  \vertex  (i5) at (2.4,.3);   
\diagram*{
   (i1) -- [  thick] (i4) -- [  thick] (i3)  -- [thick] (i2)  -- [  thick] (i5) ,  
};
  \end{feynman}
\end{tikzpicture}
  \end{minipage}
   +
     \begin{minipage}[h]{0.14\linewidth}
\begin{tikzpicture}
\begin{feynman}[small]
\vertex (i1) at (0,0);
\node [dot] (i2) at (.5,.15);
 \node [dot] (i3) at (1.2,.3);
 \node [dot] (i4) at (1.9,.0);
  \vertex  (i5) at (2.4,.3);   
\diagram*{
   (i1) -- [  thick] (i4) -- [  thick] (i2)  -- [thick] (i3)  -- [  thick] (i5) ,  
};
  \end{feynman}
\end{tikzpicture}
  \end{minipage} 
   +
     \begin{minipage}[h]{0.14\linewidth}
\begin{tikzpicture}
\begin{feynman}[small]
\vertex (i1) at (0,0);
\node [dot] (i2) at (.5,.3);
 \node [dot] (i3) at (1.2,0);
 \node [dot] (i4) at (1.9,.15);
  \vertex  (i5) at (2.4,.3);   
\diagram*{
   (i1) -- [  thick] (i3) -- [  thick] (i4)  -- [thick] (i2)  -- [  thick] (i5) ,  
};
  \end{feynman}
\end{tikzpicture}
  \end{minipage} 
   =    3 \frac{g_2^3}{8}    \frac{1}{(1-3)^2}     =   3 \frac{g_2^3}{32  } \label{ttcite2}
      \ee
Finally, there  are several connected contributions at $O(g^4)$: 
   \bea
  &&    \begin{minipage}[h]{0.175\linewidth}
\begin{tikzpicture}
\begin{feynman}[small]
\vertex (i1) at (0,0);
\node [dot] (i2) at (.5,.6);
 \node [dot] (i3) at (1.2,.4);
 \node [dot] (i4) at (1.9,.2);
  \node [dot]  (i5) at (2.6,0);   
  \vertex (i6) at (3.1,.6);   
\diagram*{
   (i1) -- [  thick] (i5) -- [  thick] (i4)  -- [thick] (i3)  -- [  thick] (i2) -- [  thick] (i6) ,  
};
  \end{feynman}
\end{tikzpicture}
  \end{minipage} 
  +       
  \begin{minipage}[h]{0.175\linewidth}
\begin{tikzpicture}
\begin{feynman}[small]
\vertex (i1) at (0,0);
\node [dot] (i2) at (.5,.6);
 \node [dot] (i3) at (1.2,.2);
 \node [dot] (i4) at (1.9,.4);
  \node [dot]  (i5) at (2.6,0);   
  \vertex (i6) at (3.1,.6);   
\diagram*{
   (i1) -- [  thick] (i5) -- [  thick] (i3)  -- [thick] (i4)  -- [  thick] (i2) -- [  thick] (i6) ,  
};
  \end{feynman}
\end{tikzpicture}
  \end{minipage} 
 +
  2 \begin{minipage}[h]{0.175\linewidth}
\begin{tikzpicture}
\begin{feynman}[small]
\vertex (i1) at (0,0);
\node [dot] (i2) at (.5,.2);
 \node [dot] (i3) at (1.2,.6);
 \node [dot] (i4) at (1.9,.4);
  \node [dot]  (i5) at (2.6,0);   
  \vertex (i6) at (3.1,.6);   
\diagram*{
   (i1) -- [  thick] (i5) -- [  thick] (i2)  -- [thick] (i4)  -- [  thick] (i3) -- [  thick] (i6) ,  
};
  \end{feynman}
\end{tikzpicture}
  \end{minipage} 
  \nonumber \\[.2cm]
  &&+
 2     \begin{minipage}[h]{0.175\linewidth}
\begin{tikzpicture}
\begin{feynman}[small]
\vertex (i1) at (0,0);
\node [dot] (i2) at (.5,.4);
 \node [dot] (i3) at (1.2,.6);
 \node [dot] (i4) at (1.9,.2);
  \node [dot]  (i5) at (2.6,0);   
  \vertex (i6) at (3.1,.6);   
\diagram*{
   (i1) -- [  thick] (i5) -- [  thick] (i4)  -- [thick] (i2)  -- [  thick] (i3) -- [  thick] (i6) ,  
};
  \end{feynman}
\end{tikzpicture}
  \end{minipage} 
  \nonumber
+
   2   \begin{minipage}[h]{0.175\linewidth}
\begin{tikzpicture}
\begin{feynman}[small]
\vertex (i1) at (0,0);
\node [dot] (i2) at (.5,.2);
 \node [dot] (i3) at (1.2,.4);
 \node [dot] (i4) at (1.9,.6);
  \node [dot]  (i5) at (2.6,0);   
  \vertex (i6) at (3.1,.6);   
\diagram*{
   (i1) -- [  thick] (i5) -- [  thick] (i2)  -- [thick] (i3)  -- [  thick] (i4) -- [  thick] (i6) ,  
};
  \end{feynman}
\end{tikzpicture}
  \end{minipage} +
    2  \begin{minipage}[h]{0.175\linewidth}
\begin{tikzpicture}
\begin{feynman}[small]
\vertex (i1) at (0,0);
\node [dot] (i2) at (.5,.4);
 \node [dot] (i3) at (1.2,.2);
 \node [dot] (i4) at (1.9,.6);
  \node [dot]  (i5) at (2.6,0);   
  \vertex (i6) at (3.1,.6);   
\diagram*{
   (i1) -- [  thick] (i5) -- [  thick] (i3)  -- [thick] (i2)  -- [  thick] (i4) -- [  thick] (i6) ,  
};
  \end{feynman}
\end{tikzpicture}
  \end{minipage} 
  \nonumber
  \\[.2cm]
  & & +
        \begin{minipage}[h]{0.175\linewidth}
\begin{tikzpicture}
\begin{feynman}[small]
\vertex (i1) at (0,0);
\node [dot] (i2) at (.5,.2);
 \node [dot] (i3) at (1.2,.6);
 \node [dot] (i4) at (1.9,0);
  \node [dot]  (i5) at (2.6,.4);   
  \vertex (i6) at (3.1,.6);   
\diagram*{
   (i1) -- [  thick] (i4) -- [  thick] (i2)  -- [thick] (i5)  -- [  thick] (i3) -- [  thick] (i6) ,  
};
  \end{feynman}
\end{tikzpicture}
  \end{minipage} 
  +
        \begin{minipage}[h]{0.175\linewidth}
\begin{tikzpicture}
\begin{feynman}[small]
\vertex (i1) at (0,0);
\node [dot] (i2) at (.5,.4);
 \node [dot] (i3) at (1.2,.6);
 \node [dot] (i4) at (1.9,0);
  \node [dot]  (i5) at (2.6,.2);   
  \vertex (i6) at (3.1,.6);   
\diagram*{
   (i1) -- [  thick] (i4) -- [  thick] (i5)  -- [thick] (i2)  -- [  thick] (i3) -- [  thick] (i6) ,  
};
  \end{feynman}
\end{tikzpicture}
  \end{minipage}   
      \begin{minipage}[h]{0.175\linewidth}
\begin{tikzpicture}
\begin{feynman}[small]
\vertex (i1) at (0,0);
\node [dot] (i2) at (.5,.4);
 \node [dot] (i3) at (1.2,.0);
 \node [dot] (i4) at (1.9,.6);
  \node [dot]  (i5) at (2.6,.2);   
  \vertex (i6) at (3.1,.6);   
\diagram*{
   (i1) -- [  thick] (i3) -- [  thick] (i5)  -- [thick] (i2)  -- [  thick] (i4) -- [  thick] (i6) ,  
};
  \end{feynman}
\end{tikzpicture}
  \end{minipage}  \, . 
  \eea
It is straightforward to compute these diagrams and we are led to
\be
 \cE_1^{(4)}|_\text{Conn.} =  -g_2^4 \frac{11}{128} \label{treeconc}
\ee
   
   \noindent      \underline{Disconnected diagrams}.  There is a single diagram at $O(g_2^2)$ and it is equal to the $O(g_2^2)$ vacuum correction.
At $O(g_2^3)$,  we have various kinds of pieces arising from $V_{1k}E_{1k}^{-1}V_{kk^\prime}E^{-1}_{1k^\prime}V_{k^\prime 1}$. 
 The first type is
 \be
 \begin{minipage}[h]{0.1\linewidth}
\begin{tikzpicture}
\begin{feynman}[small]
\vertex (i1) at (0,0);
\node [blob , opacity = .35, radius=1.5cm] (i2) at (.8,.5) {\scriptsize{$\cE_0^{(3)}$}} ;
\node [blob, fill=none ] (i4) at (.8,.5) {\scriptsize{$\cE_0^{(3)}$}} ;
 \vertex  (i3) at (1.6,0);
\diagram*{
   (i1) -- [  thick] (i3) ,  
};
  \end{feynman}
\end{tikzpicture}
  \end{minipage}     \, , 
 \ee
where the blob is fully connected and given in \reef{I2}.
Then we can also have 
 \be
  \begin{minipage}[h]{0.1\linewidth}
\begin{tikzpicture}
\begin{feynman}[small]
\vertex (i1) at (0,0);
\node [dot] (i2) at (.8,0) ;
 \vertex  (i3) at (1.6,0);
\node [dot] (j1) at (.4,.5) ;
\node [dot] (j2) at (1.2,.5) ;
\diagram*{
   (i1) -- [  thick] (i3) ,  
   (j1) -- [ half left, looseness=.8,  thick] (j2)  -- [ half left, looseness=.8 ,  thick] (j1) ,
};
  \end{feynman}
\end{tikzpicture}
  \end{minipage} 
      = \frac{g_2}{2}\frac{g_2^2}{2 } \int_2^{\infty} \frac{dE}{2\pi} \frac{\Phi_2(E)}{E^2}    = \frac{g_2^4}{256\pi}  \label{ttcite} \, ,
 \ee
  which cancels with the  disconnected 1-loop bubble of
  \be
    - V_{11}  \times V_{1k}(m-E_k)^{-2}V_{k1} =  -\frac{g_2}{2} \times \Bigg(   \underbrace{     \frac{g_2^2}{4 } \frac{1}{(1- 3)^2} }_\text{tree}+  \underbrace{\frac{g_2^2}{2 } \int_2^{\infty} \frac{dE}{2\pi} \frac{\Phi_2(E)}{E^2}}_{1-\text{loop}}\Bigg) \, , \label{disc3} 
    \ee
    while the tree-level piece of the former equation combines with \reef{ttcite2} to give $-g_2^3/16$.

At  $O(g_2^4)$ the term $V \left[ 1/(m-H_0) V \right]^3$ gives rise to two types of disconnected contribution.    
   \be
  \begin{minipage}[h]{0.1\linewidth}
\begin{tikzpicture}
\begin{feynman}[small]
\vertex (i1) at (0,0);
\node [blob , opacity = .35, radius=1.5cm] (i2) at (.8,.5) {\scriptsize{$\cE_0^{(4)}$}} ;
\node [blob, fill=none ] (i4) at (.8,.5) {\scriptsize{$\cE_0^{(4)}$}} ;
 \vertex  (i3) at (1.6,0);
\diagram*{
   (i1) -- [  thick] (i3) ,  
};
  \end{feynman}
\end{tikzpicture}
  \end{minipage}       \, ,  \label{incpart}
 \ee
  where the bubble is fully connected $O(g^4)$ vacuum  given in \reef{I3}  --  once the  fully disconnected contribution  from   $-\cE_1^{(2) } \, V_{1k}E_k^{-2}V_{k1}$ [proportional to $\pi^{-2}$] are added up.
Meanwhile  there are a number  disconnected  diagrams of the second type
\be
  \begin{minipage}[h]{0.12\linewidth}
\begin{tikzpicture}
\begin{feynman}[small]
\vertex (i1) at (0,0);
 \node [dot] (i2) at (.4,.2);
  \node [dot] (i3) at (1.6,0);
 \vertex  (i4) at (2.,.2);     
 \node [dot] (j1) at (.8,.5) ;
\node [dot] (j2) at (1.2,.5) ;
\diagram*{
   (i1) -- [  thick] (i3) -- [  thick] (i2)  -- [  thick] (i4) ,  
     (j1) -- [ half left, looseness=.8,  thick] (j2)  -- [ half left, looseness=.8 ,  thick] (j1) ,
};
  \end{feynman}
\end{tikzpicture}
  \end{minipage} 
  +
     \begin{minipage}[h]{0.12\linewidth}
\begin{tikzpicture}
\begin{feynman}[small]
\vertex (i1) at (0,0);
 \node [dot] (i2) at (.8,.2);
  \node [dot] (i3) at (1.6,0);
 \vertex  (i4) at (2.,.2);     
 \node [dot] (j1) at (.4,.5) ;
\node [dot] (j2) at (1.2,.5) ;
\diagram*{
   (i1) -- [  thick] (i3) -- [  thick] (i2)  -- [  thick] (i4) ,  
     (j1) -- [ half left, looseness=.35,  thick] (j2)  -- [ half left, looseness=.35 ,  thick] (j1) ,
};
  \end{feynman}
\end{tikzpicture}
  \end{minipage} 
+
     \begin{minipage}[h]{0.12\linewidth}
\begin{tikzpicture}
\begin{feynman}[small]
\vertex (i1) at (0,0);
 \node [dot] (i2) at (.8,.2);
  \node [dot] (i3) at (1.2,0);
 \vertex  (i4) at (2.,.2);     
 \node [dot] (j1) at (.4,.5) ;
\node [dot] (j2) at (1.6,.5) ;
\diagram*{
   (i1) -- [  thick] (i3) -- [  thick] (i2)  -- [  thick] (i4) ,  
     (j1) -- [ half left, looseness=.2,  thick] (j2)  -- [ half left, looseness=.2 ,  thick] (j1) ,
};
  \end{feynman}
\end{tikzpicture}
  \end{minipage}
   +
        \begin{minipage}[h]{0.12\linewidth}
\begin{tikzpicture}
\begin{feynman}[small]
\vertex (i1) at (0,0);
 \node [dot] (i2) at (.4,.2);
  \node [dot] (i3) at (1.2,0);
 \vertex  (i4) at (2.,.2);     
 \node [dot] (j1) at (.8,.5) ;
\node [dot] (j2) at (1.6,.5) ;
\diagram*{
   (i1) -- [  thick] (i3) -- [  thick] (i2)  -- [  thick] (i4) ,  
     (j1) -- [ half left, looseness=.35,  thick] (j2)  -- [ half left, looseness=.35 ,  thick] (j1) ,
};
  \end{feynman}
\end{tikzpicture}
  \end{minipage}
    +
          \begin{minipage}[h]{0.12\linewidth}
\begin{tikzpicture}
\begin{feynman}[small]
\vertex (i1) at (0,0);
 \node [dot] (i2) at (.8,0);
  \node [dot] (i3) at (1.2,0);
 \vertex  (i4) at (2.,0);     
 \node [dot] (j1) at (.4,.3) ;
\node [dot] (j2) at (1.6,.3) ;
\diagram*{
   (i1) -- [  thick] (i2) -- [  thick] (i3)  -- [  thick] (i4) ,  
     (j1) -- [ half left, looseness=.2,  thick] (j2)  -- [ half left, looseness=.2 ,  thick] (j1) ,
};
  \end{feynman}
\end{tikzpicture}
  \end{minipage}
  +%
   \begin{minipage}[h]{0.12\linewidth}
\begin{tikzpicture}
\begin{feynman}[small]
 \node [dot] (i1) at (0,0);
 \node [dot, right = .5cm of i1] (i2);
 \node [dot, right = .5cm  of i2] (i3);  
 \vertex (j1) at (-.5,-.2);
 \node [dot] (j2) at (.25,-.2);
 \vertex (j3) at (1.5,-.2);
\diagram*{
   (i1) -- [ thick] (i2) ,  
    (i2) -- [ thick] (i3) ,
        (i3) -- [ half right, looseness=.8 ,  thick] (i1) ,
        (j1) -- [ thick] (j2) -- [ thick] (j3) ,
};
  \end{feynman}
\end{tikzpicture}
  \end{minipage} 
  +
      \begin{minipage}[h]{0.12\linewidth}
\begin{tikzpicture}
\begin{feynman}[small]
 \node [dot] (i1) at (0,0);
 \node [dot, right = .5cm of i1] (i2);
 \node [dot, right = .5cm  of i2] (i3);  
 \vertex (j1) at (-.5,-.2);
 \node [dot] (j2) at (.75,-.2);
 \vertex (j3) at (1.5,-.2);
\diagram*{
   (i1) -- [ thick] (i2) ,  
    (i2) -- [ thick] (i3) ,
        (i3) -- [ half right, looseness=.8 ,  thick] (i1) ,
        (j1) -- [ thick] (j2) -- [ thick] (j3) ,
};
  \end{feynman}
\end{tikzpicture}
  \end{minipage}
  \ , \label{ldiags}
\ee
 given by
\be
 - g_2^4  \int_{2}^{\infty} \frac{dE  }{2\pi}\Phi_2(E)  \left( \frac{1}{8}\left[ {\roig \frac{1}{4}  \frac{1 }{ 2+E} +  \frac{1}{2E}   \frac{1}{ 2+E} }  +  { \blu\frac{1}{E^2}  \frac{1}{ 2+E}   +  \frac{1}{2E}    \frac{1}{ 2+E}} + \dblu{ \frac{1}{2 E^3}}\right]+{\gray \frac{1}{E^4}}\right)    = - \frac{ 17  g_2^4}{3072 \pi}  \, , \nonumber
     \ee
     where each summand in the $[\cdots]$ piece of the integrand corresponds to the five first diagrams in  \reef{ldiags}; while the last two in \reef{ldiags} correspond to the $1/E^4$ piece.
 Next we need to compute the fully disconnected pieces, there are three of them [the last three pieces of $\cE_i^{(4)}$ in \reef{basicpert}]:
\bea
- \cE_1^{(2)} \times [V_{1k}(1-E_k)^{-2}V_{k1}] &=&   \left[     \frac{g_2^2}{8}+ {\roig \frac{g_2^2}{32 \pi} } \right] \times \left[     \frac{g_2^2}{16}+  \blu{\frac{g_2^2}{128 \pi}}\right] \, , \label{dcanc1} \\
-2 V_{11}   \times V_{1k}(1-E_k)^{-2}V_{kk^\prime }  (1-E_{k^\prime})^{-1}  V_{k^\prime 1} &=&   g_2 \times  \left[3\frac{g_2^3}{64}  +{\dblu\frac{g_2^3}{768\pi}}+  {\gray\frac{g_2^3}{512\pi }} \right] \, \label{dcanc2}\\
 V_{11}^2   \times  V_{1k}(1-E_k)^{-3} V_{k1} &=&  -  \frac{g_2^2}{4} \times \left[  \frac{g_2^2}{32}+{\dblu \frac{g_2^2}{384 \pi} }\right] \, .  \label{dcanc3}
 \eea
Thus, we  have that 
 \be
 \reef{dcanc1}+ \reef{dcanc2}+ \reef{dcanc3} = \underbrace{\frac{3}{64}}_\text{ tree}+\underbrace{\frac{17}{3072\pi}}_{1-\text{loop}}+\underbrace{\frac{1}{4096\pi^2}}_{2-\text{loop}}   \, .
 \ee
The two loop piece has  already been accounted for in \reef{incpart};  the one-loop piece cancels against  \reef{ldiags} [we have  coloured terms that cancel each other]; while the tree-level piece combines with the connected tree-level diagrams in \reef{treeconc} to give $
\cE_1^{(4)}|_\text{tree} =  -g_2^4 \frac{5}{128} $. This again shows the importance of the subtraction terms in \reef{subte}. 

Adding all the contributions, we have 
 \be
 \cE_1 =
 m + \frac{g_2}{2m}-\frac{g_2^2}{8m^3} +\frac{g_2^3}{16m^5} -\frac{5 g_2^4}{128m^7}
   +  
  \begin{minipage}[h]{0.1\linewidth}
\begin{tikzpicture}
\begin{feynman}[small]
\vertex (i1) at (0,0);
\node [blob ] (i4) at (.8,.4)  ;
 \vertex  (i3) at (1.6,0);
\diagram*{
   (i1) -- [  thick] (i3) ,  
};
  \end{feynman}
\end{tikzpicture}
  \end{minipage}  + O(g^5),
 \ee
 where the blob involves fully connected vacuum diagrams.

The cancellation of disconnected bubbles is expected from Lorentz covariant perturbation theory. It is however interesting to note that in the non-covariant calculation, the cancellation of the disconnected bubbles  appears  non-trivial. 
Additionally, although  disconnected bubble diagrams cancel out as $E_T\rightarrow \infty$, note that in HPT not all disconnected diagrams cancel. Indeed, the tree-level pieces of $O(g_2^3)$ and $O(g_2^4)$  involve disconnected tree-level diagrams [this is also expected since we are computing the mass gap instead of the mass-gap squared].

All in all,   the mass gap is given by   
   \be
   \Delta \equiv \cE_1-\cE_0  =   m + \frac{g_2}{2m}-\frac{g_2^2}{8m^3} +\frac{g_2^3}{16m^5} -\frac{5 g_2^4}{128m^7}+ O(g_2^5) = \sqrt{m^2+g_2} + O(g_2^5)   \, ,  \label{pertmass}
   \ee  
which of course reproduces  the result of a Lorentz covariant  calculation.
Our derivation of \reef{pertmass} with HPT is a very  inefficient way to solve for the harmonic oscillator! 
    However, Hamiltonian Truncation [which will require a detailed understanding of the cancelation of the  disconnected bubbles]  will turn out to be a very efficient way to solve for other theories that we do not know how to solve analytically.

\subsection{Hamiltonian Truncation solution}

\begin{figure}[t]\begin{center}
\includegraphics[height=5.2cm]{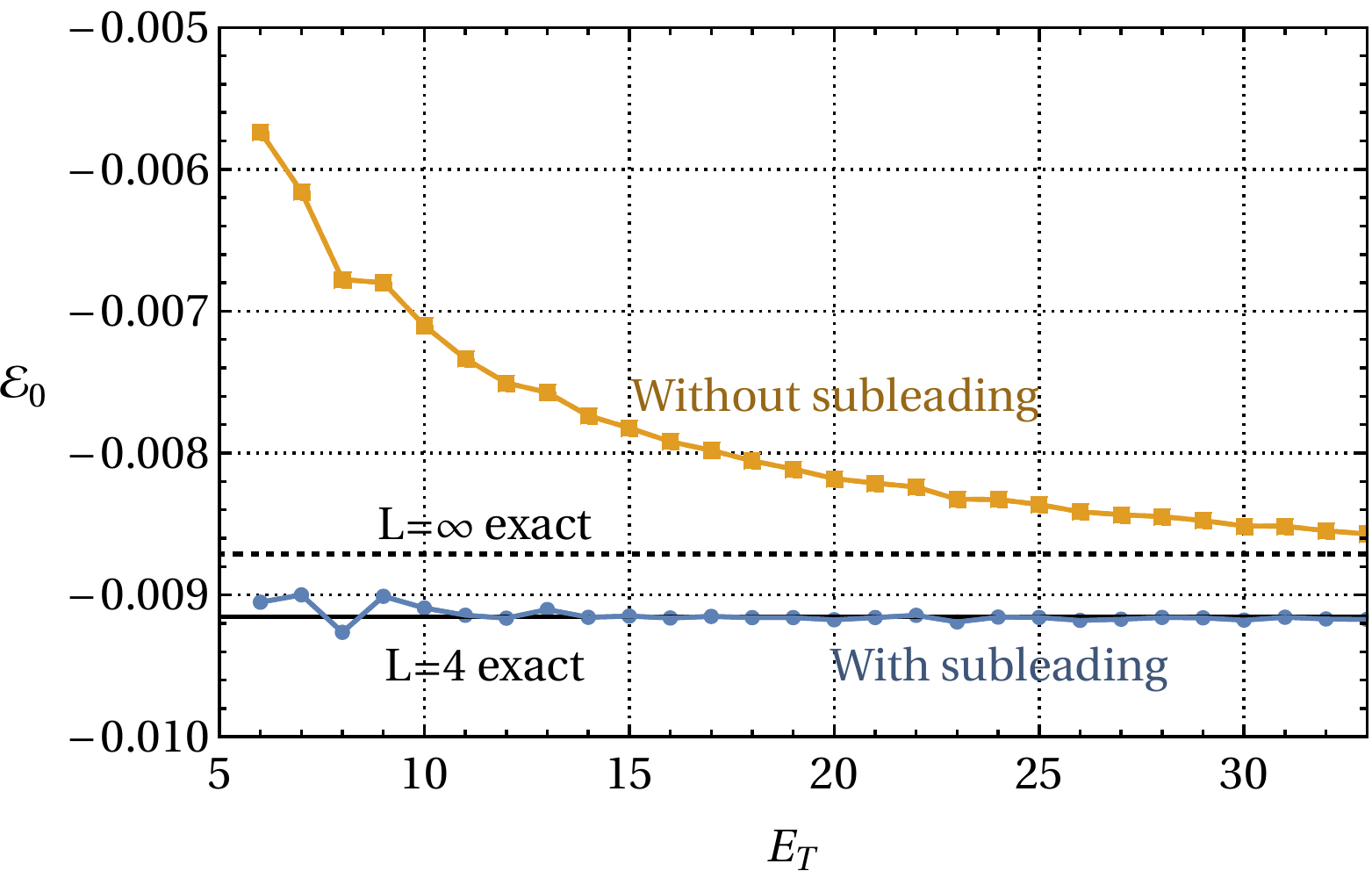}\qquad
\includegraphics[height=5.2cm]{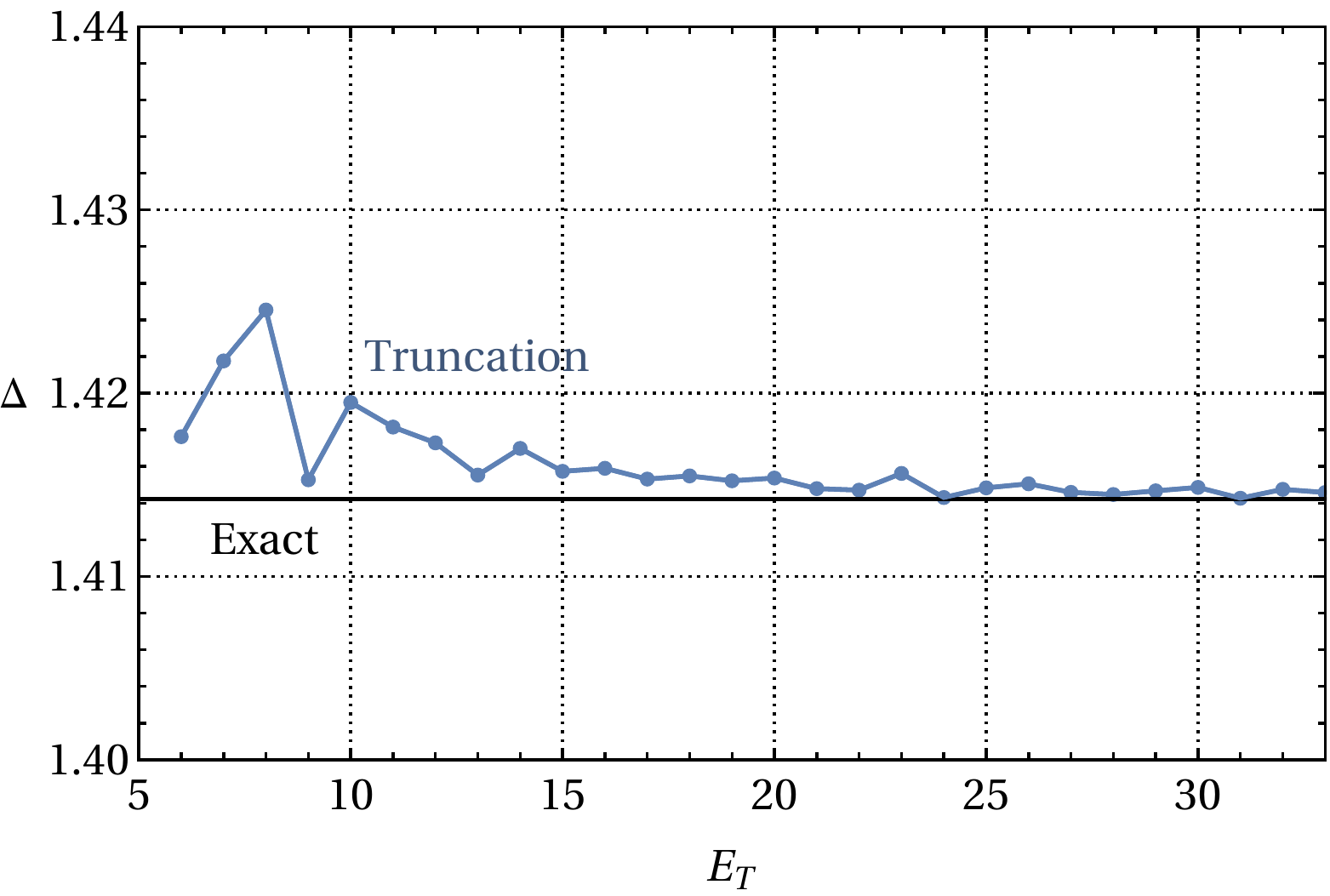} \end{center}
\caption{The vacuum energy density [left] and mass gap [right] of the theory defined by Eq.~\eqref{free1} deformed by Eq.~\eqref{eq:phi2def} with $g_2=1$ as a function of the cutoff energy $E_T$. The analytic predictions both in the infinite volume limit and at finite volume are also indicated. The results for the vacuum energy are shown with and without including the sub-leading correction \reef{phi2sublead}, which improves the convergence.}
\label{fig:phi21}
\end{figure}

As well as providing a useful theoretical warm up for what will follow, we can use the $\phi^2$ perturbation as a first test of the Hamiltonian Truncation method. This is analogous to the analysis carried out in $1+1$ dimension in \cite{Rychkov:2014eea}. Starting from a Fock space basis of the theory in \reef{free1} we introduce a perturbation 
\be
\Delta H=   g_2  V_2 \, . \label{eq:phi2def}
\ee
This simply corresponds to turning off $g_4$ in \reef{ham1} with $m=1$. Since the theory is UV finite, and the cancellation of bubble diagrams discussed around \reef{toinc1} is restored when taking the limit $E_T \rightarrow \infty$,~\footnote{A  proof of this statement at all orders in perturbation theory will be presented    in section \ref{probHPT} and appendix \ref{2ptfact}. } the counter-terms in \eqref{ham1} can be set to zero in this case. Extrapolating the values of the vacuum energy and the mass gap obtained from the diagonalisation to $E_T \rightarrow \infty$ the exact analytic results should be recovered.

In Figure~\ref{fig:phi21} left we plot the vacuum energy density obtained from HT calculations as a function of $E_T$ for a theory with $g_2=1$ and $L=4$. For such a box the winding corrections to the exact vacuum energy are significant, and we show the analytic result both for an infinitely large box and including finite volume effects. The direct result from the truncation calculation is shown in orange [labeled ``without sub-leading"]. Although this appears to be approaching the analytic result, at the accessible values of $E_T$ it is not yet fully converged.

\begin{figure}[t]\begin{center}
\includegraphics[height=5.2cm]{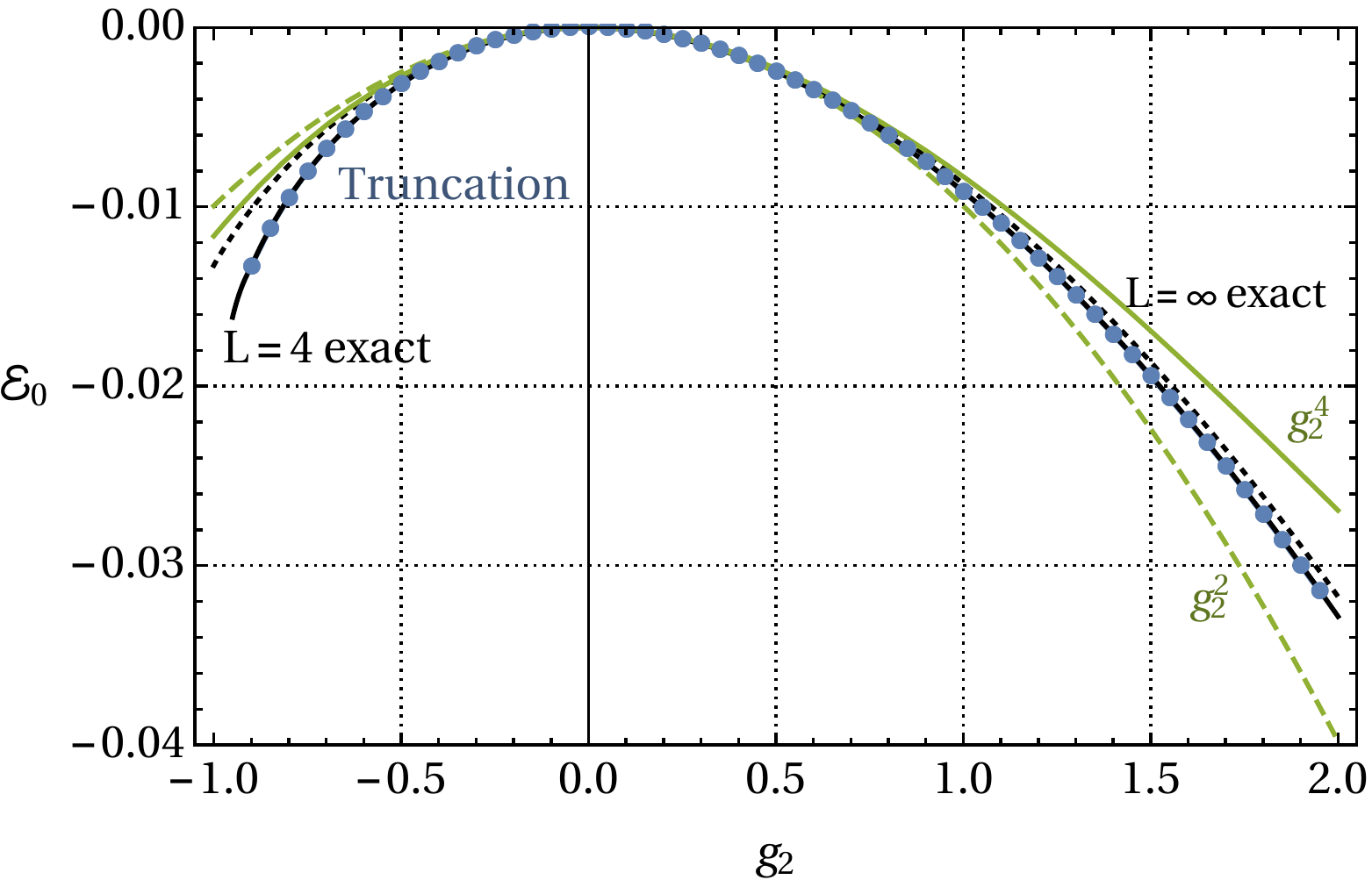}\qquad
\includegraphics[height=5.2cm]{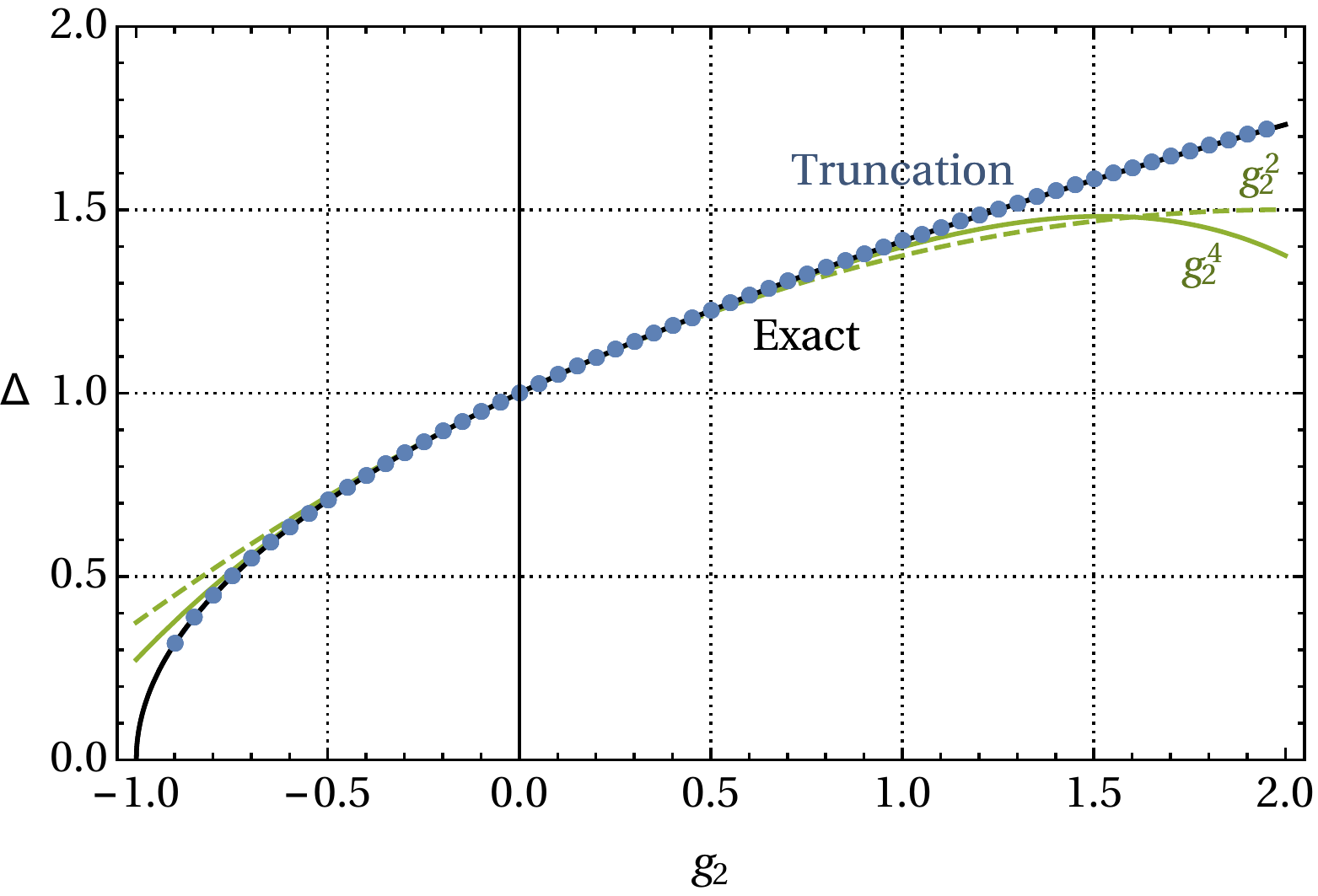} \end{center}
\caption{The vacuum energy density [left] and mass gap [right] as a function of $g_2$ obtained from Hamiltonian Truncation calculations after extrapolation to $E_T \rightarrow \infty$. We also show the exact analytic results [for the vacuum energy density with and without including winding corrections, solid and dashed respectively] and the prediction from perturbation theory [labeled PT] at order $g_2^2$ and $g_2^4$.}
\label{fig:phi22}
\end{figure}

The convergence of the vacuum energy as a function of $E_T$ can be improved by adding corrections to the truncated Hamiltonians. These account for the leading effects of states with $E > E_T$ that are missing from the computation, and vanish in the $E_T \rightarrow \infty$ limit. Such corrections have been studied extensively, starting with  \cite{Feverati:2006ni,Watts:2011cr,Lencses:2014tba}, continuing with the full systematic calculation in \cite{Rychkov:2014eea,Hogervorst:2014rta}, while further developments and the state of the art perspective can be found in \cite{Elias-Miro:2015bqk,Elias-Miro:2017xxf,Elias-Miro:2017tup,Rutter:2018aog}. For our purposes we simply include the first correction, which scales as $1/E_T$ and arises due to the integral corresponding to the left diagram of \reef{I2} being cut off at $E_T$. The contribution ``missing" from such a diagram due to the truncation is a shift in all of the energy levels of~\cite{Elias-Miro:2015bqk}
\be
\Delta H_2 =\frac{-g_2^2}{4} \int_{E_T}^\infty \frac{dE}{2\pi}\frac{\Phi_2(E)}{E}  =  - \frac{g_2^2}{16\pi E_T}     + O(1/E_T^2) \, , \label{phi2sublead}
\ee
which is easily incorporated as an additional diagonal term in the Hamiltonian matrix.  The results with the sub-leading correction included are also plotted in Figure~\ref{fig:phi22}, and they converge much faster to the analytic result.

In Figure~\ref{fig:phi21} right we show the mass gap in the same theory. The results obtained are unaffected by whether the diagonal correction \reef{phi2sublead} is included. Instead the leading correction to the mass gap is higher order in $g_2$ and $E_T$, and is not needed for our present purposes. We have also confirmed that the energies of the next few excited states quickly converge to their expected values.

In Figure~\ref{fig:phi22} we plot the vacuum energy density and mass gap obtained in the limit $E_T \rightarrow \infty$ as the coupling $g_2$ is varied;  the results are shown  along with the analytic predictions at finite and infinite volumes.~\footnote{In particular we fit the finite $E_T$ data with a function of the form $ \alpha_0 +  \alpha_1/E_T$, where $ \alpha_i$ are constants.  Adding an extra $ \alpha_i/E_T \log E_T$ or $ \alpha_2/E_T^2$ freedom in the fit does not affect the value of $ \alpha_0$ in any significant way.  } The perturbative prediction at orders $g_2^2$ and $g_2^4$ is also plotted. 
We  find good agreement between HPT and HT for perturbative values $g_2\in(-1/2,1)$, while HT agrees precisely with the exact calculation \reef{exst} for the entire range of couplings.  Additionally we see that the truncation calculation is correctly capturing the winding corrections. 
It is encouraging that the truncation calculation is sensitive to high order diagrams of HPT, and that it gives precise results.

\section{Hamiltonian Truncation for a UV divergent perturbation}
\label{pertUV}

Now we turn to our main concern: developing a formalism to apply HT to UV divergent relevant perturbations that require renormalisation. 
To do so, we analyse  HPT with an $E_T$ cutoff for the $\phi^4$ perturbation in detail.

A straightforward inspection of the HPT  Feynman diagrams reveals the presence of  two primitive UV divergencies associated to the vacuum  
 \be
 \begin{minipage}[h]{0.058\linewidth}
\begin{tikzpicture}
\begin{feynman}[small]
 \node [dot] (i1) at (0,0);
 \node [dot] (i2) at (.8,0);  
\diagram*{
   (i1) -- [ quarter right, looseness=.8,  thick] (i2)  -- [ quarter right, looseness=.8 ,  thick] (i1) ,
     (i1) -- [ half right, looseness=1.3,  thick] (i2)  -- [ half right, looseness=1.3 ,  thick] (i1)
};
  \end{feynman}
\end{tikzpicture}
  \end{minipage} = c_0(E_T)  \sim \frac{-g^2L^2}{96(4\pi)^3}\left(E_T-8m \log\frac{E_T}{m} \right)  \,  ,
   \ \ \ 
\begin{minipage}[h]{0.088\linewidth}
\begin{tikzpicture}
\begin{feynman}[small]
 \node [dot] (i1) at (0,0);
 \node [dot] (i2) at (.7,0);  
 \node [dot] (i3) at (1.4,0);  
\diagram*{
   (i1) -- [ quarter right, looseness=.8,  thick] (i2)  -- [ quarter right, looseness=.8 ,  thick] (i1) ,
     (i2) -- [ quarter right, looseness=.8,  thick] (i3)  -- [ quarter right, looseness=.8 ,  thick] (i2) ,
     (i1) -- [ half right, looseness=.8,  thick] (i3)  -- [ half right, looseness=.8 ,  thick] (i1)
};
  \end{feynman}
\end{tikzpicture}
  \end{minipage}
  = d_0(E_T) \sim \frac{g^3L^2}{3072 (4\pi)^2}\log\frac{E_T}{m}   \,   , \label{vac1}
         \ee
and one to the computation of the first excited energy level  at $O(g^2)$, 
 \be
\begin{minipage}[h]{0.085\linewidth}
\begin{tikzpicture}
\begin{feynman}[small]
 \vertex (i0) at (-.3,0);
 \node [dot] (i1) at (0,0);
 \node [dot] (i2) at (.8,0);  
 \vertex (i3) at (1.1,0);
\diagram*{
   (i0) -- [thick]   (i3) ,
     (i1) -- [ half right, looseness=1.,  thick] (i2)  -- [ half right, looseness=1. ,  thick] (i1)
};
  \end{feynman}
\end{tikzpicture}
  \end{minipage}  \sim  c_2(E_T)= -  \frac{g^2}{6(4\pi)^2}  \log\frac{E_T}{m}    \, , \label{mass1}
       \ee 
       we call the latter the sunset diagram.~\footnote{Note that $c_0$ and $d_0$ are taken to be equal to the diagrams, while for $c_2$ we take the leading log. See appendix \ref{acts} for further details on the calculation of \reef{vac1} and \reef{mass1}.}
Thus we must add counter-terms to define a theory with a finite $E_T\rightarrow \infty$ limit. 
Tentatively,  one may consider the following potential as a perturbation of the free massive theory
\be
V \stackrel{?}{=} V_4 - c_0(E_T)  - d_0(E_T)    -c_2(E_T)  V_2   \ ,  \label{toamend}
\ee
as one would do in a covariant calculation of correlation functions or scattering amplitudes. 
However,  this theory does not have a finite $E_T\rightarrow \infty$ limit. Next we will prove this claim and we will 
identify  the problems in doing a perturbative calculation with an $E_T$ cutoff regulator.

We start  by   explaining  in detail the  calculation of the  vacuum energy up to $O(g^4)$ in section \ref{probHPT}. 
This analysis parallels  the explanation in section \ref{anagphi2} for the $\phi^2$ perturbation, but with some key differences that will be stressed below. 
Subsequently,  in section \ref{gencase5} we generalise this analysis  to arbitrary order in perturbation theory.  
Following this, in section~\ref{ssPatch}   we will solve the problems with perturbation theory, which will enable us to formulate  Hamiltonian Truncation in section~\ref{subHT}.

\subsection{Problems with naive perturbation theory}
\label{probHPT}

The calculation of the vacuum  at $O(g^2)$ and $O(g^3)$ is straightforward.  There are only two connected diagrams, given in \reef{vac1}. In the theory of \reef{toamend}, the UV divergences associated to these diagrams are readily subtracted by lower order diagrams involving the  counter-terms $c_0$ and $d_0$. 

At $O(g^4)$, the calculation gets much more interesting. 
We discuss the disconnected diagrams first and then the connected diagrams. 
We have two disconnected pieces. The first piece, arising from the first term of $\cE_0^{(4)}$ in  \reef{basicpert},  is given by 
\be
V_{0k_1} E_{0k_1}^{-1} V_{k_1 k_2}   E_{0k_2}^{-1}  V_{k_2 k_3}  E_{0k_3}^{-1}  V_{k_3 0} |_\text{Disc.} 
= \  \begin{minipage}[h]{0.085\linewidth}
\begin{tikzpicture}
\begin{feynman}[small]
 \node [dot] (i1) at (0,0);
 \node [dot] (i2) at (.8,0);  
 \node [dot] (j1) at (.4,.6);
 \node [dot] (j2) at (1.2,.6);  
\diagram*{
   (i1) -- [ quarter right, looseness=.3,  thick] (i2)  -- [ quarter right, looseness=.3 ,  thick] (i1) ,
     (i1) -- [ half right, looseness=.8,  thick] (i2)  -- [ half right, looseness=.8 ,  thick] (i1) ,
   (j1) -- [ quarter right, looseness=.3,  thick] (j2)  -- [ quarter right, looseness=.3 ,  thick] (j1) ,
     (j1) -- [ half right, looseness=.8,  thick] (j2)  -- [ half right, looseness=.8 ,  thick] (j1)
};
  \end{feynman}
\end{tikzpicture}
  \end{minipage}
  + \ 
  \begin{minipage}[h]{0.075\linewidth}
\begin{tikzpicture}
\begin{feynman}[small]
 \node [dot] (i1) at (0,0);
 \node [dot] (i2) at (1.2,0);  
 \node [dot] (j1) at (.3,.6);
 \node [dot] (j2) at (.9,.6);  
\diagram*{
   (i1) -- [ quarter right, looseness=.2,  thick] (i2)  -- [ quarter right, looseness=.2 ,  thick] (i1) ,
     (i1) -- [ half right, looseness=.5,  thick] (i2)  -- [ half right, looseness=.5 ,  thick] (i1) ,
   (j1) -- [ quarter right, looseness=.3,  thick] (j2)  -- [ quarter right, looseness=.3 ,  thick] (j1) ,
     (j1) -- [ half right, looseness=.8,  thick] (j2)  -- [ half right, looseness=.8 ,  thick] (j1)
};
  \end{feynman}
\end{tikzpicture}
  \end{minipage} \label{phi4vac} \ ,
\ee
and formally cancels against the fully disconnected piece [i.e. second term of $\cE^{(4)}_0$ in \reef{basicpert}]:
\be
  -\cE_0^{(2) } \times  V_{0k}E_k^{-2}V_{k0}
  =    \frac{(g L)^2}{24}  \int_{4} \frac{dE}{2\pi}\frac{\Phi_4(E)}{E}    \times    \frac{(g L)^2}{24}  \int_{4} \frac{dE}{2\pi}\frac{\Phi_4(E)}{E^2}    \, .  \label{fact1}
\ee
Indeed, the diagrams in \reef{phi4vac} are given by
\be
 \begin{minipage}[h]{0.085\linewidth}
\begin{tikzpicture}
\begin{feynman}[small]
 \node [dot] (i1) at (0,0);
 \node [dot] (i2) at (.8,0);  
 \node [dot] (j1) at (.4,.6);
 \node [dot] (j2) at (1.2,.6);  
\diagram*{
   (i1) -- [ quarter right, looseness=.3,  thick] (i2)  -- [ quarter right, looseness=.3 ,  thick] (i1) ,
     (i1) -- [ half right, looseness=.8,  thick] (i2)  -- [ half right, looseness=.8 ,  thick] (i1) ,
   (j1) -- [ quarter right, looseness=.3,  thick] (j2)  -- [ quarter right, looseness=.3 ,  thick] (j1) ,
     (j1) -- [ half right, looseness=.8,  thick] (j2)  -- [ half right, looseness=.8 ,  thick] (j1)
};
  \end{feynman}
\end{tikzpicture}
  \end{minipage}  =   -\frac{(g L)^4}{576} \int_4 \frac{dE_1}{2\pi} \frac{dE_2}{2\pi} \frac{\Phi_4(E_1)}{E_1+E_2}   \frac{\Phi_4(E_2)}{E_1E_2}      \ , \quad 
  \begin{minipage}[h]{0.075\linewidth}
\begin{tikzpicture}
\begin{feynman}[small]
 \node [dot] (i1) at (0,0);
 \node [dot] (i2) at (1.2,0);  
 \node [dot] (j1) at (.3,.6);
 \node [dot] (j2) at (.9,.6);  
\diagram*{
   (i1) -- [ quarter right, looseness=.2,  thick] (i2)  -- [ quarter right, looseness=.2 ,  thick] (i1) ,
     (i1) -- [ half right, looseness=.5,  thick] (i2)  -- [ half right, looseness=.5 ,  thick] (i1) ,
   (j1) -- [ quarter right, looseness=.3,  thick] (j2)  -- [ quarter right, looseness=.3 ,  thick] (j1) ,
     (j1) -- [ half right, looseness=.8,  thick] (j2)  -- [ half right, looseness=.8 ,  thick] (j1)
};
  \end{feynman}
\end{tikzpicture}
  \end{minipage}  = -\frac{(g L)^4}{576} \int_{4} \frac{dE_1}{2\pi} \frac{dE_2}{2\pi} \frac{\Phi_4(E_1)}{E_1+E_2}   \frac{\Phi_4(E_2)}{E_2^2}   \, ,  \label{discoo}
  \ee
  and upon adding them the integrals in \reef{phi4vac} neatly  factorise as in \reef{fact1}.
  Note however that both \reef{phi4vac} and \reef{fact1} are UV divergent in $d=2+1$ dimensions because 
  \be
  \Phi_4(x)= \frac{x}{128\pi^2}+ O(x^0) \, . 
  \ee
  Therefore, the former expressions require regularisation. If we proceed with a covariant regulator, e.g. cutting the momenta circulating in the loops, the cancellation of  $\reef{phi4vac}+ \reef{fact1}=0$ still takes place [alternatively we can perform dimensional regularisation in position space \reef{wicks}, $D(x^\mu) = (2\pi)^{-d/2} (m/\sqrt{x_\mu^2})^{d-2}K_{(d-2)/2}(m\sqrt{x_\mu^2})$]. 
  The problem arises if we insist on regularising the theory by restricting the Hilbert space  to the states  with a maximal   $H_0$-energy $E_T$. 
  Then, we must replace \reef{fact1} and \reef{discoo} with 
\bea
  -\cE_0^{(2) } \times  V_{0k}E_k^{-2}V_{k0}
 &=&   \frac{(g L)^2}{24}  \int_{4}^{E_T} \frac{dE}{2\pi}\frac{\Phi_4(E)}{E}  \times    \frac{(g L)^2}{24}  \int_{4}^{E_T} \frac{dE}{2\pi}\frac{\Phi_4(E)}{E^2}  \label{toadd1}  \, , \\[.2cm]
 \begin{minipage}[h]{0.085\linewidth}
\begin{tikzpicture}
\begin{feynman}[small]
 \node [dot] (i1) at (0,0);
 \node [dot] (i2) at (.8,0);  
 \node [dot] (j1) at (.4,.6);
 \node [dot] (j2) at (1.2,.6);  
\diagram*{
   (i1) -- [ quarter right, looseness=.3,  thick] (i2)  -- [ quarter right, looseness=.3 ,  thick] (i1) ,
     (i1) -- [ half right, looseness=.8,  thick] (i2)  -- [ half right, looseness=.8 ,  thick] (i1) ,
   (j1) -- [ quarter right, looseness=.3,  thick] (j2)  -- [ quarter right, looseness=.3 ,  thick] (j1) ,
     (j1) -- [ half right, looseness=.8,  thick] (j2)  -- [ half right, looseness=.8 ,  thick] (j1)
};
  \end{feynman}
\end{tikzpicture}
  \end{minipage}   + \ 
  \begin{minipage}[h]{0.075\linewidth}
\begin{tikzpicture}
\begin{feynman}[small]
 \node [dot] (i1) at (0,0);
 \node [dot] (i2) at (1.2,0);  
 \node [dot] (j1) at (.3,.6);
 \node [dot] (j2) at (.9,.6);  
\diagram*{
   (i1) -- [ quarter right, looseness=.2,  thick] (i2)  -- [ quarter right, looseness=.2 ,  thick] (i1) ,
     (i1) -- [ half right, looseness=.5,  thick] (i2)  -- [ half right, looseness=.5 ,  thick] (i1) ,
   (j1) -- [ quarter right, looseness=.3,  thick] (j2)  -- [ quarter right, looseness=.3 ,  thick] (j1) ,
     (j1) -- [ half right, looseness=.8,  thick] (j2)  -- [ half right, looseness=.8 ,  thick] (j1)
};
  \end{feynman}
\end{tikzpicture}
  \end{minipage} & =& -\frac{(g L)^4}{576}\int_4^{E_T-4} \frac{dE_2}{2\pi}   \frac{\Phi_4(E_2)}{E_2^2}  \int_{4}^{E_T-E_2} \frac{dE_1}{2\pi}\frac{\Phi_4(E_1)}{E_1}   \, .  \label{toadd2}
\eea
Now, when we add  the previous  two expressions we get
\be
\reef{toadd1}+\reef{toadd2} = (gL)^4  \frac{E_T-8\log E_T}{144 (8\pi)^6} + O(E_T^0) \, ,  \label{problemon}
\ee
namely, disconnected diagrams do not cancel at finite $E_T$, and most importantly the effect does not decouple and diverges when the cutoff is removed!

In the connected sector  there are three diagrams plus all their possible vertex orderings:
\vspace{-.8cm}
\be
\cE_0^{(4)}|_\text{Conn.} =   
  \begin{minipage}[h]{0.115\linewidth}
\begin{tikzpicture}
\begin{feynman}[small]
   \vertex(Y) at (.6,1.62)  ;
      \vertex(YY) at (1,1.62)  ;
 \node [dot] (i1) at (0,0);
 \node [dot] (i2) at (1.2,0);  
 \node [dot] (j1) at (.3,.6);
 \node [dot] (j2) at (.9,.6);  
  \vertex (j0) at (0,.6) ; 
   \vertex (j3) at (1.3,.6) ;
   \vertex(X) at (.6,1.05)  ;
      \vertex(XX) at (.6,-.85) {\scriptsize $-{\roig E_l}-E_\text{ext}$}; 
\diagram*{
     (i1) -- [ half right, looseness=.5,  thick] (i2)  -- [ half right, looseness=.5 ,  thick] (i1) ,
     (j1) -- [ red, half right, looseness=.8,  thick] (j2)  -- [  red,half right, looseness=.8 ,  thick] (j1) ,
(j1) --[thick] (j0) -- [half right, looseness=1.5, thick] (i1) -- [thick]  (i2) -- [half right, looseness=1.5, thick] (j3) --[thick] (j2)  -- [red,  thick] (j1),
(X) -- [thick, red, scalar] (XX),
(Y) -- [white] (YY)
};
  \end{feynman}
\end{tikzpicture}
  \end{minipage} 
 \,  +  \
\begin{minipage}[h]{0.085\linewidth}
\begin{tikzpicture}
\begin{feynman}[small]
 \node [dot] (i1) at (0,0);
 \node [dot] (i2) at (1.2,0);  
 \node [dot] (j1) at (.3,.6);
 \node [dot] (j2) at (.9,.6);  
\diagram*{
     (i1) -- [ thick] (i2)  -- [ quarter left, looseness=.5 ,  thick] (i1) ,
     (j1) -- [ quarter right, looseness=.6,  thick] (j2)  -- [ quarter right, looseness=.6 ,  thick] (j1) ,
(j1) -- [quarter right, looseness=.7, thick] (i1) -- [quarter right, looseness=.7, thick] (j1)  , 
 (i2) -- [quarter right, looseness=.7, thick] (j2)  -- [quarter right, looseness=.7, thick] (i2)  

};
  \end{feynman}
\end{tikzpicture}
  \end{minipage} 
  + \hspace{-.09cm}
\begin{minipage}[h]{0.1\linewidth}
\begin{tikzpicture}
\begin{feynman}[small]
 \node [dot] (i1) at (0,0);
 \node [dot] (i2) at (1.2,0);  
 \node [dot] (j1) at (.3,.4);
 \node [dot] (j2) at (.9,.4);  
 \vertex (X) at (0,-.4);
 \vertex (XX) at (1,-.4);
\diagram*{
     (i1) -- [   thick] (i2)  -- [ quarter left, looseness=.5 ,  thick] (i1) ,
     (j1) -- [   thick] (j2)  -- [ quarter right, looseness=.7 ,  thick] (j1) ,
(j1) -- [ thick] (i1) , 
 (i2) -- [quarter right, looseness=.8, thick] (j2) ,
 (i2) -- [thick] (j1)  ,
 (j2) -- [half right, looseness = 1.4 , thick] (i1) ,
 (X) -- [white] (XX)
};
  \end{feynman}
\end{tikzpicture}
  \end{minipage} 
  + 
  \text{all vertex re-orderings}
  \vspace{-.2cm} \label{sunsetdiv}
\ee
Only the first  diagram and its vertex re-orderings can be divergent. 
Indeed,   the first diagram contains a UV divergent sub-sunset diagram [painted in red] for energies $E_l\sim E_T$. Note that the lower sunset sub-diagram propagating  a state of energy $E_\text{ext}$   is rendered convergent by the two vertex insertions [from the upper red sunset]  in between its ends. 
The divergence in the upper sunset sub-diagram is taken care of by the counter-term $c_2$ at lower order in perturbation theory, and the divergences in the the vertex reordered diagrams are similarly cancelled. More generally, the connected vacuum diagrams are finite for the theory \reef{toamend} at all orders in perturbation theory: \emph{the only divergent sub-diagrams are sunset diagrams and these are subtracted by $c_2$.}

To sum up, the vacuum energy is equal to the sum of all connected vacuum Feynman diagrams. 
In HPT this is true because of  delicate cancelations of the type $\reef{phi4vac}+\reef{fact1}=0$. We have found that this cancelation is spoiled when we introduce the non-covariant  regulator $E_T$.

Note that the problem of UV divergences due to disconnected bubbles discussed in this section is not present in $\phi^4$ in $d=1+1$ dimensions. Equations \reef{toadd1} and \reef{toadd2} are formally valid in any spacetime dimension, provided we use the appropriate phase space function,
\be
\Phi_4^{d=1+1}(x)  = \frac{H_1(4x^{-2})-i H_2(4x^{-2})+3 H_3(4x^{-2})}{8\pi^2} = \frac{3}{2\pi^2 }\frac{1}{x^2} \left[\log^2 (x)-\pi^2/12\right]+O(x^{-4}) \, , 
\ee
where the functions are $H_i$ are  given in terms of the Bessel function $K$.\footnote{One finds $H_1=K(t_+(x))K(t_-(x))$,  $H_2=K(t_+(x))K(1-t_-(x))$,  $H_3=-K(1-t_+(x))K(1-t_-(x))/3$ and $t_\pm(x)=1/(4x) (2x+(1-2x)\sqrt{(x-1)/x}\pm\sqrt{(4x-1)/x})$.}
We find
\be
\reef{toadd1}+\reef{toadd2} \stackrel{d=1+1}{=}   O(E_T^{-1}) \, . 
\ee
Therefore in previous $\phi_2^4$ Hamiltonian Truncation studies  \cite{Rychkov:2014eea} the problem in \reef{problemon}  did not arise.

\subsection{The general case}
\label{gencase5}

The problem found in \reef{problemon} is ubiquitous to all orders $O(g^n)$ in perturbation theory for $n\geq 4$. 
To identify all the  diagrams that, as a result of the $E_T$ cutoff,  diverge in the  $E_T\rightarrow \infty$ extrapolation we generalise the 
two separate  UV divergent contributions in \reef{toadd2}. 
In section \ref{genebub}, we discuss the generalisation of the second diagram in   \reef{phi4vac}. Then, in  section \ref{genebub2} we will deal with the generalisation of the first diagram in \reef{phi4vac}.~\footnote{Most of the formulas of this section are valid in either finite or infinite volume (and $d$ dimensions), provided the correct phase-space function $\Phi_4$ is used. }

 The perturbative correction to the $E_i$ state at $O(g^n)$ is given by
 \be
 \cE_i^{(n)} = \underbrace{\frac{V_{ik_1} V_{k_1k_2}  \cdots  V_{k_{n}i}}{E_{ik_1}E_{ik_2}\cdots E_{ik_{n-1}}} }_{\text{1st contribution}} - \underbrace{ \cE_0^{(2)} \,    \frac{V_{ik_1} V_{k_1k_2}  \cdots  V_{k_{n-2}i}}{E_{ik_1}E_{ik_2}\cdots E_{ik_{n-3}}}\sum_{s=1}^{n-3}\frac{1}{E_{ik_s}}    -  \cE_i^{(n-2)}
\, V_{ik} E_{ik}^{-2}V_{ki}}_{\text{2nd contribution}} +  \ldots  \label{1stc}
  \ee
  where a sum over $k \neq i$ is implicit. In \eqref{1stc}  $\ldots$   denotes further corrections that are irrelevant for the current discussion -- these are terms proportional to  either  $V_{nn}$ or  $\cE_i^{(s)}$ with $3\leq s  \leq n-3 $. The diagrams that we are presently interested in are obtained from the first contribution in  \reef{1stc} by considering each  $O(g^{n-2})$  diagram [either  connected or disconnected]  with  $i$  external lines on each side of the diagram     and dressing   such diagram with two-point bubbles. 
We denote a generic diagram  by 
 $\begin{minipage}[h]{0.19\linewidth}
\begin{tikzpicture}
\begin{feynman}[small]
 \node [dot] (i1) at (0,0);
 \node [dot] (i2) at (.5,0);  
 \node [dot] (i3) at (1.1,0);   
 \node  (i4) at (1.8,0);  
 \node  (i5) at (1.8,0);  
 \node [dot] (i6) at (2.,0);  
 \node [dot] (i7) at (2.6,0);  
  \node  (x1) at (-.3,0);  
    \node  (x2) at (-.3,.3); 
    \node  (x3) at (-.3,-.3);  
  \node  (y1) at (2.9,0);  
    \node  (y2) at (2.9,.3); 
    \node  (y3) at (2.9,-.3);  
\diagram*{
   (i1) -- [line width=.7mm, red] (i2)-- [line width=.7mm, red] (i3)  -- [line width=.7mm, red, scalar]  (i6)-- [line width=.7mm,red] (i7) ,
     (x1) -- [thick, red] (i1),
    (x2) -- [thick, red] (i1),
        (x3) -- [thick, red] (i1) ,
     (y1) -- [thick, red] (i7),
    (y2) -- [thick, red] (i7),
        (y3) -- [thick, red] (i7)
};
  \end{feynman}
\end{tikzpicture}
  \end{minipage}  $, where the thick red lines represent any number of regular black lines.

The diagrams with a two-point bubble inserted  in between  every two consecutive vertices are represented by
\be
\hspace{-.4cm}
\begin{minipage}[h]{0.27\linewidth}
\begin{tikzpicture}
\begin{feynman}[small]
 \node [dot] (i1) at (0,0);
 \node [dot] (i2) at (1.2,0);  
 \node [dot] (i3) at (1.8,0);   
 \node  (i4) at (2.5,0);  
 \node  (i5) at (2.9,0);  
 \node [dot] (i6) at (3.6,0);  
 \node [dot] (i7) at (4.2,0);  
 \node [dot] (j1) at (.4,.3);
 \node [dot] (j2) at (.8,.3); 
  \node  (x1) at (-.3,0);  
    \node  (x2) at (-.3,.3); 
    \node  (x3) at (-.3,-.3);  
  \node  (y1) at (4.5,0);  
    \node  (y2) at (4.5,.3); 
    \node  (y3) at (4.5,-.3);  
\diagram*{
   (i1) -- [line width=.7mm, red] (i2)-- [line width=.7mm, red] (i3)  -- [line width=.7mm, red, scalar] (i4), 
   (i5)-- [scalar, line width=.7mm, red] (i6)-- [line width=.7mm,red] (i7) ,
   (j1) -- [ quarter right, looseness=.3,  thick] (j2)  -- [ quarter right, looseness=.3 ,  thick] (j1) ,
     (j1) -- [ half right, looseness=.8,  thick] (j2)  -- [ half right, looseness=.8 ,  thick] (j1) ,
     (x1) -- [thick, red] (i1),
    (x2) -- [thick, red] (i1),
        (x3) -- [thick, red] (i1) ,
     (y1) -- [thick, red] (i7),
    (y2) -- [thick, red] (i7),
        (y3) -- [thick, red] (i7)
};
  \end{feynman}
\end{tikzpicture}
  \end{minipage}   
 \ \  + 
  \begin{minipage}[h]{0.27\linewidth}
\begin{tikzpicture}
\begin{feynman}[small]
 \node [dot] (i1) at (0,0);
 \node [dot] (i2) at (.6,0);  
 \node [dot] (i3) at (1.8,0);   
 \node  (i4) at (2.5,0);  
 \node  (i5) at (2.9,0);  
 \node [dot] (i6) at (3.6,0);  
 \node [dot] (i7) at (4.2,0);  
 \node [dot] (j1) at (1.,.3);
 \node [dot] (j2) at (1.4,.3);  
  \node  (x1) at (-.3,0);  
    \node  (x2) at (-.3,.3); 
    \node  (x3) at (-.3,-.3);  
  \node  (y1) at (4.5,0);  
    \node  (y2) at (4.5,.3); 
    \node  (y3) at (4.5,-.3);  
\diagram*{
   (i1) -- [line width=.7mm, red] (i2)-- [line width=.7mm, red] (i3)  -- [line width=.7mm, red, scalar] (i4), 
   (i5)-- [scalar, line width=.7mm, red] (i6)-- [line width=.7mm,red] (i7) ,
   (j1) -- [ quarter right, looseness=.3,  thick] (j2)  -- [ quarter right, looseness=.3 ,  thick] (j1) ,
     (j1) -- [ half right, looseness=.8,  thick] (j2)  -- [ half right, looseness=.8 ,  thick] (j1) ,
     (x1) -- [thick, red] (i1),
    (x2) -- [thick, red] (i1),
        (x3) -- [thick, red] (i1) ,
     (y1) -- [thick, red] (i7),
    (y2) -- [thick, red] (i7),
        (y3) -- [thick, red] (i7)

};
  \end{feynman}
\end{tikzpicture}
  \end{minipage} 
 \ \  + \cdots +  
  \begin{minipage}[h]{0.27\linewidth}
\begin{tikzpicture}
\begin{feynman}[small]
 \node [dot] (i1) at (0,0);
 \node [dot] (i2) at (.6,0);  
 \node [dot] (i3) at (1.2,0);   
 \node  (i4) at (1.9,0);  
 \node  (i5) at (2.2,0);  
 \node [dot] (i6) at (2.9,0);  
 \node [dot] (i7) at (4.1,0);  
 \node [dot] (j1) at (3.3,.3);
 \node [dot] (j2) at (3.7,.3);  
  \node  (x1) at (-.3,0);  
    \node  (x2) at (-.3,.3); 
    \node  (x3) at (-.3,-.3);  
  \node  (y1) at (4.5,0);  
    \node  (y2) at (4.5,.3); 
    \node  (y3) at (4.5,-.3);  
\diagram*{
   (i1) -- [line width=.7mm, red] (i2)-- [line width=.7mm, red] (i3)  -- [line width=.7mm, red, scalar] (i4), 
   (i5)-- [scalar, line width=.7mm, red] (i6)-- [line width=.7mm,red] (i7) ,
   (j1) -- [ quarter right, looseness=.3,  thick] (j2)  -- [ quarter right, looseness=.3 ,  thick] (j1) ,
     (j1) -- [ half right, looseness=.8,  thick] (j2)  -- [ half right, looseness=.8 ,  thick] (j1) ,
     (x1) -- [thick, red] (i1),
    (x2) -- [thick, red] (i1),
        (x3) -- [thick, red] (i1) ,
     (y1) -- [thick, red] (i7),
    (y2) -- [thick, red] (i7),
        (y3) -- [thick, red] (i7)

};
  \end{feynman}
\end{tikzpicture}
  \end{minipage}  \label{discoo2}
\vspace{-.1cm} \ee 
and the corresponding expression is   given by
 \be
 \int \prod_{a=1}^{n-3} \frac{d x_a}{x_{ia}} \,  g_{E_i}( \vec x) 
\,  \sum_{s=1}^{n-3}   \frac{1}{x_{is}}\underbrace{ \frac{(gL)^2}{24}\int_4^{\infty} \frac{dE}{(2\pi)} \frac{\Phi_4(E)}{E_i-(E+x_s)}}_{\text{two-point bubble +  a free $|x_s\rangle$ state}}    \, ,  \label{toreg2}
 \ee
where  $x_s$ is the energy of the state being propagated in between any two consecutive vertices of the lower diagrams, and $x_{is}\equiv E_i-x_s$. There are $(n-2)-1$ such variables for a diagram with $n-2$  vertices. The function $  g_{E_n}( \vec x) $   depends on the specific lower diagram and   includes a  symmetry factor that  is the same for each of the diagrams  in \reef{discoo2}. 
 Eq.~\reef{toreg2} is UV divergent because $\Phi_4(x)\sim x/(128\pi^2)$, thus  we are assuming that \reef{toreg2} is properly regulated with e.g. a momentum cutoff. 
 
 Next we should add all the diagrams with the bubble's vertices in every other possible location. For instance, we must add diagrams such as
 \be
 \begin{minipage}[h]{0.2\linewidth}
\begin{tikzpicture}
\begin{feynman}[small]
 \node [dot] (i1) at (0,0);
 \node [dot] (i2) at (.5,0);  
 \node [dot] (i3) at (1.1,0);   
 \node  (i4) at (1.8,0);  
 \node  (i5) at (1.8,0);  
 \node [dot] (i6) at (2.,0);  
 \node [dot] (i7) at (2.6,0);  
  \node  (x1) at (-.3,0);  
    \node  (x2) at (-.3,.3); 
    \node  (x3) at (-.3,-.3);  
  \node  (y1) at (2.9,0);  
    \node  (y2) at (2.9,.3); 
    \node  (y3) at (2.9,-.3);  
 \node [dot] (j1) at (-.3,.4);
 \node [dot] (j2) at (.3,.4); 
\diagram*{
   (i1) -- [line width=.7mm, red] (i2)-- [line width=.7mm, red] (i3)  -- [line width=.7mm, red, scalar]  (i6)-- [line width=.7mm,red] (i7) ,
     (x1) -- [thick, red] (i1),
    (x2) -- [thick, red] (i1),
        (x3) -- [thick, red] (i1) ,
     (y1) -- [thick, red] (i7),
    (y2) -- [thick, red] (i7),
        (y3) -- [thick, red] (i7),
          (j1) -- [ quarter right, looseness=.2,  thick] (j2)  -- [ quarter right, looseness=.2 ,  thick] (j1) ,
     (j1) -- [ half right, looseness=.6,  thick] (j2)  -- [ half right, looseness=.6 ,  thick] (j1) ,
};
  \end{feynman}
\end{tikzpicture}
  \end{minipage}
+
 \begin{minipage}[h]{0.2\linewidth}
\begin{tikzpicture}
\begin{feynman}[small]
 \node [dot] (i1) at (0,0);
 \node [dot] (i2) at (.5,0);  
 \node [dot] (i3) at (1.1,0);   
 \node  (i4) at (1.8,0);  
 \node  (i5) at (1.8,0);  
 \node [dot] (i6) at (2.,0);  
 \node [dot] (i7) at (2.6,0);  
  \node  (x1) at (-.3,0);  
    \node  (x2) at (-.3,.3); 
    \node  (x3) at (-.3,-.3);  
  \node  (y1) at (2.9,0);  
    \node  (y2) at (2.9,.3); 
    \node  (y3) at (2.9,-.3);  
 \node [dot] (j1) at (-.3,.4);
 \node [dot] (j2) at (.8,.4); 
\diagram*{
   (i1) -- [line width=.7mm, red] (i2)-- [line width=.7mm, red] (i3)  -- [line width=.7mm, red, scalar]  (i6)-- [line width=.7mm,red] (i7) ,
     (x1) -- [thick, red] (i1),
    (x2) -- [thick, red] (i1),
        (x3) -- [thick, red] (i1) ,
     (y1) -- [thick, red] (i7),
    (y2) -- [thick, red] (i7),
        (y3) -- [thick, red] (i7),
          (j1) -- [ quarter right, looseness=.1,  thick] (j2)  -- [ quarter right, looseness=.1 ,  thick] (j1) ,
     (j1) -- [ half right, looseness=.4,  thick] (j2)  -- [ half right, looseness=.4 ,  thick] (j1) ,
};
  \end{feynman}
\end{tikzpicture}
  \end{minipage}
+
 \begin{minipage}[h]{0.2\linewidth}
\begin{tikzpicture}
\begin{feynman}[small]
 \node [dot] (i1) at (0,0);
 \node [dot] (i2) at (.5,0);  
 \node [dot] (i3) at (1.1,0);   
 \node  (i4) at (1.8,0);  
 \node  (i5) at (1.8,0);  
 \node [dot] (i6) at (2.,0);  
 \node [dot] (i7) at (2.6,0);  
  \node  (x1) at (-.3,0);  
    \node  (x2) at (-.3,.3); 
    \node  (x3) at (-.3,-.3);  
  \node  (y1) at (2.9,0);  
    \node  (y2) at (2.9,.3); 
    \node  (y3) at (2.9,-.3);  
 \node [dot] (j1) at (.8,.4);
 \node [dot] (j2) at (2.9,.4); 
\diagram*{
   (i1) -- [line width=.7mm, red] (i2)-- [line width=.7mm, red] (i3)  -- [line width=.7mm, red, scalar]  (i6)-- [line width=.7mm,red] (i7) ,
     (x1) -- [thick, red] (i1),
    (x2) -- [thick, red] (i1),
        (x3) -- [thick, red] (i1) ,
     (y1) -- [thick, red] (i7),
    (y2) -- [thick, red] (i7),
        (y3) -- [thick, red] (i7),
          (j1) -- [ quarter right, looseness=.04,  thick] (j2)  -- [ quarter right, looseness=.04 ,  thick] (j1) ,
     (j1) -- [ half right, looseness=.18,  thick] (j2)  -- [ half right, looseness=.18 ,  thick] (j1) ,
};
  \end{feynman}
\end{tikzpicture}
  \end{minipage} + \cdots \label{dings}
  \ee
   After we add up all such diagrams, i.e. after dressing 
   $\begin{minipage}[h]{0.19\linewidth}
\begin{tikzpicture}
\begin{feynman}[small]
 \node [dot] (i1) at (0,0);
 \node [dot] (i2) at (.5,0);  
 \node [dot] (i3) at (1.1,0);   
 \node  (i4) at (1.8,0);  
 \node  (i5) at (1.8,0);  
 \node [dot] (i6) at (2.,0);  
 \node [dot] (i7) at (2.6,0);  
  \node  (x1) at (-.3,0);  
    \node  (x2) at (-.3,.3); 
    \node  (x3) at (-.3,-.3);  
  \node  (y1) at (2.9,0);  
    \node  (y2) at (2.9,.3); 
    \node  (y3) at (2.9,-.3);  
\diagram*{
   (i1) -- [line width=.7mm, red] (i2)-- [line width=.7mm, red] (i3)  -- [line width=.7mm, red, scalar]  (i6)-- [line width=.7mm,red] (i7) ,
     (x1) -- [thick, red] (i1),
    (x2) -- [thick, red] (i1),
        (x3) -- [thick, red] (i1) ,
     (y1) -- [thick, red] (i7),
    (y2) -- [thick, red] (i7),
        (y3) -- [thick, red] (i7)
};
  \end{feynman}
\end{tikzpicture}
  \end{minipage}  $ with a two-point bubble in all possible ways, we find  the following expression
   \be
  \underbrace{ \frac{(gL)^2}{24}\int_4^{\infty} \frac{dE}{(2\pi)} \frac{\Phi_4(E)}{E_i-E}}_{\text{two-point bubble}}  \, \times   \,   \sum_{s=1}^{n-3}   \int \prod_{a=1}^{n-3} \frac{ d x_{ia}}{x_{ia}}  \,    g_{E_i}( \vec x)  \Big( \sum_{s=1}^{n-1}\frac{1}{x_{is} } + \frac{1}{E_i-E}  \Big) \, , 
 \ee
 i.e.  the two-point vacuum bubble  factors out.
Therefore, the sum of  all the dressings of each $O(g^{n-2})$ diagram by a two point bubble cancels with the second contribution in \reef{1stc}!
In appendix \ref{dvb} we show how the factorisation and cancelation of the disconnected bubble takes place in detail. Although this is expected on general grounds~\footnote{See e.g. Chapter 16 of Weinberg vol. II \cite{Weinberg:1996kr}.}, it is   interesting to see how it  happens in HPT and how it fails to happen when we regulate our theory with the $E_T$ cutoff. 

\subsubsection{Two-point  bubbles I}
\label{genebub}

When we regulate  \reef{toreg2} with an $E_T$ cutoff we get
 \be
   \int \prod_{a=1}^{n-3} \frac{d x_a}{x_{ia}} \,   g_{E_i}( \vec x) 
\,   \sum_{s=1}^{n-3}  \frac{1}{x_{is}} \frac{(gL)^2}{24}\int_4^{E_T-X_s} \frac{dE}{(2\pi)} \frac{\Phi_4(E)}{E_i-( E+x_s)} \, , \label{toexp1} 
 \ee
 where $X_s=x_s$. It is  useful to keep  $X_s$ and $x_s$ as independent variables because, while  the $X_s$ dependence is due to the $E_T$ regularisation, the $x_s$ dependence is physical. 
Then, $X_s$ will allow us to track by how much \reef{toexp1}, once added to the rest of the two-point bubble dressings \reef{dings}, fails to cancel against the fully disconnected  \emph{second contribution} of \reef{1stc}.

The important question however is by how much the former expression spoils the  cancelation of disconnected two point bubbles as $E_T\rightarrow \infty$. 
To answer this, we can focus on a particular diagram contribution to \reef{toexp1} where the bubble is inserted on top of the $s$th propagating state. Then expanding at large $E_T$, we get 
 \be
   \int \prod_{a=1 }^{n-3}  \frac{d x_a}{x_{ia}}  
\,     \frac{ g_{E_i}( \vec x) }{x_{is}} \frac{(gL)^2}{24} \Big[  \underbrace{ \frac{X_s-E_T}{256\pi^3} }_\text{first term}+ \alpha_{0}+ \underbrace{ \frac{X_s(-4-\alpha_{-1})+\alpha_{-1}^2}{  256\pi^3 \, E_T}}_\text{second term}  + O( E_T^{-2})\Big] \, , \label{insp}
 \ee
 where $\alpha_0$ and $\alpha_{-1}$ are functions of $x_{is}$ and are readily   computed using the phase-space formula \reef{vacu22}.~\footnote{We get $\alpha_{-1}=4-E_i+x_s$ and $\alpha_0=\frac{4-\alpha_{-1}-4\log 4+(4-\alpha_{-1}^2/4)\log E_T+[\alpha_{-1}^2/4]\log \alpha_{-1}}{(4\pi)^3(4-\alpha_{-1})}$.}

Therefore we are led to analyse which diagrams  $\begin{minipage}[h]{0.125\linewidth}
\begin{tikzpicture}
\begin{feynman}[small]
 \node [dot] (i1) at (0,0);
 \node [dot] (i2) at (.25,0);  
 \node [dot] (i3) at (.5,0);    
 \node [dot] (i6) at (1.25,0);  
 \node [dot] (i7) at (1.5,0);  
  \node  (x1) at (-.3,0);  
    \node  (x2) at (-.3,.3); 
    \node  (x3) at (-.3,-.3);  
  \node  (y1) at (1.8,0);  
    \node  (y2) at (1.8,.3); 
    \node  (y3) at (1.8,-.3);  
\diagram*{
   (i1) -- [line width=.7mm, red] (i2)-- [line width=.7mm, red] (i3)  -- [line width=.7mm, red, scalar]  (i6)-- [line width=.7mm,red] (i7) ,
     (x1) -- [thick, red] (i1),
    (x2) -- [thick, red] (i1),
        (x3) -- [thick, red] (i1) ,
     (y1) -- [thick, red] (i7),
    (y2) -- [thick, red] (i7),
        (y3) -- [thick, red] (i7)
};
  \end{feynman}
\end{tikzpicture}
  \end{minipage}  $, when dressed with two-point bubbles in between two consecutive vertices,  are sensitive to $X_s$. 
Nicely,  in light of \reef{insp}, the list of such type of diagrams is rather short. 
Indeed, only  those diagrams  ${\cal D}$ that are UV divergent -- when the energy of one of the propagating states $X_s\sim E_T$ -- will probe the  $X_s$ dependence in  \reef{insp} when   dressed with a two-point bubble in between two consecutive vertices. The non-cancellation of such diagrams against the subtraction terms leads to UV divergences not accounted by the counter-terms in \reef{vac1}-\reef{mass1}. This is clear because, having  identified the parts of \reef{toexp1} that are sensitive to the $E_T$ regulator, we can restore $X_s=x_s$, and the  dependence cancels off as
\be
\reef{insp}  =   \int \prod_{a=1}^{n-3} \frac{d x_a}{x_{ia}} \,     g_{E_i}( \vec x)
\,      \Big[   \frac{ \cancel{x_{is}^{-1} X_s} }{256\pi^3}   \frac{(gL)^2}{24}  + O( X_s^{0}, X_s/E_T) \Big]    \, ,
\ee
and we are left with the diagram ${\cal D}$ without the two-point bubble dressing, i.e.  $ \int \prod_{a=1}^{n-3}  d x_a \, x_{ia}^{-1} \,     g_{E_i}( \vec x)$. 

For instance the following  diagram probes the \emph{first term} in \reef{insp}, 
\be
 \begin{minipage}[h]{0.3\linewidth}
\begin{tikzpicture}
\begin{feynman}[small]
 \vertex (i0) at (.4,-.45);
 \node [dot] (i1) at (0,-.2);
 \node [dot] (i2) at (1.2,-.2);  
 \vertex (i3) at (-.4,0);
 \vertex (i4) at (-.2,0);
  \vertex (i5) at (-.3,0);
   \node [dot] (j1) at (.3,.23);
 \node [dot] (j2) at (.9,.23);  
  \node [dot] (a1) at (-1.6,-.6);
 \node [dot] (a2) at (-1,-.6);  
 \node [dot] (a22) at (-.4,-.6);  
 \node [dot] (a3) at (1.6,-.6);   
 \node [dot] (a6) at (2.2,-.6);  
 \node [dot] (a7) at (2.8,-.6);  
  \node  (b1) at (-.3-1.6,-.6);  
    \node  (b2) at (-.3-1.6,-.3); 
    \node  (b3) at (-.3-1.6,-.9);  
  \node  (c1) at (.3+2.8,-.6);  
    \node  (c2) at (.3+2.8,-.3); 
    \node  (c3) at (.3+2.8,-.9); 
\diagram*{
        (a1) -- [line width=.7mm, red] (a2)-- [line width=.7mm, red, scalar] (a22)-- [line width=.7mm, red] (a3)  -- [line width=.7mm, red, scalar]  (a6)-- [line width=.7mm,red] (a7) ,
     (b1) -- [thick, red] (a1),
    (b2) -- [thick, red] (a1),
        (b3) -- [thick, red] (a1) ,
     (c1) -- [thick, red] (a7),
    (c2) -- [thick, red] (a7),
        (c3) -- [thick, red] (a7),
(a22) --  [quarter left, looseness=.7, thick] (i1) --[thick]   (i2) 
   --  [thick, quarter left, looseness= .7] (a3)      ,
     (i1) -- [ half right, looseness=.4,  thick] (i2)  -- [ half right, looseness=.4 ,  thick] (i1),
       (j1) -- [ quarter right, looseness=.2,  thick] (j2)  -- [ quarter right, looseness=.2 ,  thick] (j1) ,
     (j1) -- [ half right, looseness=.6,  thick] (j2)  -- [ half right, looseness=.6 ,  thick] (j1),
   };
  \end{feynman}
\end{tikzpicture}
  \end{minipage}  \, .  \label{cdfgs}
    \vspace{-.28cm} 
  \ee
  Indeed, the sunset diagram involves $ g_{E_i}( \vec x) \sim O(x_s^0)$ because $\Phi_3(x)\sim O(x^0)$.
Following the same logic,  all those   diagrams ${\cal D}$ that   produce $g_{E_i}( \vec x) \sim O(x_s^1)$
will be sensitive to  the \emph{second term}, as well as the \emph{first term},  in \reef{insp}. For instance,
\be
 \begin{minipage}[h]{0.3\linewidth}
\begin{tikzpicture}
\begin{feynman}[small]
 \node [dot] (i1) at (0,0);
 \node [dot] (i2) at (1.2,0);  
 \node [dot] (j1) at (.3,.43);
 \node [dot] (j2) at (.9,.43);  
  \node [dot] (a1) at (-1.6,-.4);
 \node [dot] (a2) at (-1,-.4);  
 \node [dot] (a22) at (-.4,-.4);  
 \node [dot] (a3) at (1.6,-.4);   
 \node [dot] (a6) at (2.2,-.4);  
 \node [dot] (a7) at (2.8,-.4);  
  \node  (b1) at (-.3-1.6,-.4);  
    \node  (b2) at (-.3-1.6,-.1); 
    \node  (b3) at (-.3-1.6,-.7);  
  \node  (c1) at (.3+2.8,-.4);  
    \node  (c2) at (.3+2.8,-.1); 
    \node  (c3) at (.3+2.8,-.7); 
\diagram*{
   (i1) -- [ quarter right, looseness=.1,  thick] (i2)  -- [ quarter right, looseness=.1 ,  thick] (i1) ,
     (i1) -- [ half right, looseness=.4,  thick] (i2)  -- [ half right, looseness=.4 ,  thick] (i1) ,
   (j1) -- [ quarter right, looseness=.2,  thick] (j2)  -- [ quarter right, looseness=.2 ,  thick] (j1) ,
     (j1) -- [ half right, looseness=.6,  thick] (j2)  -- [ half right, looseness=.6 ,  thick] (j1),
        (a1) -- [line width=.7mm, red] (a2)-- [line width=.7mm, red, scalar] (a22)-- [line width=.7mm, red] (a3)  -- [line width=.7mm, red, scalar]  (a6)-- [line width=.7mm,red] (a7) ,
     (b1) -- [thick, red] (a1),
    (b2) -- [thick, red] (a1),
        (b3) -- [thick, red] (a1) ,
     (c1) -- [thick, red] (a7),
    (c2) -- [thick, red] (a7),
        (c3) -- [thick, red] (a7)
};
  \end{feynman}
\end{tikzpicture}
  \end{minipage} 
    \vspace{-.18cm} \label{detdiag1} \, ,
  \ee 
produces  $g_{E_i}( \vec x) \sim O(x_s^1)$ because of the phase space growth $\Phi_4(x)\sim x/(128\pi^2)$ at large $x$. 
  The instance in \reef{detdiag1} is  a straightforward generalisation of the right hand diagram in \reef{discoo}.~\footnote{
  Note  that since $c_0\sim E_T-8m\log E_T$,  the second term in \reef{insp} spoils the cancelation of disconnected bubbles by finite $O(E_T^0)$ pieces. }

So far we have analysed all  the extra  UV divergent terms that are produced when we regulate  diagrams that  
contain a two-point vacuum bubble in between two given vertices with the $E_T$ cutoff. 
To conclude the analysis we need to analyse the cases where the disconnected   vacuum bubble spans one or more vertices as in for instance  the diagrams in \reef{dings}. This is the topic of next section.

\subsubsection{Two-point  bubbles II}
\label{genebub2}

We continue by  analysing    all diagrams consisting  of a bubble  crossing a single vertex,
\be
\begin{minipage}[h]{0.25\linewidth}
\begin{tikzpicture}
\begin{feynman}[small]
 \node [dot] (i1) at (.6,0);
 \node [dot] (i2) at (1.2,0);  
 \node [dot] (i3) at (1.8,0);   
 \node  (i4) at (2.5,0);  
 \node  (i5) at (2.9,0);  
 \node [dot] (i6) at (3.6,0);  
 \node [dot] (i7) at (4.2,0);  
 \node [dot] (j1) at (.3,.4);
 \node [dot] (j2) at (.9,.4); 
  \node  (x1) at (.3,0);  
    \node  (x2) at (.3,.3); 
    \node  (x3) at (.3,-.3);  
  \node  (y1) at (4.5,0);  
    \node  (y2) at (4.5,.3); 
    \node  (y3) at (4.5,-.3);  
\diagram*{
   (i1) -- [line width=.7mm, red] (i2)-- [line width=.7mm, red] (i3)  -- [line width=.7mm, red, scalar] (i4), 
   (i5)-- [scalar, line width=.7mm, red] (i6)-- [line width=.7mm,red] (i7) ,
   (j1) -- [ quarter right, looseness=.25,  thick] (j2)  -- [ quarter right, looseness=.25 ,  thick] (j1) ,
     (j1) -- [ half right, looseness=.7,  thick] (j2)  -- [ half right, looseness=.7 ,  thick] (j1) ,
     (x1) -- [thick, red] (i1),
    (x2) -- [thick, red] (i1),
        (x3) -- [thick, red] (i1) ,
     (y1) -- [thick, red] (i7),
    (y2) -- [thick, red] (i7),
        (y3) -- [thick, red] (i7)
};
  \end{feynman}
\end{tikzpicture}
  \end{minipage}  
  +
  \begin{minipage}[h]{0.25\linewidth}
\begin{tikzpicture}
\begin{feynman}[small]
 \node [dot] (i1) at (.6,0);
 \node [dot] (i2) at (1.2,0);  
 \node [dot] (i3) at (1.8,0);   
 \node  (i4) at (2.5,0);  
 \node  (i5) at (2.9,0);  
 \node [dot] (i6) at (3.6,0);  
 \node [dot] (i7) at (4.2,0);  
 \node [dot] (j1) at (.9,.4);
 \node [dot] (j2) at (1.5,.4); 
  \node  (x1) at (.3,0);  
    \node  (x2) at (.3,.3); 
    \node  (x3) at (.3,-.3);  
  \node  (y1) at (4.5,0);  
    \node  (y2) at (4.5,.3); 
    \node  (y3) at (4.5,-.3);  
\diagram*{
   (i1) -- [line width=.7mm, red] (i2)-- [line width=.7mm, red] (i3)  -- [line width=.7mm, red, scalar] (i4), 
   (i5)-- [scalar, line width=.7mm, red] (i6)-- [line width=.7mm,red] (i7) ,
   (j1) -- [ quarter right, looseness=.25,  thick] (j2)  -- [ quarter right, looseness=.25 ,  thick] (j1) ,
     (j1) -- [ half right, looseness=.7,  thick] (j2)  -- [ half right, looseness=.7 ,  thick] (j1) ,
     (x1) -- [thick, red] (i1),
    (x2) -- [thick, red] (i1),
        (x3) -- [thick, red] (i1) ,
     (y1) -- [thick, red] (i7),
    (y2) -- [thick, red] (i7),
        (y3) -- [thick, red] (i7)
};
  \end{feynman}
\end{tikzpicture}
  \end{minipage}
  + \cdots +
  \begin{minipage}[h]{0.23\linewidth}
\begin{tikzpicture}
\begin{feynman}[small]
 \node [dot] (i1) at (.6,0);
 \node [dot] (i2) at (1.2,0);  
 \node [dot] (i3) at (1.8,0);   
 \node  (i4) at (2.5,0);  
 \node  (i5) at (2.9,0);  
 \node [dot] (i6) at (3.6,0);  
 \node [dot] (i7) at (4.2,0);  
 \node [dot] (j1) at (3.9,.4);
 \node [dot] (j2) at (4.5,.4); 
  \node  (x1) at (.3,0);  
    \node  (x2) at (.3,.3); 
    \node  (x3) at (.3,-.3);  
  \node  (y1) at (4.5,0);  
    \node  (y2) at (4.5,.3); 
    \node  (y3) at (4.5,-.3);  
\diagram*{
   (i1) -- [line width=.7mm, red] (i2)-- [line width=.7mm, red] (i3)  -- [line width=.7mm, red, scalar] (i4), 
   (i5)-- [scalar, line width=.7mm, red] (i6)-- [line width=.7mm,red] (i7) ,
   (j1) -- [ quarter right, looseness=.25,  thick] (j2)  -- [ quarter right, looseness=.25 ,  thick] (j1) ,
     (j1) -- [ half right, looseness=.7,  thick] (j2)  -- [ half right, looseness=.7 ,  thick] (j1) ,
     (x1) -- [thick, red] (i1),
    (x2) -- [thick, red] (i1),
        (x3) -- [thick, red] (i1) ,
     (y1) -- [thick, red] (i7),
    (y2) -- [thick, red] (i7),
        (y3) -- [thick, red] (i7)
};
  \end{feynman}
\end{tikzpicture}
  \end{minipage}  
    \label{discoo3}
\vspace{-.2cm} \, . 
 \ee 
This  expression is given by 
 \be
 \int \prod_{a=1}^{n-3} \frac{d x_a}{x_{ia}} \,  g_{E_i}( \vec x) 
\,  \sum_{s=1}^{n-3}    \frac{(gL)^2}{24}\int_4^{E_T-X_s} \frac{dE}{(2\pi)} \frac{\Phi_4(E)}{[E_i-(E+x_{s-1})][E_i-(E+x_{s})]}     \, ,  \label{toreg22}
 \ee
 where $x_{0}=0$ and  $X_s=\text{max}\{x_{s-1},x_s\}$, and we have regulated the divergent  $dE$ integral
 with the $E_T$ cutoff. 
 Again, we analyse the spurious divergences that are introduced with this regularisation by expanding at large $E_T$.
\be
\reef{toreg22}= \int \prod_{a=1}^{n-3} \frac{d x_a}{x_{ia}} \,  g_{E_i}( \vec x)  \, \sum_{s=1}^{n-3}
\,    \frac{g^2}{24} \left[\beta_0 + \frac{\left(\beta_{-1}-X_s\right)}{256 \pi^3 \, E_T}  + O(E_T^{-2})\right] \, , \label{crossvertex}
\ee
where $\beta_i$ are functions of $E_T$ and $x_s$ but independent of $X_s$. 
In order to  probe the coefficient of  $X_s$  in \reef{crossvertex} 
and generate a UV divergence,  the lower diagram must diverge as $\sim E_T^\alpha$ with $\alpha\geq1$, i.e. we need  $g_{E_i}( \vec x)\gtrsim x_s$. 
This for instance can be achieved  if the vertex below the bubble in \reef{discoo3} is part of another two point bubble, 
\be
 \begin{minipage}[h]{0.12\linewidth}
\begin{tikzpicture}
\begin{feynman}[small]
 \node [dot] (i1) at (-.4,-.05);  
  \node [dot] (i2) at (.8,-.05);  
 \node [dot] (j1) at (.4,.43);
 \node [dot] (j2) at (1.2,.43);  
  \node [dot] (a1) at (-2,-.4);
 \node [dot] (a2) at (-1.4,-.4);  
 \node [dot] (a222) at (-.8,-.4);   
 \node [dot] (a3) at (1.6,-.4);   
 \node [dot] (a6) at (2.2,-.4);  
 \node [dot] (a7) at (2.8,-.4);  
  \node  (b1) at (-.3-2,-.4);  
    \node  (b2) at (-.3-2,-.1); 
    \node  (b3) at (-.3-2,-.7);  
  \node  (c1) at (.3+2.8,-.4);  
    \node  (c2) at (.3+2.8,-.1); 
    \node  (c3) at (.3+2.8,-.7); 
\diagram*{
          (a1) -- [line width=.7mm, red] (a2)-- [line width=.7mm, red, scalar] (a222)-- [line width=.7mm, red] (a3)  -- [line width=.7mm, red, scalar]  (a6)-- [line width=.7mm,red] (a7) ,
     (b1) -- [thick, red] (a1),
    (b2) -- [thick, red] (a1),
        (b3) -- [thick, red] (a1) ,
     (c1) -- [thick, red] (a7),
    (c2) -- [thick, red] (a7),
        (c3) -- [thick, red] (a7),
  (i1) -- [ quarter right, looseness=.1,  thick] (i2)  -- [ quarter right, looseness=.1 ,  thick] (i1) ,
     (i1) -- [ half right, looseness=.4,  thick] (i2)  -- [ half right, looseness=.4 ,  thick] (i1),
   (j1) -- [ quarter right, looseness=.2,  thick] (j2)  -- [ quarter right, looseness=.2 ,  thick] (j1) ,
     (j1) -- [ half right, looseness=.6,  thick] (j2)  -- [ half right, looseness=.6 ,  thick] (j1)
     };
  \end{feynman}
\end{tikzpicture}
  \end{minipage}  \, . \hspace{2.8cm}
    \vspace{-.28cm} \label{detdiag2}
  \ee 
More generally iterated bubble diagrams also generates a linear divergence
 \vspace{-.1cm}\be
 \begin{minipage}[h]{0.19\linewidth}
\begin{tikzpicture}
\begin{feynman}[small]
 \node [dot] (i1) at (0,0);  
  \node [dot] (i2) at (1,0);  
 \node [dot] (j1) at (.7,-.43);
 \node [dot] (j2) at (1.3,-.43);  
 \vertex (X1) at (-.8,-.1) {\scriptsize{\bf{\dots}}};  
 \node [dot] (k1) at (-.7,-.43);
 \node [dot] (k2) at (.3,-.43);  
  \node [dot] (a1) at (-2-.8,-.9);
 \node [dot] (a2) at (-1.4-.8,-.9);  
 \node [dot] (a222) at (-.8-.8,-.9);  
 \node [dot] (a3) at (1.6,-.9);   
 \node [dot] (a6) at (2.2,-.9);  
 \node [dot] (a7) at (2.8,-.9);  
  \node  (b1) at (-.3-2-.8,-.9);  
    \node  (b2) at (-.3-2-.8,-.9+.3); 
    \node  (b3) at (-.3-2-.8,-.9-.3);  
  \node  (c1) at (.3+2.8,-.9);  
    \node  (c2) at (.3+2.8,-.9+.3); 
    \node  (c3) at (.3+2.8,-.9-.3); 
     \vertex (v1) at (1.45,.35);
     \vertex (v2) at (1.45,-1.25);
     \vertex (v3) at (1.15,.35);
     \vertex (v4) at (1.15,-1.25);
     \vertex (v5) at (.85,.35);
     \vertex (v6) at (.85,-1.25);
     \vertex (v7) at (.5,.35);
     \vertex (v8) at (.5,-1.25);
     \vertex (v9) at (.15,.35);
     \vertex (v10) at (.15,-1.25);
     \vertex (v11) at (-.15,.35);
     \vertex (v12) at (-.15,-1.25);
\diagram*{
          (a1) -- [line width=.7mm, red] (a2)-- [line width=.7mm, red, scalar] (a222)-- [line width=.7mm, red] (a3)  -- [line width=.7mm, red, scalar]  (a6)-- [line width=.7mm,red] (a7) ,
     (b1) -- [thick, red] (a1),
    (b2) -- [thick, red] (a1),
        (b3) -- [thick, red] (a1) ,
     (c1) -- [thick, red] (a7),
    (c2) -- [thick, red] (a7),
        (c3) -- [thick, red] (a7),
  (i1) -- [ quarter right, looseness=.1,  thick] (i2)  -- [ quarter right, looseness=.1 ,  thick] (i1) ,
     (i1) -- [ half right, looseness=.4,  thick] (i2)  -- [ half right, looseness=.4 ,  thick] (i1),
   (j1) -- [ quarter right, looseness=.2,  thick] (j2)  -- [ quarter right, looseness=.2 ,  thick] (j1) ,
     (j1) -- [ half right, looseness=.6,  thick] (j2)  -- [ half right, looseness=.6 ,  thick] (j1),
   (k1) -- [ quarter right, looseness=.1,  thick] (k2)  -- [ quarter right, looseness=.1 ,  thick] (k1) ,
     (k1) -- [ half right, looseness=.4,  thick] (k2)  -- [ half right, looseness=.4 ,  thick] (k1),
     (v1) -- [scalar, red, thick] (v2),
     (v3) -- [scalar, red, thick] (v4),
     (v5) -- [scalar, red, thick] (v6),
     (v7) -- [scalar, red, thick] (v8),
     (v9) -- [scalar, red, thick] (v10),
     (v11) -- [scalar, red, thick] (v12)
     };
  \end{feynman}
\end{tikzpicture}
  \end{minipage} \hspace{2.8cm}
    \vspace{-.2cm}  \label{arg}
  \ee 
   where vertical lines are drawn to guide the eye to the  propagating states. 
     Iterated bubble diagrams of the form \reef{arg} are the only type of  diagrams that diverge as   $\sim E_T^\alpha$ with $\alpha\geq1$, in particular they have $\alpha=1$. This completes the generalisation of the left hand diagram in \reef{discoo}.

  Note that two-point bubbles spanning more than one vertex  do not introduce new UV divergences. In such case, the bubble is rendered $\log$ divergent at most. 
  Then, the difference between cutting the integrals at $E_T$ or $E_T-X_s$ is suppressed by  $E_T$.

Thus, we have concluded the identification of all those diagrams 
for which the $E_T$ regulator leads to extra UV divergences, which would not be present in a covariant scheme. 
All in all, these are diagrams of the form  \reef{discoo2} and  \reef{arg}.~\footnote{
Strictly speaking we have only discussed those diagrams arising from the \emph{first contribution} of \reef{1stc}. Note however that those diagrams arising form the \emph{subtraction terms} in \reef{subte} have either equal or softer UV properties due to the extra factors $[E_i-E_j+i\eps]^{-1}$.}

\subsubsection{Further comments}

From the previous discussion it is clear that there are further diagrams that have a sensitivity to the non-covariant cutoff that scales as $E_T^0$, i.e. remains finite, in the large $E_T$ limit. Such finite pieces are potentially problematic since they will make it challenging to match a Hamiltonian Truncation calculation with a covariant calculation.
Consequently, we now examine some examples of diagrams that do and do not lead to finite corrections.

For instance, a  source of finite $O(E_T^0)$ terms comes from the the correction to the energy of an $n$-particle state [with unperturbed energy $E_n$] by the disconnected $g^2$ bubble of \eqref{vac1}, i.e.
 \be
 \cE_n^{(2)} \supset  \, \begin{minipage}[h]{0.09\linewidth}
\begin{tikzpicture}
\begin{feynman}[small]
 \node [dot] (i1) at (0,1);
 \node [dot] (i2) at (.8,1);  
 \vertex (j1) at (-.4,.6);
  \vertex (j2) at (1.2,.6);  
 \vertex (j3) at (-.4,.5);
 \vertex (j4) at (1.2,.5);  
 \vertex (j5) at (-.4,0);
 \vertex (j6) at (1.2,0);  
 \vertex (k) at (.4,.35) {$\vdots$};
\diagram*{
   (i1) -- [ quarter right, looseness=.3,  thick] (i2)  -- [ quarter right, looseness=.3 ,  thick] (i1) ,
     (i1) -- [ half right, looseness=.8,  thick] (i2)  -- [ half right, looseness=.8 ,  thick] (i1) ,
   (j1) -- [   thick] (j2),
      (j3) -- [   thick] (j4),
         (j5) -- [   thick] (j6)  
   };
  \end{feynman}
\end{tikzpicture}
  \end{minipage}\,  - c_0(E_T) L^2 \, ,  \label{tuc1}
  \ee
where the   diagram shifts the energy level by
$
\Delta E_n = \frac{(g L)^2}{24} \int_{4}^{E_T-E_n}  dE \Phi_4(E)/(2\pi E )     
 $. 
The UV divergence is canceled by $V_{nn} = c_0\left(E_T\right)L^2$,  leaving a finite piece 
\be
\Delta E_n -c_0 L^2 = \left(-\frac{\log 2}{384 \pi^3}+ \frac{1}{6144 \pi^3}E_n\right) g^2 L^2 ~.
\ee
This is due to the linear divergence present in the two-point bubble. 
Such a finite piece is not expected if a covariant scheme is used, namely a regulator that cuts the loop  in  \reef{tuc1} independently of the state flowing below.  

Note that the sunset diagram does not lead to an analogous effect because it is only $\log$ divergent. For example, consider the diagrams
 \be
 \cE_n^{(2)} \supset  \,  \begin{minipage}[h]{0.085\linewidth}
\begin{tikzpicture}
\begin{feynman}[small]
 \vertex (i0) at (-.4,1);
 \node [dot] (i1) at (0,1);
 \node [dot] (i2) at (.8,1);  
 \vertex (i3) at (1.2,1);
  \vertex (j3) at (-.4,.5);
 \vertex (j4) at (1.2,.5);  
 \vertex (j5) at (-.4,0);
 \vertex (j6) at (1.2,0);  
 \vertex (k) at (.4,.35) {$\vdots$};
\diagram*{
   (i0) -- [thick]   (i3) ,
     (i1) -- [ half right, looseness=1.,  thick] (i2)  -- [ half right, looseness=1. ,  thick] (i1) ,
      (j3) -- [   thick] (j4),
         (j5) -- [   thick] (j6) 
};
  \end{feynman}
\end{tikzpicture}
  \end{minipage}  \
  +    \ 
  \begin{minipage}[h]{0.085\linewidth}
\begin{tikzpicture}
\begin{feynman}[small]
 \vertex (im2) at (1.1,1);
  \vertex (im1) at (1,1);
 \vertex (i0) at (.9,.9);
 \node [dot] (i1) at (0,1);
 \node [dot] (i2) at (.8,1);  
 \vertex (i3) at (-.1,1.1);
 \vertex (i4) at (-.2,1);
  \vertex (i5) at (-.3,1);
   \vertex (j3) at (-.4,.5);
 \vertex (j4) at (1.2,.5);  
 \vertex (j5) at (-.4,0);
 \vertex (j6) at (1.2,0);  
 \vertex (k) at (.4,.35) {$\vdots$};
 \vertex (g1) at (.4,-.24);
 \vertex (g2) at (.5,-.24);
\diagram*{
(im2)--[thick](im1)--[quarter right, thick]   (i0) -- [half left, looseness=1.3, thick] (i1) --[thick]   (i2) 
   --  [thick, half right, looseness= 1.3] (i3)     -- [thick, quarter left , looseness = 1] (i4) --[thick] (i5)  ,
     (i1) -- [ half right, looseness=1.,  thick] (i2)  -- [ half right, looseness=1. ,  thick] (i1)   ,
           (j3) -- [   thick] (j4),
         (j5) -- [   thick] (j6) ,
         (g1) -- [   thick, white] (g2) 
};
  \end{feynman}
\end{tikzpicture}
  \end{minipage}
  \  -  \,  c_2(E_T) \, ,  \label{tsh}
   \ee 
which are given by
\be
  \frac{1}{6} \int_{3}^{E_T-X} \frac{dE}{2\pi} \frac{\Phi_3\left(E\right)}{E_n-\left(E+x\right)}  +
    \frac{1}{6} \int_{3}^{E_T-2-X} \frac{dE}{2\pi} \frac{\Phi_3\left(E\right)}{E_n-\left(E+2+x\right)} ~, \label{tsh2}
\ee
where $x$ is the energy of the state below the loop, and  $X=x$, but it is useful to treat them as independent variables. 
We are led to 
\be
\reef{tsh2}- c_2(E_T) =-   \frac{\log(E_T/[3-E_n+x])}{96\pi^2}-     \frac{\log([3-E_n+x]/3)}{32\pi^2 (E_n-x)}  + O(X E_T^{-1}, E_T^{-1}) \, , \label{tsh0}
\ee
namely there is no dependence on $X \times O(E_T^0)$. 
This is in accord with covariant regularisation. For instance, we may regulate the integrals in dimensional regularisation, or  simply by $dE\rightarrow dE \, (\mu/E)^\eps$ with $\eps>0$.
Then, the
$c_2$ counter-term is given by  
 $
c_2(\eps) =  -1/(96\pi^2\eps) 
 $, 
 such that the calculation of 
\be
\begin{minipage}[h]{0.085\linewidth}
\begin{tikzpicture}
\begin{feynman}[small]
 \vertex (i0) at (-.3,0);
 \node [dot] (i1) at (0,0);
 \node [dot] (i2) at (.8,0);  
 \vertex (i3) at (1.1,0);
\diagram*{
   (i0) -- [thick]   (i3) ,
     (i1) -- [ half right, looseness=1.,  thick] (i2)  -- [ half right, looseness=1. ,  thick] (i1)
};
  \end{feynman}
\end{tikzpicture}
  \end{minipage}+ \begin{minipage}[h]{0.085\linewidth}
\begin{tikzpicture}
\begin{feynman}[small]
 \vertex (im2) at (1.1,0);
  \vertex (im1) at (1,0);
 \vertex (i0) at (.9,-.1);
 \node [dot] (i1) at (0,0);
 \node [dot] (i2) at (.8,0);  
 \vertex (i3) at (-.1,.1);
 \vertex (i4) at (-.2,0);
  \vertex (i5) at (-.3,0);
\diagram*{
(im2)--[thick](im1)--[quarter right, thick]   (i0) -- [half left, looseness=1.3, thick] (i1) --[thick]   (i2) 
   --  [thick, half right, looseness= 1.3] (i3)     -- [thick, quarter left , looseness = 1] (i4) --[thick] (i5)  ,
     (i1) -- [ half right, looseness=1.,  thick] (i2)  -- [ half right, looseness=1. ,  thick] (i1)
};
  \end{feynman}
\end{tikzpicture}
  \end{minipage}- c_2 \label{tmatch}
\ee
is matched in either  the $E_T$ cutoff or $dE \, (\mu/E)^\eps$ scheme with $\mu=m$ [and using respectively $c_2(E_T)$ or $c_2(\eps)$]. 
But, the point is that once we fix the $c_2$ countertem such that \reef{tmatch} matches in the $E_T$ and $\eps$ schemes,   other diagrams like the ones in \reef{tsh} automatically match in the two schemes. 
Indeed,  upon performing the following integrals 
\be
  \frac{1}{6} \int_{3}^{\infty} \frac{dE}{2\pi} \frac{\mu^\eps}{E^\eps} \frac{\Phi_3\left(E\right)}{E_n-\left(E+x\right)]}  +
    \frac{1}{6} \int_{3}^{\infty} \frac{dE}{2\pi}\frac{\mu^\eps}{E^\eps}  \frac{\Phi_3\left(E\right)}{E_n-\left(E+2+x\right) }    - c_2(\eps) ~, \label{tsh3}
\ee
we find   $
\lim_{ET\rightarrow \infty}\reef{tsh0} = \lim_{\eps\rightarrow 0}\reef{tsh3}|_{\mu=m} 
$. 

We note however, that 
the sunset diagram can lead to finite $O(E_T^0)$ pieces when a two-point vacuum bubble spans one of or both its vertices.
For example, consider embedding the  sunset diagram  in a  diagram [either connected or disconnected] such that its two vertices are consecutive in time. These type of diagrams are given by 
\be
 \int \prod_{i=1}^{n-3} \frac{d x_i}{x_i} \,  g_{E_n}( \vec x) 
\,   \frac{1}{x_s} \frac{g^2}{6} \int_3^{E_T-X_s} \frac{dE}{2\pi}\frac{\Phi_3\left(E\right)}{E_e - \left(E+x_s \right)} ~ ,
\ee
where again $X_s=x_s$ but it is again useful to keep them independent to distinguish the scheme dependent pieces. 
The  sensitivity to $X_s$ for large $E_T$ is  
\be
\Delta_s= \int \prod_{i=1}^{n-3} \frac{d x_i}{x_i} \,  g_{E_n}( \vec x) 
\,   \frac{1}{x_s} \frac{g^2}{192\pi^2} \left( \frac{X_s}{E_T} +O\left(E_T^{-2}\right)  \right) ~.
\ee
If  a disconnected two-point bubble with energy $x_s$ spans the sunset diagram $g_{E_n}(x_s) \sim x_s$, which leads to a finite piece in the large $E_T$ limit. Meanwhile if, with the sunset bubble removed, the diagram is convergent or log divergent  $\Delta_s$ vanishes in the large $E_T$ limit.

If the vertices that comprise the sunset diagram are not consecutive in time, the degree of divergence is lowered. No such diagrams lead to new UV divergences, although those in which a two-point  disconnected bubble spans one of the sunset vertices give a finite regulator dependent piece.

 By virtue of the fairly weak divergences in our chosen theory there are only a handful of types of diagrams where  an $O(E_T^0)$ piece, not present in covariant regularisation, is generated after removing the cutoff $E_T\rightarrow \infty$. 
Further details are provided in  appendix~\ref{app:finite}.

\subsection{Patching up perturbation theory with $E_T$ regularisation} \label{ssPatch}

At this point we have nailed down the problems with perturbation theory in the $E_T$ regularisation scheme precisely. All of the issues have a common origin: the problem of  missing states! Namely, if a loop of energy $E_l$   appears   as a sub-diagram of HPT,  then it belongs  to a state propagating between two vertices that contains other particles of total energy $E_\text{ext}$. The $E_T$ cutoff restricts the energy of the propagating state to $E_l+E_\text{ext}\leq E_T$. Therefore $E_l$ explores energies strictly below $E_T$, which does not coincide with the energy being probed by the primitive UV divergencies that we identified in \reef{vac1} and \reef{mass1}. Instead, if all loops were democratically cut at energy $E_T$ the counter-terms \eqref{vac1} and \eqref{mass1} would be sufficient to a obtain a finite theory in the $E_T\rightarrow \infty$ limit.  

We will now fix all these problems.  
We will do so by ensuring  that all loop diagrams that have a UV divergent sensitivity to the regulator are effectively cut equally at $E_T$. Such diagrams involve the two point bubble inserted with its vertices sequential in time \reef{discoo2} and towers of the two point bubbles \reef{arg}. 
As we argued above, if the loops of the disconnected two point bubbles are cut at $E_T$, the cancelation of such disconnected diagrams against the subtraction terms in \reef{subte} is automatic.

\subsubsection{Patch I}

First, we show how the effects of the states that are missing from the two point bubble diagram with vertices that are consecutive in time can be accounted for. We introduce a state dependent counterterm that  is given by
\be
\delta V_{nn, i}^I = 
\
 \begin{minipage}[h]{0.1\linewidth}
\begin{tikzpicture}
\begin{feynman}[small]
 \node [ crossed dot,rotate=45 ](i1) at (.4,.8);
 \vertex (j1) at (-.4,.6);
  \vertex (j2) at (1.2,.6);  
 \vertex (j3) at (-.4,.5);
 \vertex (j4) at (1.2,.5);  
 \vertex (j5) at (-.4,0);
 \vertex (j6) at (1.2,0);  
 \vertex (k) at (.4,.37) {$\vdots$};
\diagram*{
   (i1),
   (j1) -- [   thick] (j2),
      (j3) -- [   thick] (j4),
         (j5) -- [   thick] (j6)  
   };
  \end{feynman}
\end{tikzpicture}
  \end{minipage}
= \frac{g^2}{24} \int_{E_T-x_n}^{E_T} \frac{dE}{2\pi} \frac{\Phi_4\left(E\right)}{E_{i} - \left(x_n + E\right)} ~. \label{patchvac}
\ee
Note that, due to the appearance of $E_i$, the counter-term depends on which energy level is being calculated.

Considering the \emph{1st contribution} at each order in perturbation theory in \eqref{1stc}: For any diagram involving a two point bubble we can immediately write a diagram at the next order down in perturbation theory where the two vertices that make up the bubble are replaced by a single insertion of $V_{nn}$:
\be
 \begin{minipage}[h]{0.27\linewidth}
\begin{tikzpicture}
\begin{feynman}[small]
 \node [dot] (j1) at (0,.-.1);
 \node [dot] (j2) at (.6,.-.1);  
  \node [dot] (a1) at (-1.6,-.4);
 \node [dot] (a2) at (-1,-.4);  
 \node [dot] (a22) at (-.4,-.4);  
 \node [dot] (a3) at (1.,-.4);   
 \node [dot] (a6) at (1.6,-.4);  
 \node [dot] (a7) at (2.2,-.4);  
  \node  (b1) at (-.3-1.6,-.4);  
    \node  (b2) at (-.3-1.6,-.1); 
    \node  (b3) at (-.3-1.6,-.7);  
  \node  (c1) at (.3+2.2,-.4);  
    \node  (c2) at (.3+2.2,-.1); 
    \node  (c3) at (.3+2.2,-.7); 
\diagram*{
   (j1) -- [ quarter right, looseness=.2,  thick] (j2)  -- [ quarter right, looseness=.2 ,  thick] (j1) ,
     (j1) -- [ half right, looseness=.6,  thick] (j2)  -- [ half right, looseness=.6 ,  thick] (j1),
        (a1) -- [line width=.7mm, red] (a2)-- [line width=.7mm, red, scalar] (a22)-- [line width=.7mm, red] (a3)  -- [line width=.7mm, red, scalar]  (a6)-- [line width=.7mm,red] (a7) ,
     (b1) -- [thick, red] (a1),
    (b2) -- [thick, red] (a1),
        (b3) -- [thick, red] (a1) ,
     (c1) -- [thick, red] (a7),
    (c2) -- [thick, red] (a7),
        (c3) -- [thick, red] (a7)
};
  \end{feynman}
\end{tikzpicture}
  \end{minipage}
  +
   \begin{minipage}[h]{0.27\linewidth}
\begin{tikzpicture}
\begin{feynman}[small]
 \node [crossed dot,rotate=45] (j1) at (.3,-.2);  
  \node [dot] (a1) at (-1.6,-.4);
 \node [dot] (a2) at (-1,-.4);  
 \node [dot] (a22) at (-.4,-.4);  
 \node [dot] (a3) at (1.,-.4);   
 \node [dot] (a6) at (1.6,-.4);  
 \node [dot] (a7) at (2.2,-.4);  
  \node  (b1) at (-.3-1.6,-.4);  
    \node  (b2) at (-.3-1.6,-.1); 
    \node  (b3) at (-.3-1.6,-.7);  
  \node  (c1) at (.3+2.2,-.4);  
    \node  (c2) at (.3+2.2,-.1); 
    \node  (c3) at (.3+2.2,-.7); 
\diagram*{
        (a1) -- [line width=.7mm, red] (a2)-- [line width=.7mm, red, scalar] (a22)-- [line width=.7mm, red] (a3)  -- [line width=.7mm, red, scalar]  (a6)-- [line width=.7mm,red] (a7) ,
     (b1) -- [thick, red] (a1),
    (b2) -- [thick, red] (a1),
        (b3) -- [thick, red] (a1) ,
     (c1) -- [thick, red] (a7),
    (c2) -- [thick, red] (a7),
        (c3) -- [thick, red] (a7)
};
  \end{feynman}
\end{tikzpicture}
  \end{minipage}
    \vspace{-.28cm}   \  .  \label{cittt}
  \ee 
The former two diagrams replace  \eqref{toexp1} by
 \be
\reef{cittt}=   \int \prod_{a=1}^{n-3} \frac{d x_a}{x_{ia}} \,   g_{E_i}( \vec x) 
\,   \sum_{s=1}^{n-3} \left( \frac{1}{x_{is}} \frac{(gL)^2}{24}\int_4^{E_T-X_s} \frac{dE}{(2\pi)} \frac{\Phi_4(E)}{E_i-( E+x_s)} + \frac{1}{x_{is}} \delta V^I_{ss}  \right) \, .   \label{ssct}
 \ee
  The role of  $\delta V^I_{nn}$ is to  account for the states that are missing due to the non-covariant regulator, so that the $X_s$ dependence in \reef{ssct} is removed. 
Note that $V_{00} =0$ since no states are missing in this case.

The counterterm  \reef{patchvac} also patches up the same problem in the \emph{2nd contribution}  of \eqref{1stc}, i.e. the subtraction terms in \reef{subte}. At order $n$ in perturbation theory this includes expressions of the form
\be
\cE_0^{(j)} \,    \frac{V_{ik_1} V_{k_1k_2}  \cdots  V_{k_{n-j-1}i}}{E_{ik_1}E_{ik_2}\cdots E_{ik_{n-j-1}}}\sum_{s=1}^{n-j-1}\frac{1}{E_{ik_s}} ~. \label{eq2ndform}
\ee 
States missing from two-point bubbles inside $\cE_0^{(j)}$ are corrected for by a similar term at order $n-1$ that has a prefactor $\cE_0^{(j-1)}$ containing $\delta V_{nn}^I$. For example, for $j=2$ this amounts to simply replacing $\cE_0^{(2)}$ with $\delta V_{nn}^I$. Two-point bubbles appearing in the second factor in \eqref{eq2ndform}, i.e. the part involving $V_{ik_1}\cdots V_{k_{n-j-1}i}$ , are only dangerous if they do not have the extra propagator on them. These are patched up by insertions of $\delta V_{nn}^{I}$ in the second factor of terms at the next order down.\footnote{Two point bubbles with an extra propagator on are not patched, but these are logarithmically divergent so only lead to finite corrections due to the $E_T$ regulator.} At high order there are subtraction terms similar to \eqref{eq2ndform} involving multiple factors of $\cE_0^k$, but these are still patched up in the same way.

This exhausts all of the places that a two point bubble [with vertices consecutive in time] can be missing states in the perturbation theory, and also all of the places that the counterterm \reef{patchvac} can be inserted. Hence, \eqref{patchvac} is sufficient to completely fix the problem we laid down in \ref{genebub}.

\subsubsection{Patch II}

The second issue to address  is how to obtain a theory without a UV divergent sensitivity to  the missing states of diagrams involving towers of overlapping two point bubbles \reef{arg},
\vspace{-.2cm}\be
 \begin{minipage}[h]{0.19\linewidth}
\begin{tikzpicture}
\begin{feynman}[small]
 \node [dot] (i1) at (0,0);  
  \node [dot] (i2) at (1,0);  
 \node [dot] (j1) at (.7,-.43);
 \node [dot] (j2) at (1.3,-.43);  
 \vertex (X1) at (-.8,-.1) {\scriptsize{\bf{\dots}}};  
 \node [dot] (k1) at (-.7,-.43);
 \node [dot] (k2) at (.3,-.43);  
  \node [dot] (a1) at (-2-.8,-.9);
 \node [dot] (a2) at (-1.4-.8,-.9);  
 \node [dot] (a222) at (-.8-.8,-.9);  
 \node [dot] (a3) at (1.6,-.9);   
 \node [dot] (a6) at (2.2,-.9);  
 \node [dot] (a7) at (2.8,-.9);  
  \node  (b1) at (-.3-2-.8,-.9);  
    \node  (b2) at (-.3-2-.8,-.9+.3); 
    \node  (b3) at (-.3-2-.8,-.9-.3);  
  \node  (c1) at (.3+2.8,-.9);  
    \node  (c2) at (.3+2.8,-.9+.3); 
    \node  (c3) at (.3+2.8,-.9-.3); 
\diagram*{
          (a1) -- [line width=.7mm, red] (a2)-- [line width=.7mm, red, scalar] (a222)-- [line width=.7mm, red] (a3)  -- [line width=.7mm, red, scalar]  (a6)-- [line width=.7mm,red] (a7) ,
     (b1) -- [thick, red] (a1),
    (b2) -- [thick, red] (a1),
        (b3) -- [thick, red] (a1) ,
     (c1) -- [thick, red] (a7),
    (c2) -- [thick, red] (a7),
        (c3) -- [thick, red] (a7),
  (i1) -- [ quarter right, looseness=.1,  thick] (i2)  -- [ quarter right, looseness=.1 ,  thick] (i1) ,
     (i1) -- [ half right, looseness=.4,  thick] (i2)  -- [ half right, looseness=.4 ,  thick] (i1),
   (j1) -- [ quarter right, looseness=.2,  thick] (j2)  -- [ quarter right, looseness=.2 ,  thick] (j1) ,
     (j1) -- [ half right, looseness=.6,  thick] (j2)  -- [ half right, looseness=.6 ,  thick] (j1),
   (k1) -- [ quarter right, looseness=.1,  thick] (k2)  -- [ quarter right, looseness=.1 ,  thick] (k1) ,
     (k1) -- [ half right, looseness=.4,  thick] (k2)  -- [ half right, looseness=.4 ,  thick] (k1)
     };
  \end{feynman}
\end{tikzpicture}
  \end{minipage} \hspace{2.8cm}
     \label{ite2}
  \ee 
As we argued above, these diagrams are problematic if we regulate with an $E_T$ cutoff.
The bubble integrals of  former diagram are schematically given by
\be
\int_4^{E_T-x_s} dE_1 \int_4^{E_T-x_s-E_1} dE_2 \int_{4}^{E_T-x_s-E_1-E_2} dE_3 
\frac{\Phi_4(E_1)}{E_1} \frac{1}{E_1+E_2}\frac{\Phi_4(E_2)}{E_2} \frac{1}{E_2+E_3}\frac{\Phi_4(E_3)}{E_3} \cdots \, , \label{integbub}
\ee
where $x_s$ is the energy of the state  below the iterated bubbles, and we have dropped the dependence on the eigenstate $E_i$ and $x_s$ on the energy denominators.

A moment of thought reveals that this problem is hard to solve with a simple counter-term that can be written in closed form.~\footnote{
For instance, one may be tempted to correct the quartic coupling in a way that modifies all of its $a^{4}$, $\left(a^{\dagger}\right)^{4}$ etc. components, but we have found this   not to work. }
We instead offer a pragmatic solution, while we look forward to more elegant solutions in the future. 
We proceed in a similar fashion as we did in the previous section, i.e.  we  introduce a further state-dependent vacuum counter-term, 
to account for the missing states in \reef{ite2}
\be
\delta V^{II}_{nn}  =
 \frac{(gL)^4}{24^2}
\left(\int_4^{E_T} dE_1
 \int_4^{E_T}  dE_2  - \int_4^{E_T-x_s}  dE_1 
 \int_4^{E_T-E_1-x_s}  dE_2   \right) I(E_1,E_2) + \cdots    \label{itect} 
\ee
where we have only  explicitly written  the expression corresponding to  two iterated bubbles, and the integrand is given by
\be
I(E_1,E_2)= 
  \frac{\Phi_4(E_1)\Phi_4(E_2)}{[E_i-(E_1+x_s)][E_i-(E_1+E_2+x_{s})][E_i-(E_2+x_{s})]}  \, . 
\ee
It is straightforward to write down the integral expression for a tower containing $n$ iterated bubbles, denoted by the dots in \reef{itect}.  

The counter-term in \reef{itect} can be  computed in a  numerical code up to  arbitrarily large numbers of bubbles. In our implementation below we patch up diagrams involving up to five overlapping bubbles, i.e. order $g^{10}$. The results that we obtain are not sensitive to including corrections corresponding to patching up larger numbers of bubbles. This indicates that the remaining cutoff dependence is not dominating our results.


 \subsubsection{Comment on off-diagonal counter-terms  \cite{Rutter:2018aog}}

Ref.~\cite{Rutter:2018aog} stressed a potential problem with off-diagonal counter-terms.  
The main point is the following. Let us divide our potential as $V\rightarrow V+C^{(2)}+C^{(3)}$, where $C^{(n)}$ is a counter-term operator of $O(g^n)$ and $V=O(g)$.
On one hand, the  order $O(g^2)$ correction to the energy levels  is
\be
 V_{i k} E_{ik}^{-1}V_{k j}+ C^{(2)}_{i j } \ , \quad \text{with}\quad  i=j  \ ,  \label{off1}
\ee
where recall that a sum over $k\neq i$ is implicit. The off-diagonal terms of $C^{(2)}$  are not fixed by demanding that \reef{off1} is finite when the regulator is removed. 
On the other hand the third order correction to the energy levels reads
\be
V_{i k} E_{ik}^{-1}V_{k k^\prime}E_{ik^\prime} V_{k^\prime i} +C^{(2)}_{i k} E_{ik}^{-1}V_{k j} + V_{i k} E_{ik}^{-1}C^{(2)}_{k j} +C^{(3)}_{ii} - \text{subtraction terms} \, . \label{off2}
\ee
Thus the $O(g^3)$ correction to the energy levels is sensitive to the off-diagonal terms of $C^{(2)}$.
Now, a natural question arises about whether we should take care about the off-diagonal terms of $C_{ij}^{(2)}$. Or more generally, what is the structure of UV divergences in \reef{off2}.~\footnote{The \emph{subtraction terms} in \reef{off2} are independent of $C^{(n)}$ and thus irrelevant for our main point. }

Let us rephrase the problem from the perspective of our detailed diagrammatic understanding. We consider the $g_2\phi^2+g_4\phi^4$ theory in $d$ dimensions as an instance, although similar comments apply more generally. 
As we have argued above if a hard cutoff in the Hilbert space is used [like $E_T$], divergent loops embedded in larger diagrams are cut at energies strictly below the cutoff [for instance see the first diagram in \reef{sunsetdiv}]. 
Now it is not hard to imagine a sub-divergence arising from an off-diagonal term of $C^{(2)}$. For instance,  consider the following $O(g_2 g_4^2)$ correction to the  one particle state's mass
\vspace{-.3cm}\be
  \begin{minipage}[h]{0.085\linewidth}
\begin{tikzpicture}
\begin{feynman}[small]
 \vertex (im1) at (.8,.9);
 \node [dot] (i0) at (-.25,1);
 \node [dot] (i1) at (.1,1);
 \node [dot] (i2) at (.5,1);  
 \vertex (i3) at (.25,1.3);
 \vertex (i4) at (-.5,1.3);
 \vertex (X1) at (.3,1.5);
 \vertex (X2) at (.3,.5);
 \vertex (X3) at (.3,.4){\scriptsize $-{\roig E_l}-E_\text{ext}$};
 \vertex (Y2) at (.3,1.87);
  \vertex (Y3) at (.3,1.87);
\diagram*{
 (im1) -- [ thick, quarter left] (i0) -- [ thick] (i1) --[thick]   (i2)    --  [thick, quarter  right, looseness= 1.3] (i3)    --  [thick] (i4)   ,
     (i1) -- [ half right, looseness=1.,  thick, red] (i2)  -- [ half right, looseness=1. ,  thick, red] (i1)   ,
     (X1) -- [scalar, thick, red] (X2),
      (Y1) -- [scalar, thick, white] (Y2)
};
  \end{feynman}
\end{tikzpicture}
  \end{minipage}+ \,
    \begin{minipage}[h]{0.07\linewidth}
\begin{tikzpicture}
\begin{feynman}[small]
 \vertex (im1) at (.8,.9);
 \node [dot] (i0) at (0,1);
 \node [crossed dot] (i2) at (.5,1);  
 \vertex (i3) at (.25,1.3);
 \vertex (i4) at (-.25,1.3);
\diagram*{
 (im1) -- [ thick, quarter left] (i0) --[thick]   (i2)    --  [thick, quarter  right, looseness= 1.3] (i3)    --  [thick] (i4)   ,
   ,
};
  \end{feynman}
\end{tikzpicture}
  \end{minipage} \, .  \label{offd1}
  \vspace{-.2cm}
\ee
where $\Va$ denotes the counter-term. 
Indeed, $C^{(2)}$   involves a diagonal correction to cancel the divergence from diagrams like
 \be
  \begin{minipage}[h]{0.06\linewidth}
\begin{tikzpicture}
\begin{feynman}[small]
 \vertex (i0) at (-.15,1);
 \node [dot] (i1) at (.1,1);
 \node [dot] (i2) at (.5,1);  
 \vertex (i3) at (.75,1.);
 \vertex (X1) at (.3,1.5);
 \vertex (X2) at (.3,.5);
 \vertex (X3) at (.3,.4){\scriptsize $-{  E_l}$};
\diagram*{
 (i0) -- [ thick] (i1) --[thick]   (i2)    --  [thick ] (i3)    ,
     (i1) -- [ half right, looseness=1.,  thick] (i2)  -- [ half right, looseness=1. ,  thick] (i1)   ,
     (X1) -- [scalar, thick, red] (X2),
};
  \end{feynman}
\end{tikzpicture} \label{offd2}
  \end{minipage}   \, ,
    \vspace{-.4cm}
\ee correcting the one particle state at $O(g_4^2)$. 

We see the problem of off-diagonal counter-terms as an instance of the problem of missing states. 
Namely if the energy being propagated in between two consecutive vertices is cut at $E_T$, then the energy of the state being propagated in a loop depends on where the lines external to the loop end [these external lines  are either or not attached to the loop].  

For instance, returning to the example above, the loop in \reef{offd1} is cut at $E_l\leq E_T-E_\text{ext}$ with $E_\text{ext}=2m$, while the loop in \reef{offd2} is cut at $E_l\leq E_T$. 
At this order this difference amounts to a finite scheme piece. However, after embedding these diagrams in higher order ones, $E_\text{ext}$ may belong to an energy of another loop. Then, if  $d\geq 7/2$  the sunset diagram is power like divergent;  and the remaining $E_\text{ext}$ that is left un-cancelled probes $E_T$ energies and  introduces new UV divergences. 
In this case, one must define a  counter-term that captures this effect, i.e. that adds up the missing chunk of  states $E_\text{ext}$ to the loop in \reef{offd1}.
For the theory we study in this paper [$\phi^4$ in $d=3$], the sunset is  logarithmically divergent, thus the differences between $E_l \leq E_T-E_\text{ext}$ and $E_l \leq E_T$ are $O(1/E_T)$; while the vacuum diagrams  do generally  suffer from the problem of missing states [although not necessarily due to off-diagonal contributions of $C^{(2)}$] as was carefully analysed in the sections above. 
Finally, we stress that  if a regulator   cuts all  the loops equally [like  covariant regulators do] the  problem of missing states does not arise.


\subsection{Hamiltonian Truncation  formulation  } \label{subHT}

Having understood perturbation theory with $E_T$ cutoff regularisation and its large $E_T$ extrapolation,  we are now finally in the position to  uplift the perturbative formulation into  Hamiltonian Truncation.
Thus, we will perturb the free massive theory by 
\be
V  =   V_4 -c_2(E_T) V_2  - \Omega(E_T)  \quad \text{with} \quad \Omega(E_T)=c_0   +d_0    - \delta V^I  - \delta V^{II}   \ .  \label{finalform}
\ee
The counter-terms $c_0$, $d_0$ and $c_2$ were given in \reef{vac1}-\reef{mass1}, for reference:
   \be
   c_0(E_T)\sim \frac{-g^2L^2}{96(4\pi)^3}\left(E_T-8m \log\frac{E_T}{m} \right)  \,  ,
   \ \ \ 
   d_0(E_T) \sim \frac{g^3L^2}{3072 (4\pi)^2}\log\frac{E_T}{m}  \, ,  \ \ \  c_2(E_T)=\frac{-g^2}{6(4\pi)^2}  \log\frac{E_T}{m}  \, .  \nonumber
       \ee 
 The $c_0$ and $d_0$ counter-term includes $1/E_T$ pieces to improve the convergence, the exact form is given in the appendix \ref{acts}.  We will denote the vacuum and mass counter-terms vertices by  $\Va$. 
 The corrections  $\delta V^I$ and $\delta V^{II}$ are given in  \reef{patchvac} and \reef{itect}, respectively. Their role is to add back missing states that are cut away by the $E_T$ regulator. We denote both such corrections by  
  $\begin{minipage}[h]{.016\linewidth} \begin{tikzpicture}
\begin{feynman}[small]
 \node [crossed dot,rotate=45] (i1) at (.4,.4);
\diagram*{
   (i1),
   };
  \end{feynman}
\end{tikzpicture}
  \end{minipage}$.

One difference compared to the $\phi^2_3$ test in section~\ref{secphi2} is that the counter-terms that add back missing states depend on the unperturbed external energy. For example, \eqref{patchvac} depends on $E_i$. Therefore, a separate diagonalisation must be carried out to determine each energy level in the perturbed theory.

We have only shown that the interaction in \reef{finalform} leads to a finite spectrum as $E_T\rightarrow \infty$ in perturbation theory.
However, having reached this point we will press on and implement  \reef{finalform} in a numerical routine. 
We will find  that the method gives sensible results for both  perturbative  and also moderately strong couplings.


\section{Results from truncations}\label{sNI}

Next we implement the theory described in section~\ref{subHT} in a numerical code. First we study the dependence of the results from HT calculations on $E_T$, then we compare the results for the mass gap extrapolated to $E_T\rightarrow \infty$ with the perturbative prediction.

\subsubsection*{$E_T$ dependence}

\begin{figure}[t]
\begin{center}
\includegraphics[scale=0.65]{{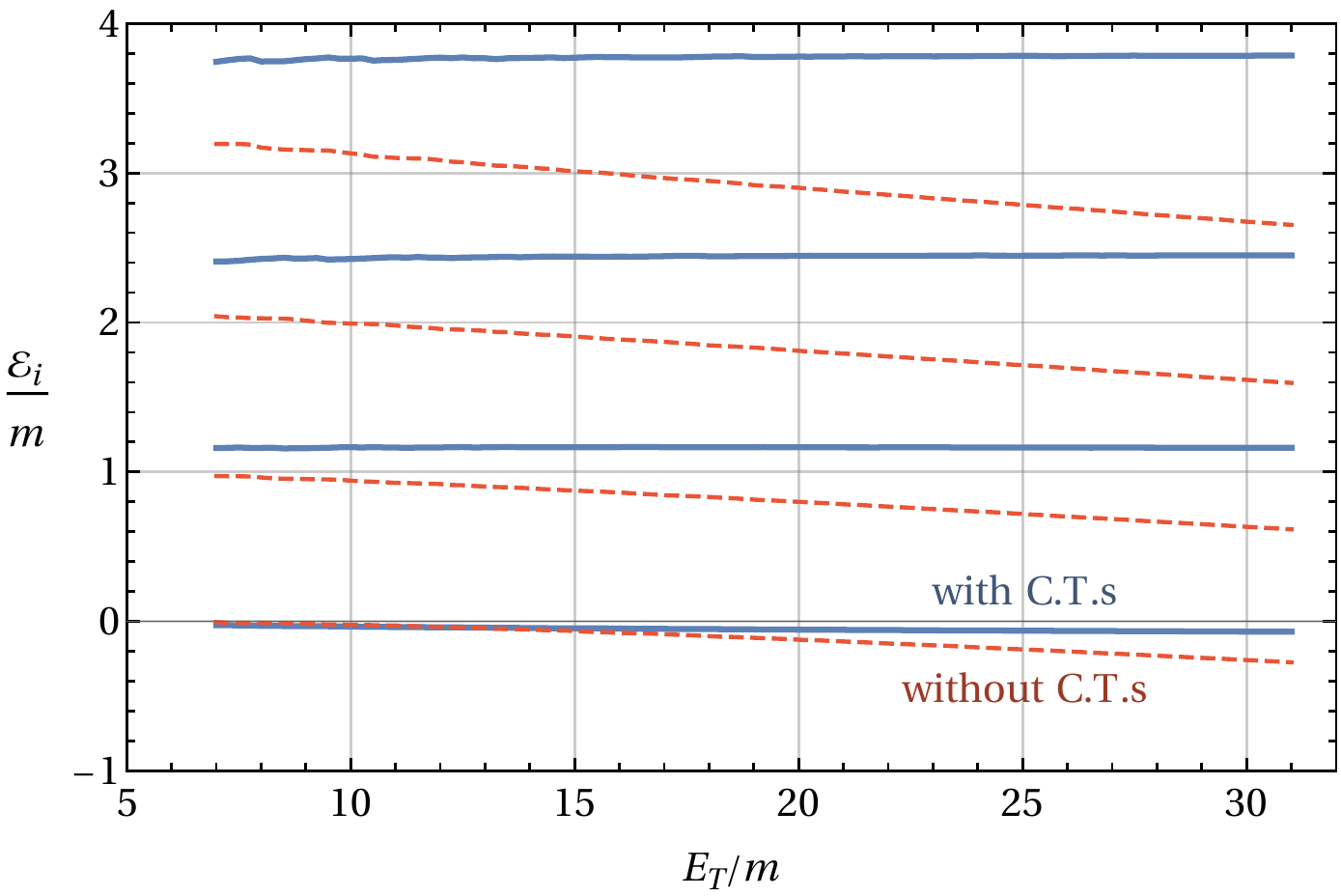}}
\end{center}
\caption{The vacuum and low lying energy levels of $\phi^4_3$ obtained from a HT calculation as a function of the cutoff $E_T$, for a theory with $g=18m$ in a box of size $L=4/m$. We show the results with the compete set of counter-terms needed to render the theory finite [``with C.T.s"], and those obtained when no counter-terms are included [``without C.T.s"].}
\label{fig:ct}
\end{figure}

In Figure~\ref{fig:ct} we show the results obtained for the vacuum energy and the energies of the first three excited states as a function of $E_T$, for the theory with $g=18m$ and $L=4/m$. It is clear that without counter-terms the energy levels of the theory do not converge, and once the counter-terms are included they are well converged. There are small fluctuations in the data, and these are only significant at $E_T \lesssim 15 m$.

Notably the vacuum energy only deviates slightly from zero in Figure~\ref{fig:ct}. This happens because the theory is fairly perturbative. We estimate the coupling $g^*$ above which the theory is strongly coupled by demanding that $\cE_0^{(3)} = \cE_0^{(2)}$ at this value, leading to $g^* = 8.3$ [see below for further details]. Since $g$ is comparable to $g^*$ and we subtract the exact vacuum diagrams up to order $g^3$, including their $E_T$ dependence, we have $\cE_0 \simeq 0$. 
 
The higher energy levels are also extremely flat as a function of $E_T$, although they differ significantly from their classical values [i.e. from their values at order $g^0$]. Such behaviour can be understood from the evaluation of the $g^2$ correction to the mass gap at finite $E_T$ [this is plotted in Figure~\ref{fig:extrap} left of appendix~\ref{aMonteCarlo}]. There is a finite correction at order $g^2$ as $E_T \rightarrow \infty$, and at $E_T \simeq 10$ this is within $10\%$ of its asymptotic value. The correction to the mass gap at $g^3$ converges much more slowly, but for $g=18m$ it is a factor of $6$ smaller than the $g^2$ piece. In section~\ref{sCM} we will show data that indicates that the mass gap also converges at strong coupling, albeit more slowly.
 
\subsubsection*{Comparison with perturbation theory}
 
As a cross check of the truncation calculations we compare the results obtained at weak coupling to the perturbative prediction. To do so we compute the mass gap up to order $g^3$ in perturbation theory. 
The diagrams that contribute to the vacuum energy at order $g^2$ are 
\be \label{deltae0g2}
\cE_0^{(2)} =  \Va +   \VVa   ~,
\ee
and the first excited level gets corrections from
\be \label{deltae1g2}
\cE_1^{(2)} = ( \VVd  +   \Vd   +  \Vdin ) + (  \VVb   +\VVc  +\Vc) ~, 
\vspace{-.1cm}
\ee
where  $\begin{minipage}[h]{0.017\linewidth}
\begin{tikzpicture}
\begin{feynman}[small]
 \node [crossed dot,rotate=45] (i1) at (.4,.8);
\diagram*{
   (i1),
   };
  \end{feynman}
\end{tikzpicture}
  \end{minipage} $ represents the counterterm that adds back in the missing states to the 2-point bubble divergence. Note that due to adding these states back in, the diagrams in the first bracket of \eqref{deltae1g2} are equal to   the shift in the vacuum energy of \eqref{deltae0g2}. Consequently the $g^2$ correction to the mass gap comes solely from the diagrams in the second bracket of \eqref{deltae1g2}.

Meanwhile at $g^3$ in perturbation theory the mass gap gets corrections from the diagrams
\bea \label{deltae0g3}
\cE_0^{(3)} &=& \VVVa  + \Va\\[.2cm]
\cE_1^{(3)} &=&   \VVVc   +  \VVVd + \VVVh  +     \Vd  + \VVVb +  \Vdin  \nonumber \\
 &  +&  \Big(    \VVe+  \VVVe   + \VVVf+\VVVg+h.c.\Big)    \,  .    \label{deltae1g3}
\eea 
We evaluate the preceding diagrams with Monte Carlo integration for a box length $L=4/m$.
 The mass gap is given by
\be \label{massgappert}
\cE_1 - \cE_0 = m + 0.62 \,  \left[g/(4!m)\right]^2 m + 0.11 \,  \left[g/(4! m ) \right]^3  m + O\left(g^4m^{-3}\right) ~.
\ee
Further details of the perturbative calculation may be found in appendix \ref{HPTapp}.

We now compare the perturbative prediction with results from HT. To do so we extract the mass gap as a function of $E_T$ from truncation calculations for different values of $g$. Our numerical power limits us to $E_{T} \leq 33 m$. At each $g$ we select data in the range $17m<E_T$ and extrapolate this to $E_T \rightarrow \infty$ using a fit of the form
\be \label{fitform}
\cE_1 - \cE_0 = \alpha_0 + \frac{ \alpha_1}{E_T} +  \frac{\alpha_2}{E_T} \log\left(\frac{E_T}{m}\right) ~,
\ee
where $ \alpha_0$, $ \alpha_1$ and $ \alpha_2$ are free parameters. This choice of functional form is motivated by naive power counting. Further, in appendix~\ref{aMonteCarlo} we show that this form gives precise extrapolations for diagrams at low order in perturbation theory.

\begin{figure}[t]
\begin{center}
\includegraphics[scale=0.7]{{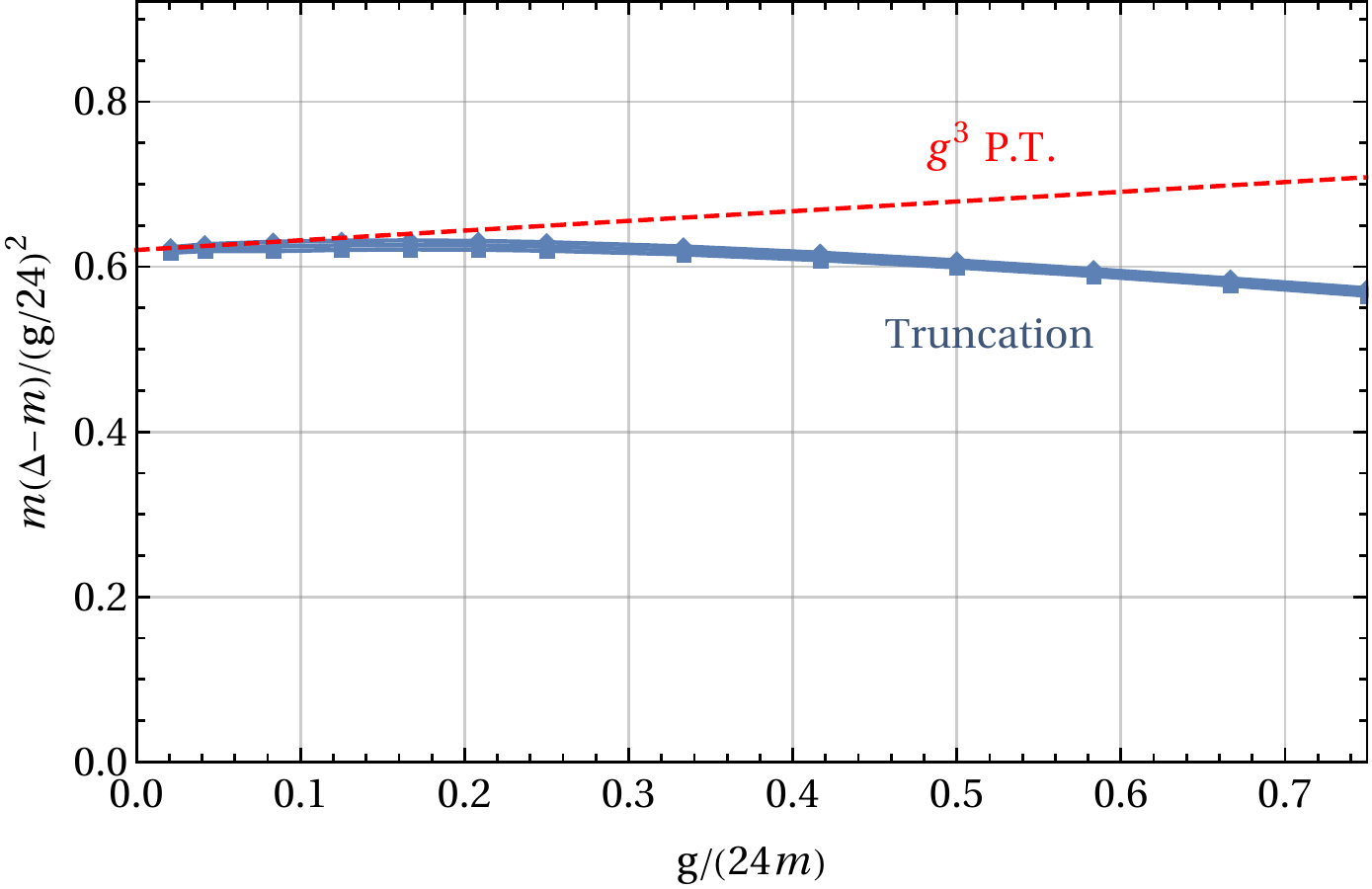}}
\end{center}
\caption{A comparison between the mass gap $\Delta$  as a function of $g$ calculated from HT calculations and the perturbative prediction. We plot $m \left(\Delta -m \right)/\left(g/24 \right)^2$ so that the perturbative prediction at $O(g^2)$ is a horizontal line and the $O(g^3)$ prediction is the straight line indicated. Details of the extrapolation of the HT data to $E_T \rightarrow \infty$ and the error estimates are given in the main text.}
\label{fig:mg}
\end{figure}

We estimate the error $\sigma$ on the $E_T \rightarrow \infty$ extrapolated mass gap $\Delta = \cE_1 - \cE_0$ by defining
\be \label{errorextrap}
\sigma = {\rm max} \left(\sigma_{\rm fit},~ \left|\Delta - \Delta_{\rm high} \right| , ~\left|\Delta - \Delta_{\rm low} \right|   \right)~.
\ee
Here $\sigma_{\rm fit}$ is the uncertainty on the fit of the full data set with $17 m < E_T \leq 33m$; $\Delta_{\rm low}$ is the extrapolated mass gap obtained when only data up to $E_T = 30m$ is used in the fit, which gives an estimate of the error from our limited numerical power; and $\Delta_{\rm high}$ is the extrapolated mass gap obtained when the fit uses only data starting at $E_T = 20 m$. Typically the error from the fit is subdominant relative to one of the other uncertainties in \reef{errorextrap}.

The extrapolated results for the mass gap are plotted as a function of $g$ in Figure~\ref{fig:mg}, where the perturbative prediction at order $g^3$ is also shown. It can be seen that, as expected, the truncation calculation asymptotes to the $g^3$ prediction as $g\rightarrow 0$. The deviation as $g$ increases is consistent with a $g^4$ correction that has a relatively large coefficient.

As a further test, we have repeated the truncation calculation with different combinations of the counter-terms to cure the primitive divergence and those to add back in missing states missing. In all cases the results obtained do not match the perturbative prediction and appear to be diverging as a function of $E_T$ [albeit slowly for sufficiently small $g$].

\section{Crosscheck through a weak/strong self-duality}
\label{sCM}

 In this section we test the numerical power of  HT at strong coupling using  a weak/strong duality  of the $\phi^4$ theory, the Magruder duality~\cite{Magruder:1976px}. 
We start  by deriving the theory and then we present our numerical results.

For the $\phi^4$ theory in $d=2$ spacetime dimensions there is a similar weak/strong duality \cite{Chang:1976ek} relating the broken and unbroken phases. The duality has been recently probed using Hamiltonian Truncation  \cite{Rychkov:2015vap} and Borel summation \cite{Serone:2019szm} techniques. 
Additionally, ref.~\cite{Windoloski:2000yb} studied the strongly coupled phase of $\phi^4$ in  $[0,T]\times S^2$ manifold and  the Magruder duality using Monte Carlo techniques [working in the basis of the $SO(3)$ harmonics, cutting on spin and sampling randomly over the Fock space states].

\subsection{Theory}
The basic idea is the following, consider the Hamiltonians~\footnote{Originally  \cite{Magruder:1976px} derived the  duality  at infinite volume and using a Lagrangian formulation. For our purposes we need to re-derive it in the Hamiltonian description and at finite volume. }
\bea
H&=& \int_0^L d^2x \,  N_m \left( \frac{1}{2} \dot\phi^2 +\frac{1}{2} \phi^{\prime 2}  +\frac{m^2}{2}\phi^2 +g\phi^4 -c(\Lambda,m) \phi^2\right)+ \Omega  \label{h1}\\[.2cm]
H^\prime &=&  \int_0^L d^2x \,  N_M \left( \frac{1}{2} \dot\phi^2 +\frac{1}{2} \phi^{\prime 2}  -\frac{M^2}{4}\phi^2 +g\phi^4- c(\Lambda,M)\phi^2\right) + \Omega^\prime \label{h2}
\eea
where $N_x$ corresponds to normal ordering with respect to mass $x$ and $\Omega$, $\Omega^\prime$ are    vacuum counter-terms.~\footnote{We normalise  $M^2/4 \phi^2$ in order to get  $M^2$ mass-square in the tree-level vacuum.   } 
The theory is regulated with a momentum cutoff $\Lambda$, but other sensible regulators are also possible. 
The mass counter-term operator is given by 
\be
c(\Lambda,x) = - \frac{g^2}{6(4\pi)^2}  \log\frac{\Lambda}{x}  \, ,  \label{cts}
\ee
The Hamiltonians  $H$ and $H^\prime$ act on the infinite dimensional Hilbert space spanned by the free Fock states $|E_i \rangle$ and the counter-terms  ensure that the spectrum of both Hamiltonians is finite in the limit  $\Lambda\rightarrow \infty$.
Both theories have the same value for the volume $L\cdot g$. 
At weak coupling, $g/m\ll 1$ and $g/M\ll 1$, the theory described by $H^\prime$ presents spontaneous symmetry breaking of the $\mathbb{Z}_2$ symmetry, while the theory described by $H$ is in the symmetric phase. 

Next we want to relate   $H$ to $H^\prime$. To do so, we need to express the normal-ordered operators in terms of not-normal-ordered ones,
\be
N_x (\phi^4) =  \phi^4- 6 z_x \phi^2 +3z_x^2 \label{nm1} \ \ ,  \quad  \quad
N_x(\phi^2) = \phi^2-z_x   \ \ ,  \quad  \quad
N_x(\dot \phi^2 +\phi^{\prime 2})=\dot \phi^2 +\phi^{\prime 2} - \,  y_x  \, . 
\ee
The functions $z_x,\, y_x$ are readily computed taking expectation values of the former expressions -- leading to
$
z_m=1/L\sum_{\vec k}\frac{1}{\om_{\vec k}}
$ and $y_m= 1/L\sum_{\vec k} \frac{2 \vec k^2+m^2}{2\om_{\vec k}}$, properly regulated.
Then from  \reef{nm1}, it follows 
\be
N_m (\phi^4) = N_M (\phi^4) + 6 Z_L \phi^2 +v_L \label{match1}
\ee
where $v_L$ and $Z_L$ are functions of the volume. We will need 
\be
 Z_L = \sum_{i, j \in\mathbb{Z}}\Delta(L\vec s_{ij};m,M)= \frac{m-M}{4 \pi}  + \frac{e^{-Lm}-e^{-LM}}{L \pi}+O(e^{-\sqrt{2}Lm},e^{-\sqrt{2}Lm}) \label{togen}
\ee
where $\Delta(L\vec s_{ij};m,M) \equiv D(L\vec s_{ij};m) -D(L\vec s_{ij};M)$ with $D(\vec r;m)=e^{-|r|m}/(4\pi |r|)$, the infinite volume propagator; while the value of  $v_L$ is unimportant for our current specific purposes. 
Using the matching condition \reef{match1}, we find that the two Hamiltonians are equivalent   when
\be
-\frac{M^2}{4} - g^2 c_2(\Lambda,M) = \frac{m^2}{2} - g^2 c_2(\Lambda,m) + 6g Z_L \, .   \label{cmL}
\ee
The regulator $\Lambda$ drops out in the former expression and thus the duality is maintained when the regulator is removed $\Lambda\rightarrow \infty$.
Let us stress that we have not been careful to  match the cosmological constant of both theories, i.e. $\Omega=p(\Omega^\prime,z_x,y_x)$. Therefore when \reef{cmL}  is satisfied  only the mass gap and spectra $\Delta_i\equiv \cE_i-\cE_0$  of the two theories coincide.

\begin{figure}[t]
\begin{center}
\includegraphics[width=0.450\textwidth]{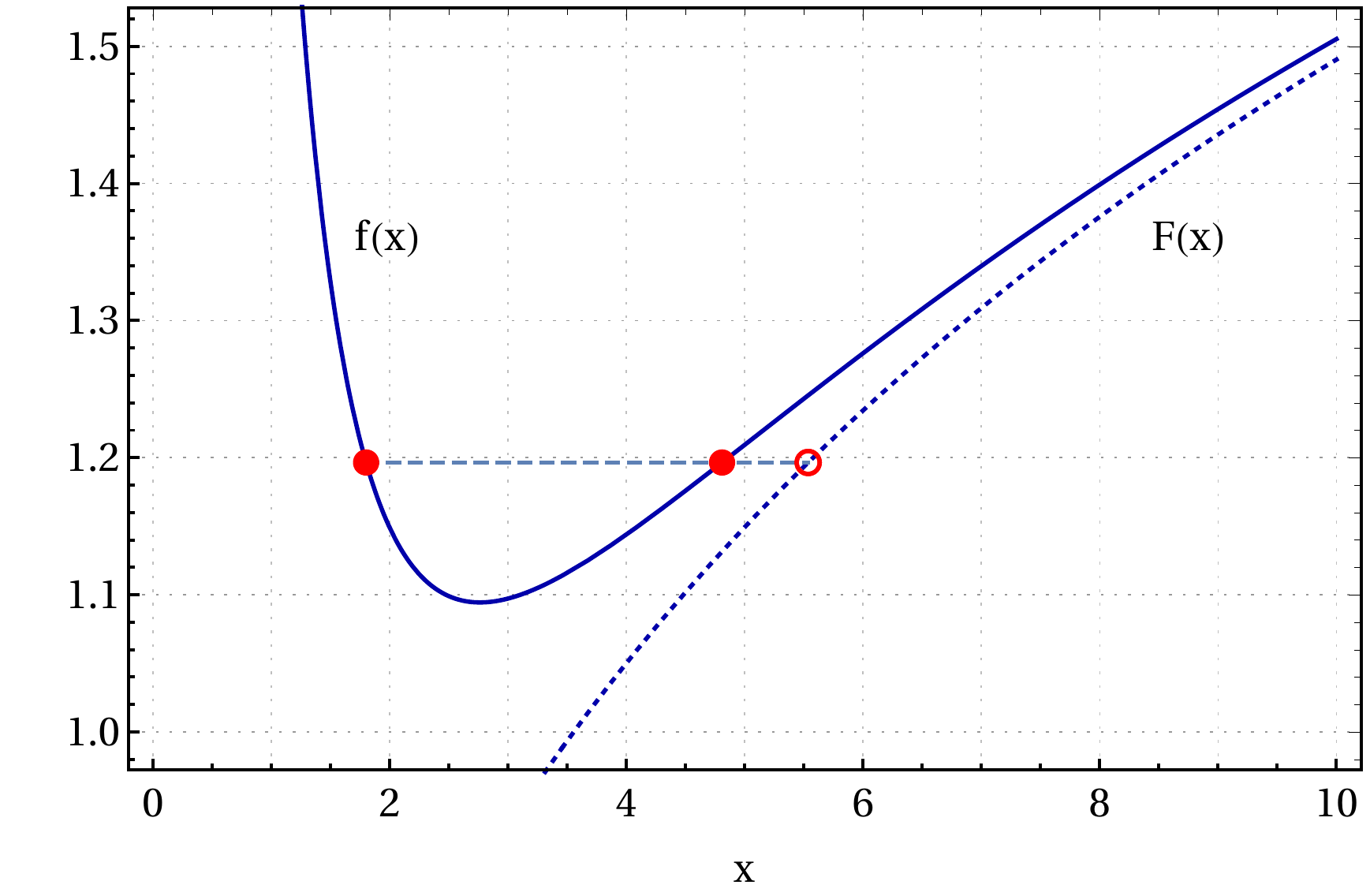}\quad\quad
\includegraphics[width=0.45\textwidth]{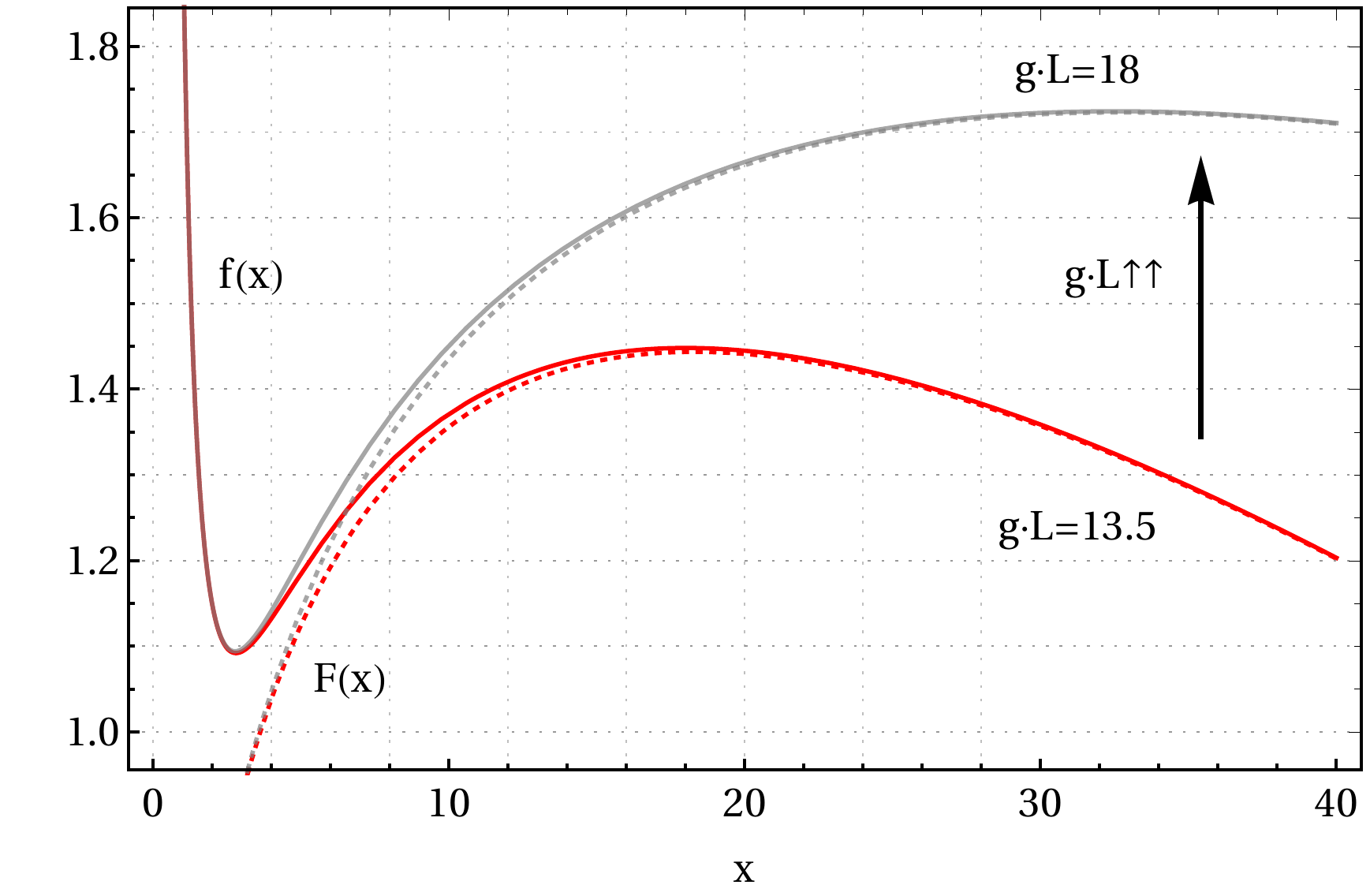}
\end{center}
\caption{Left: solutions of the duality equation \reef{solsi1} are infinite volume. Red dots correspond to two  symmetric theories that are dual to each other and to the  $H^\prime$ theory signalled with a red circle.  Right: solutions of the duality equation \reef{solsi1} for $gL\approx\{13.5,18\}$.    \label{figcmL}} 
\end{figure}

For convenience we define   
\bea
f_{g\cdot L}(x)&=&\frac{1}{x^2}+\frac{6}{\pi^2}\log x+  \frac{3}{\pi x}\left(1-\frac{4x}{L g}e^{- L g/x}+ \cdots \right) \\
 F_{g\cdot L}(x)&=&    -\frac{1}{2X^2} +\frac{6}{\pi^2}\log X+\frac{3}{\pi X}\left(1-\frac{4X}{L g}e^{- L g/X}+ \cdots \right)  \, .
\eea
where $x \equiv \frac{g}{m}$ and $   X \equiv \frac{g}{M}$ and  $\cdots$ denote further winding modes. 
Many more winding corrections $[\exp(-L g/x)]^n$ are easily computed from the definition of $Z_L$, and in our numerical explorations below we add a large number of winding modes, safely beyond those needed  for the values of $Lg$  that we will consider. 
Now,
\be
F_{g\cdot L}(X)=f_{g\cdot L}(x) \, , \label{solsi1}
\ee
 is equivalent to \reef{cmL}.
 The graph of this functions  is shown in Figure~\ref{figcmL} for various choices of the volume.
 
 Let us discuss the infinite volume limit first, i.e. the left plot of Figure~\ref{figcmL}. We define $f_\infty\equiv f$ and $F_\infty  \equiv F$. 
  The function $f$ has a ``V" shape, and for $X\gtrsim 4.43$ [and $L\rightarrow \infty$] equation \reef{solsi1} has solutions. For $X\gg1$, the  two solutions are  $X\approx x$ [following the right branch of $f$]  and $x\approx \pi/\sqrt{6 \log(X)}$ [on the left branch].
  For the first solution both $H$ and $H^\prime$ are strongly coupled with $x,X\gg 1$; while for the other  branch  $H^\prime$ at strong coupling is dual to a weakly coupled $H$ theory. Further, the two $H$ theories dual to $H^\prime$ are also dual to each other. For example, the red filled markers in Figure~\ref{figcmL} [left plot] indicated the $H$ theories with $x=1.8$ and $x\approx 4.8$ are equivalent, and both are dual to the $H^\prime$ theory at $X\approx 5.5$, marked with a red circle. 
Due to the "V" shape of $f$ there is an infinite continuum of pairs of $H$ theories that are self-dual.  We will exploit this effect by using relatively large values of $f$, so that such self-dual pairs are weak/strong dualities. We will do so at finite volume, which we describe next.

In Figure~\ref{figcmL} right, we show the graph of $f_{g\cdot L}(x)$ and $F_{g\cdot L}(X)$ for $gL\approx\{13.5, 18\}$.
For $gL\gg  X,x$ the solutions of \reef{cmL} have a similar two branch "V" shape graph as the infinite volume result. 
Therefore,  when $gL\gg  X,x$, the analysis resembles the infinite volume one, and $f_{g\cdot L}$ reaches large enough values that horizontal lines in the right handed plots of Figure~\ref{figcmL} cut the solid curve at $x\ll 1$ and $x\gg 1$, showing the existence of weak/strong self-dual $H$ theories. 
As $gL \gtrsim X,x$ the solutions of \reef{cmL} have a more intricate structure than the infinite volume counterpart duality. In the present work we will not make use of those further strongly coupled theories.

In the next section we are going to exploit the weak/strong self-duality of the $H$ theory to crosscheck our HT results at strong coupling. 
Namely, we will compute  the spectrum of two self-dual theories and check whether compatible results are obtained. Despite the two self-dual Hamiltonians being equivalent it is non-trivial that the two calculations agree. In the more weakly coupled theory the Hamiltonian matrix is close to diagonal. Meanwhile in the more strongly coupled theory there are large off diagonal terms from the $\phi^4$ operator and the $\phi^2$ counter-term. These off diagonal pieces must combine with the effects of normal ordering and winding corrections to reproduce the dynamics of a weakly coupled theory. Indeed, given that the dimensions of the Hamiltonian are different for the two dual theories at any finite $E_T$ the duality only holds in the $E_T \rightarrow \infty$ limit. We are therefore also testing the accuracy of our extrapolations.

In the future it would be interesting to study the duality between a theory with a positive mass squared parameter and one with a negative mass squared parameter. It may also be interesting to compare the vacuum energies or the energies of the higher excited states in the two theories.

\subsection{Numerical results}

Rather than simply comparing a particular pair of dual theories, we consider the family of theories with $m=1$ and varying $g$. As before we fix $L=4$, and for this value a dual theory exists for any $g \gtrsim 54$.~\footnote{Note that this is a finite volume effect due to the turn over of the right branch in Figure~\ref{figcmL} right, and  in the infinite volume limit the dual theory  exist for any $g$.} We denote the mass in the dual theory by $\tilde{m}$ and its coupling strength is characterised by $g/\tilde{m}$. The self dual point is at $g=68.7$. For $g$ smaller than this $\tilde{m}<1$, so the dual theory is more strongly coupled than the original. For larger $g$ the dual theory is more weakly coupled, and for $g \rightarrow \infty$, $\tilde{m} \rightarrow \infty$ so the dual theory reaches the perturbative regime. In this limit, the mass gap of the theory can be found by replacing $m$ with $\tilde{m}$ in \eqref{massgappert}.

For each value of $g$  we calculate the mass gap up to the limit of our numerical power $E_T = 33m$. As in section~\ref{sNI}, we select data with $E_T>17m$, and extrapolate to $E_T \rightarrow \infty $ with a fit of the form \eqref{fitform}.\footnote{Fitting the data with a function $ \alpha_0 +  \alpha_1/E_T  +\alpha_2/E_T^2$, with $ \alpha_{0}$,$ \alpha_{1}$,$ \alpha_{2}$ free parameters leads to results that are sometimes outside the error bars that we quote. However, the CM duality works less well for this choice of extrapolation.} 
We use the same fit form in the dual theories, however the growth of the Fock space with $E_T$ changes with $\tilde{m}$. As a result, we adjust the lower limit on the range of $E_T$ used for the fit, so that data with significant fluctuations is excluded. The maximum accessible $E_T / \tilde{m}$ also changes, and it is largest for the smallest $\tilde{m}$. We estimate the uncertainty on the extrapolated mass gaps as in \eqref{errorextrap}.\footnote{In the dual theory we adjust the definition of $\Delta_{\rm high}$ and $\Delta_{\rm low}$ to account for the different growth of the basis size with $E_T$.} 
  
  \begin{figure}[t]
\begin{center} 
\includegraphics[width=0.6\textwidth]{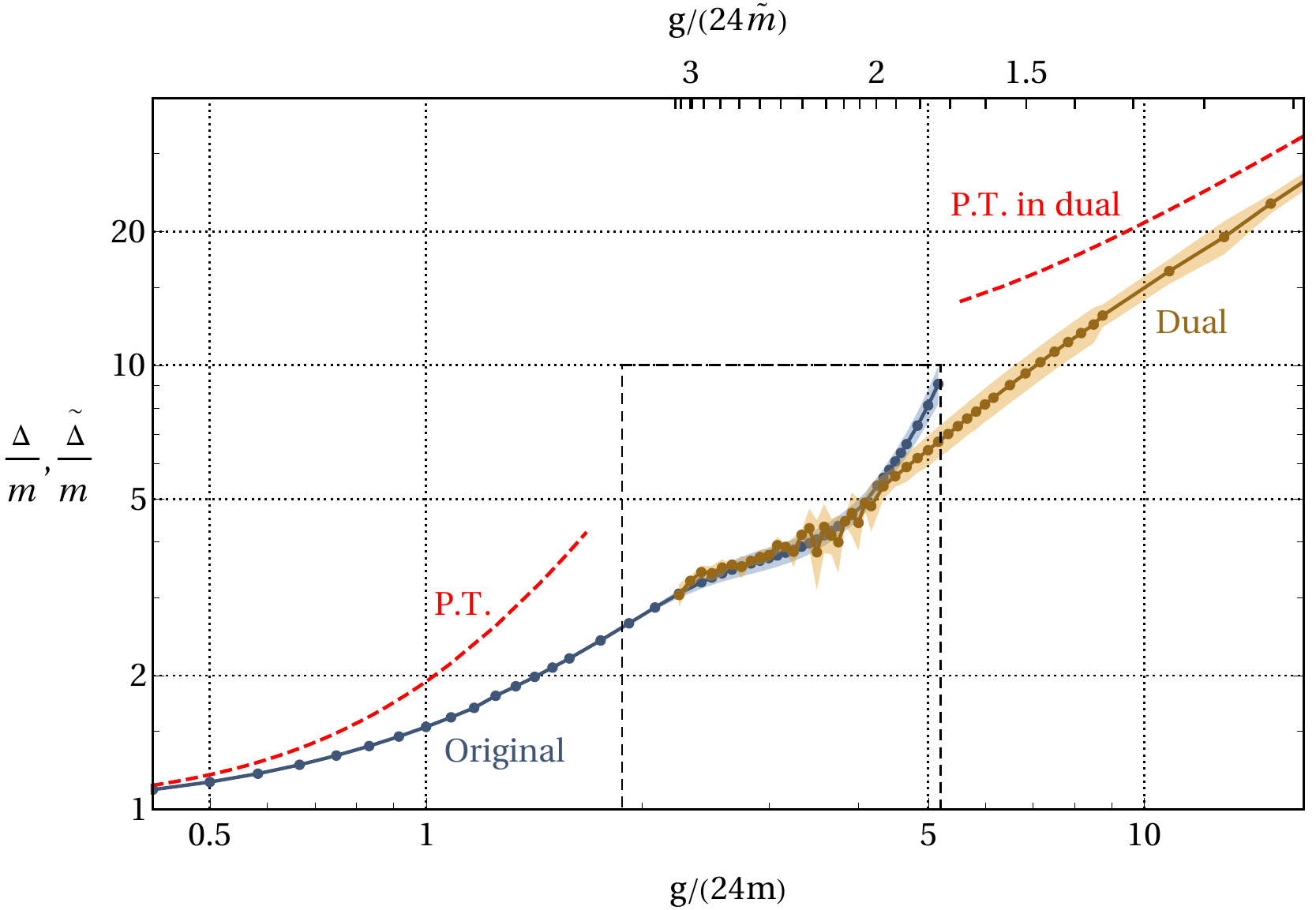}
\end{center}
\caption{The mass gap in the original [$m=1$] and dual theories as a function of $g$, calculated using HT. The strength of the coupling in the dual theory $g/\tilde{m}$ is also indicated. There is a substantial range of couplings where the two calculations agree. The perturbation theory prediction at order $g^3$ in the original and dual theories is also shown. The same data, zoomed in to the dashed black box, is plotted in Figure~\ref{figCM2} left.\label{figCM1}}
\end{figure} 

For very large $\tilde{m}$ it becomes numerically inefficient to carry out the truncation calculation, since $E_T \gg 4 \tilde{m}$ cannot be reached. Instead, by dimensional analysis we relate a theory with large $\tilde{m}$ to a theory with $\tilde{m} = 1$ in a box of size $L'=4 \tilde{m}$ with coupling $g/\tilde{m}$. For $L'>10$ we assume that this theory is close to the infinite volume limit. Its spectrum can be approximated by that of the same theory except with the box size shrunk to  $L'=10$, for which HT is computationally easier. Finally, the results are related back to the large $\tilde{m}$ theory. This trick caps the numerical difficulty as $g$ grows. Choosing the maximum $L'=10$ it applies to the dual theories with $g\gtrsim 110$.

\begin{figure}[t]
\begin{center} 
\includegraphics[width=0.475\textwidth]{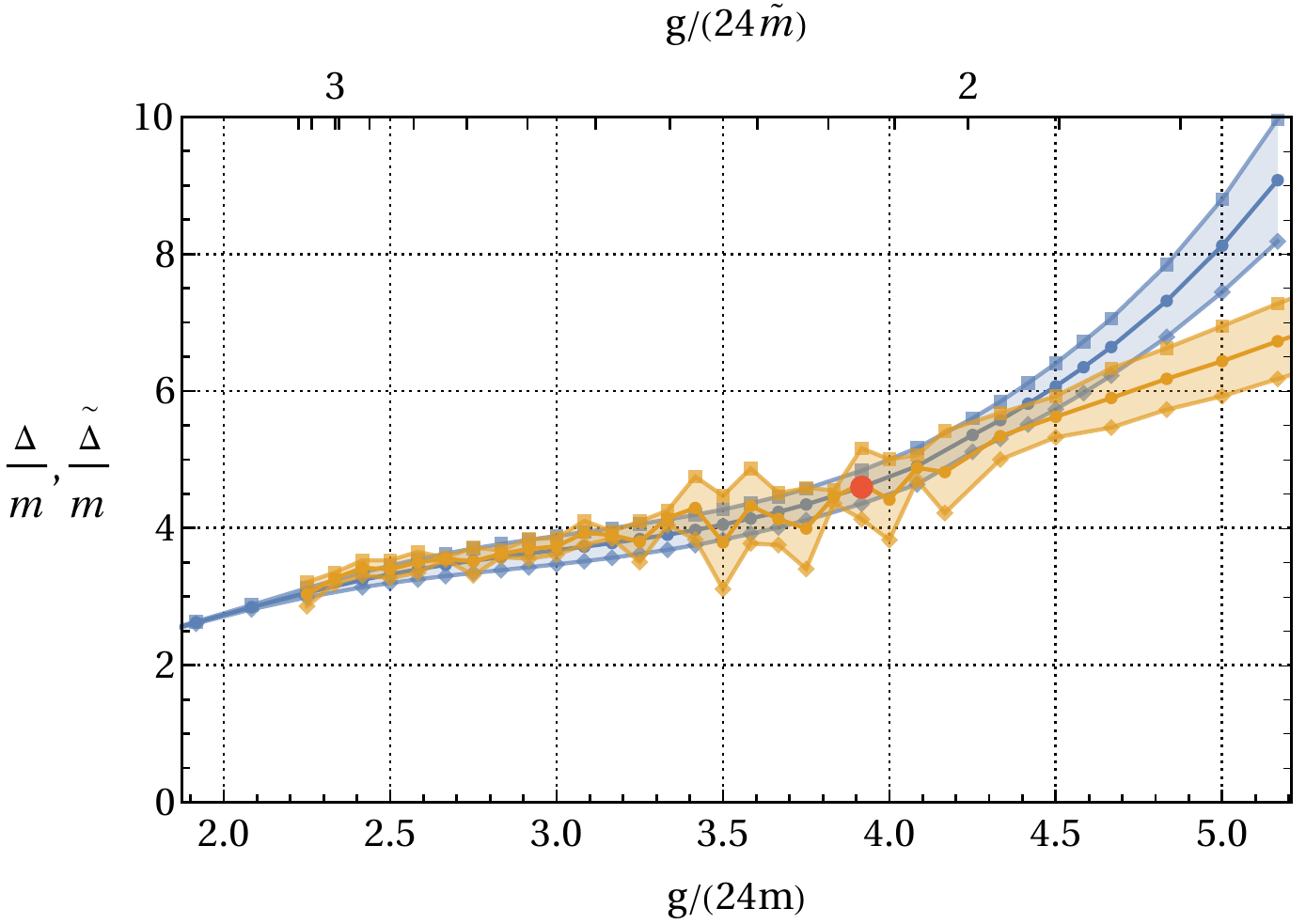}\qquad
\includegraphics[width=0.45\textwidth]{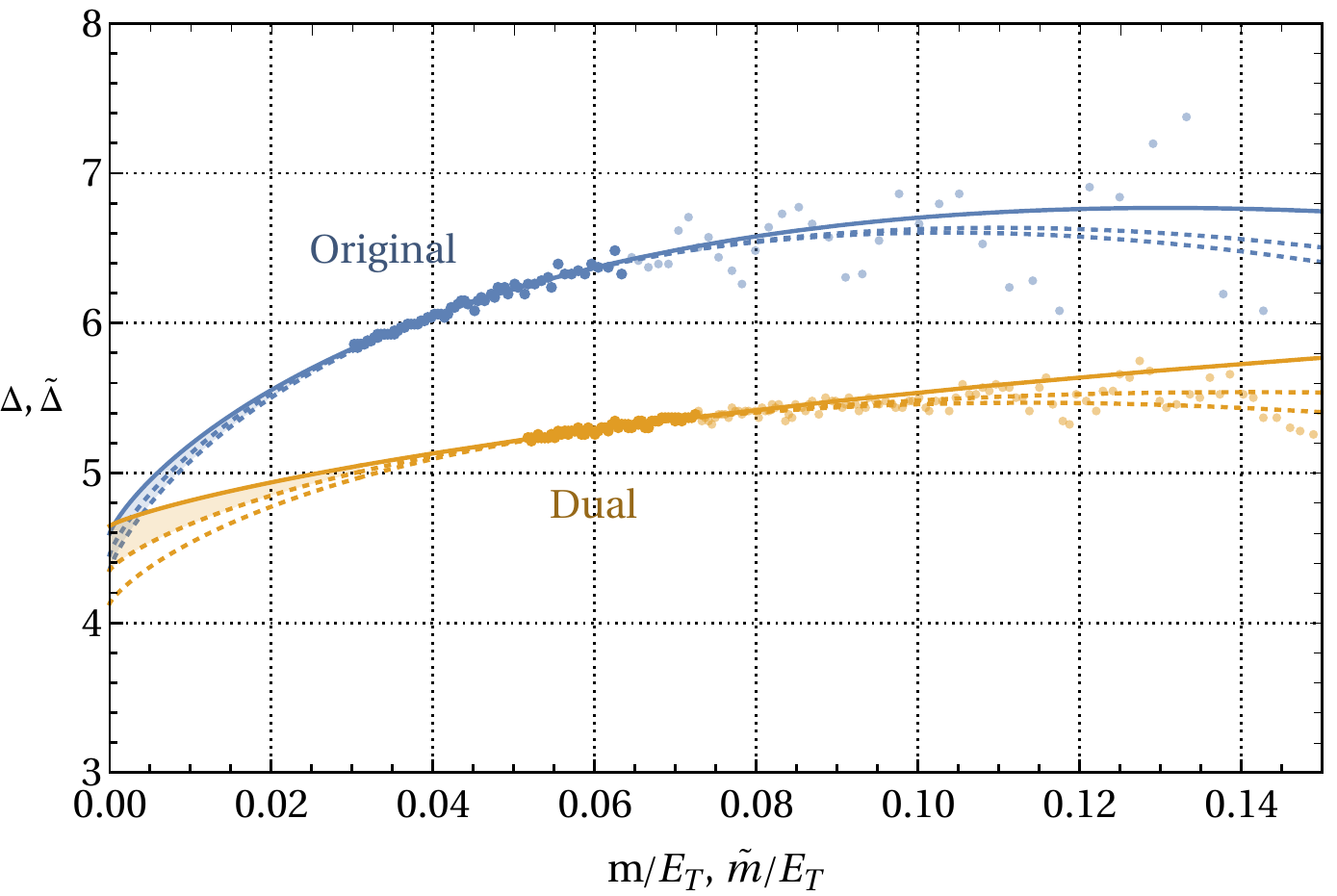}
\end{center}
\caption{Left: The same data as Figure~\ref{figCM1} focusing on the region where the original and dual theories overlap. Right: the results at finite $E_T$ obtained from HT calculations in the original and dual theories for $g=94$, corresponding to the red dot in the left figure.   The extrapolation to $E_T \rightarrow \infty$ is also shown, and the data used for this is plotted with solid points. The partially transparent points at small $E_T$ have large fluctuations and are not used for the fit. The extrapolations using data with a smaller maximum $E_T$ or with a larger minimum $E_T$ are plotted with dashed lines. As discussed below \eqref{errorextrap} these are used to estimate the uncertainty on the $E_T\rightarrow \infty$ mass gap.  \label{figCM2}}
\end{figure}

The results for the extrapolated mass gap in the original and dual theories are plotted as a function of $g$ in Figure~\ref{figCM1}. The same data, shown zoomed in, is plotted in Figure~\ref{figCM2} left. When the original theory is weakly coupled enough that it matches the perturbation theory prediction there is no dual theory. 
In the range from $g=54$, when the dual theory first exists, up to $g=100$ there is good agreement between the two theories. For $g>100$ the dual theory approaches its perturbative prediction, but the original theory deviates. We note that even at $g \simeq 300 $ in the dual theory $g/\tilde{m} \simeq 30$ is not tiny. The relative difference between the HT calculation of the mass gap in the dual theory and the perturbative prediction is comparable to the deviation from the perturbative predictions in the original theory at small $g/m$.

The agreement between the two theories is best, and the uncertainty on the result from the dual theory is smallest, when $g/\tilde{m}$ is largest. This is because $\tilde{m}$ is minimised at this point, and large values of $E_T/\tilde{m}$ can be reached computationally. As a result, the extrapolation to $E_T \rightarrow \infty$ is more precise, despite the relatively strong coupling. In the region $g/\tilde{m} \simeq 50$ the results from the dual theory have large uncertainties. Such theories are both strongly coupled and also have a large $\tilde{m}$ so only small values of $E_T/\tilde{m}$ can be accessed. At even larger $g\gtrsim 110$, the trick discussed keeps the maximum accessible $E_T/\tilde{m}$ constant. Since the dual theory becomes gradually more weakly coupled the uncertainty becomes gradually smaller as $g$ increases.

At large $g/m$ the estimated errors on the results from the original theory are not big enough to include the HT results in the dual theory [and at large enough g, the perturbation theory in the dual theory]. This indicates that our approach to estimating the errors breaks down. The reason is that at such large coupling the mass gap is not close to converged for the accessible $E_T$. Instead are are simply fitting a downward slope, and adjusting the range of the fit does not capture the qualitative change that must happen at $E_T$ beyond our numerical reach.

In Figure~\ref{figCM2} right we show the mass gap at finite $E_T$, and the extrapolation to $E_T \rightarrow \infty$, for the original and dual theories at the particular value $g= 94$. The data in the dual theory has less variation with $E_T$ than the original theory, which is expected since the dual is more weakly coupled. However, both are converging toward the same finite value in the large $E_T$ limit.   The extrapolations from the reduced data sets that are used to estimate the uncertainties are also plotted.

We note that if this calculation is attempted without including the counterterms to fill in the missing states there is no finite range of $g$ over which the dual theories agree. 
The deviation of the original theory at $g/m \gtrsim 100$ [in Figure \ref{figCM1}] could be a result of our limited numerical reach in $E_T$. Alternatively, it could be a scheme dependent effect, which we comment on in appendix~\ref{app:finite}.

The agreement between the original and dual theories over a substantial range of $g$ is an encouraging sign for the power of the truncation method.   Indeed, the agreement persists until the original theory is fairly strongly coupled with $g/m \simeq 100$ while the dual theory has $g/\tilde{m} \simeq 50$ at this point. It is also interesting that there is a small dip around $g=90$ in both theories. Preliminary investigation indicates that for larger box sizes this dip may be deeper. This could signal the infinite volume theory having a phase transition to the broken phase at some intermediate coupling.

 \section{Summary and outlook}
 \label{concs}

 In the present work we have analysed  Hamiltonian Truncation with UV divergent interactions, focusing our attention on the $\phi^4_3$ perturbation in $d=2+1$.

We began by analysing  Hamiltonian Perturbation Theory in detail. In particular, we have found that disconnected vacuum diagrams should cancel in an intricate manner but that such cancelation is spoiled if the HT regulator is used, introducing new UV divergences. 
We proposed a solution that consists of adding back the states that the $E_T$ regulator cuts away, so that all loops are cut at the same energy.
Then, we proceeded to formulate HT and we devoted the rest of the paper to inspecting the spectrum of the $\phi^4$ theory at weak and strong coupling. We have crosschecked the strongly coupled spectra through a weak/strong self-duality. 
The strategy and techniques that we have developed will be useful for many other relevant perturbation in $d>2$ spacetime dimensions. 

 Given the  good quality of our numerical results, we think it is  worthwhile to press on and further develop the theory underlying the HT idea. 
Our \emph{Patch II} may  be seen as temporary fix and thus we look forward to more elegant solutions. 
 For instance, a possible avenue consists of introducing an auxiliary momentum cutoff. Namely, consider introducing a momentum cutoff $\Lambda$ such that $m\ll \Lambda \ll E_T$. When taking the limit  $E_T\rightarrow\infty$ with $\Lambda$ kept fixed, the theory is regulated with a single momentum cutoff $\Lambda$, all loops are cut equally, and thus many of the problems we discussed associated with the non-covariant regulator are not present. Then, the idea would be to perform the two extrapolations one after the other: $E_T\rightarrow\infty$ first and $\Lambda\rightarrow\infty$ second.
 We have done some preliminary study of this idea and it seems computationally challenging, because for each fixed $\Lambda$ we must reach $E_T\gg \Lambda$.

Next we comment on other  directions that we think are worth pursuing.

A key issue for future work will be  developing further the theory of improvement terms in $d\geq 2+1$.  The state of the art analysis is \cite{Elias-Miro:2017xxf}. 
In our present work, we have shown the power of such improvement terms  in $d=2+1$ in the context of the $\phi^2$ perturbation --  see e.g. the blue line of Figure~\ref{fig:phi21} left.

  It will be fascinating to analyse conformal perturbation theory with a  TCSA like cutoff at high order. In view of our results, starting at fourth order  the effect of \emph{missing states} should  kick in for a  vanishing diagonal interaction. Meanwhile for a non-vanishing diagonal interaction, the effect arises at third order. We look forward to developing this in the near future. 
Interesting  work on conformal perturbation theory with a TCSA-like cutoff has been carried out in  \cite{Rutter:2018aog}.

Similar comments apply to the Conformal Truncation framework \cite{Katz:2016hxp}.  However, in the light cone, diagrams  in which particles are created out of the vacuum are removed \cite{Weinberg:1966jm}. Thus, we expect that in the Conformal Truncation framework  some of the problems that we have found will be ameliorated.

A tangential but interesting avenue is the one  pursued in   \cite{Serone:2018gjo}. There, precise results for the $\phi^4_2$ perturbation were reproduced using Borel series summation techniques. It would be worthwhile to carry out this analysis in $d=3$ exploiting that  the  $\phi^4_3$ perturbation is Borel summable  [for a recent reassessment see \cite{Sberveglieri:2019ccj}], and to compare with our findings.  In practice however, this  may require  a further development on the HT side that we comment on next. 

Finally, but not the least, an  important aspect that requires a dedicated study is the large volume extrapolation. 
As the coupling of the $V=g\int^\infty\phi^4$  perturbation is increased, a phase transition  where the symmetry $\mathbb{Z}_2:\phi\rightarrow -\phi$ is spontaneously broken can be reached. 
At the critical point, the theory  belongs  to the universality class of the 3D Ising model. Therefore, it would be interesting to measure observables of the critical theory using HT. 
We warn the reader of the \emph{orthogonality catastrophe} of the states as the volume is sent to infinity. Recently such effects have been studied in  the HT context for $d=2$ spacetime \cite{Elias-Miro:2017tup}.

\section*{Acknowledgements} 

We are  thankful to Lorenzo Vitale for collaboration at various  stages of this project and many useful discussions. 
We are   grateful to Slava Rychkov  for  initial collaboration, interesting   discussions and useful comments on the draft. 
We  acknowledge  Matthijs Hogervorst, Marc Montull, Ami Katz, Zuhair Khandker, Marc Riembau, Marco Serone, Giovanni Villadoro and Matthew Walters for interesting discussions and  useful comments on the draft.


\addtocontents{toc}{\protect\contentsline {chapter}{%
\vspace{0.1in}{\hspace{0.5cm}\bf Appendices:}\\
\vspace{-0.1in}
}{}{}}

\newpage 

\appendix

\section{Basis of states on a square torus}
\label{symmetriesT2}

The symmetry of a square torus is a non-abelian finite group $G$ of $O(2)$.  
There are 8 elements in this group that can be represented as 
\begin{equation}
G = \left\{ \mathbb{I} \,, X \,, Y \,, S \,, S^2 \,, S^3 \,, X S \,, Y S \right\}
\end{equation}
Where $S$ is a rotations by $\frac{\pi}{2}$, $X, Y$ are reflection with respect to horizontal (vertical) axis, and 
$X S, Y S$ are reflections with respect to the diagonals.

The  group multiplication table can be compactly summarised by
\begin{equation}
X^2 = \mathbb{I} , \quad S^4 = \mathbb{I} , \quad S^{-1} X S = Y 
\end{equation}
The maximal abelian subgroups are $\mathbb{Z}_2 \times \mathbb{Z}_2$ and  $\mathbb{Z}_4$, 
\begin{align}
\mathbb{Z}_2 \times \mathbb{Z}_2 &: \left\{ \mathbb{I} , X, Y , S^2 \right\}
\\ \mathbb{Z}_2 \times \mathbb{Z}_2 &: \left\{ \mathbb{I} , X S, Y S , S^2 \right\}
\\ \mathbb{Z}_4 &: \left\{ \mathbb{I} , S, S^2 , S^3 \right\}
\end{align}
$G$ is the semidirect product of two abelian subgroups, $G = N \rtimes H$, where $N$ is a normal subgroup 
(invariant under conjugation). Those can be chosen as 
\begin{align}
N &= \mathbb{Z}_4 = \left\{ \mathbb{I} , S, S^2 , S^3 \right\}
\\ H &= \mathbb{Z}_2 = \left\{ \mathbb{I} , X \right\}
\end{align}
Therefore, the singlets of the full group can be found by 
choosing states such that $X \ket{\psi} = S \ket{\psi} = \ket{\psi}$. It is likely that the ground state and low energy 
states are singlets of $G$, since in perturbation theory the ground states, and the lowest one-particle and two-particles 
states are all singlets. 

The three lowest energy states obtained when diagonalising the complete Fock space are all singlets up these symmetries. The energy of these states can be obtained efficiently by considering a reduced Fock space.

Given the most general non-covariant Fock-space state $\ket{\psi}$, a singlet can be simply constructed as 
\begin{equation}
\ket{\psi^S} = 
\frac{1}{\sqrt{8}} \left( \mathbb{I} + X + Y + S + S^2 + S^3 + X\cdot S + Y \cdot S  \right) \ket{\psi}
\end{equation}
However, the normalization is obviously wrong when the Fock-space state is invariant under some (or all) of the 
symmetries. There are four distinct cases we must consider:
\begin{itemize}
\item $\ket{\psi}$ is already invariant under $S$, but not invariant under $X$.
In that case, the right normalization is
\begin{equation}
\ket{\psi^S} = 
\frac{1}{\sqrt{2}} \left( \mathbb{I}  + X \right) \ket{\psi} 
\end{equation}
\item $\ket{\psi}$ is only invariant under $A = X, Y, X S, Y S$. Then the invariant state is
\begin{equation}
\ket{\psi^S} = 
\frac{1}{\sqrt{4}} \left( \mathbb{I} + S + S^2 + S^3  \right) \ket{\psi} 
\end{equation}
\item If $\ket{\psi}$ is invariant under both $A = X, Y, X S, Y S$ and $S^2$ then
\begin{equation}
\ket{\psi^S} = 
\frac{1}{\sqrt{2}} \left( \mathbb{I} + S \right) \ket{\psi} 
\end{equation}
\item If $\ket{\psi}$ is invariant only under $S^2$ then
\begin{equation}
\ket{\psi^S} = 
\frac{1}{\sqrt{4}} \left( \mathbb{I} + X + S  +X  S \right) \ket{\psi} 
\end{equation}
\end{itemize}

\section{Perturbation theory}
\label{dvb}

\subsection{Principal equations}
\label{basicHPT}

In this appendix we give a simple derivation of \reef{basicpert}. To do so we need to derive an exact effective Hamiltonian.  We separate the Hilbert space  as ${\cal H}={\cal H}_1\oplus {\cal H}_2$. The states are projected as $P_1 |y \rangle = |y_1 \rangle\in{\cal H}_1$ and $(\mathbb{I}-P_1)|y\rangle = |y_2 \rangle\in {\cal H}_2$. Operators are projected as  usual  $O_{ij}\equiv P_i O P_j$ with $i,j\in\{1,2\}$. Then, projecting the eigenvalue equation $H|\cE\rangle =  \cE|\cE\rangle$ into the two subspaces we get
\be
H_{11}|\cE_1\rangle +H_{12}|\cE_2\rangle = \cE |\cE_1\rangle \quad , \quad \quad  H_{21}|\cE_1\rangle +H_{22}|\cE_2\rangle = \cE|\cE_2\rangle \, . 
\ee
Next we substitute $|\cE_2 \rangle = (\cE-H_{22})^{-1}H_{21}|\cE_1\rangle$ from the second equation into the first one and we are led to 
\be
\left( H_{11} + V_{12} \frac{1}{\cE-H_{022}-V_{22}}V_{21}\right) \ket{ \cE_1} = \cE \ket{\cE_1} \, .  \label{hefff}
\ee

Now, restricting the ${\cal H}_1$ Hilbert space to a single state $\ket{E_i}$, the  generalised eigenvalue equation in \reef{hefff} simplifies into the following equation
\be
  [H_0 + V]_{ii} + V_{ik}  [\cE-H_0-V]_{kk}^{-1}V_{ki} - \cE =0 \, ,   \label{hefff2}
\ee
where the sum over $k \neq i$ is implicit.~\footnote{Note that indices in \reef{hefff2} denote matrix elements, while   the indices in \reef{hefff}  denote the projectors.}  All the roots $\cE$ of the former equation are the eigenvalues of  $H_0+V$. 
Then upon plugging  in the perturbative ansatz 
 $
\cE_i = \sum_n g^n \cE_i^{(n)}
 $ in \reef{hefff2}
 and solving  for $\cE_i^{(n)}$, by consistently equating powers of $V$, equation  \reef{basicpert} follows. 
Thus in general we have the form  \reef{subte}, i.e. 
$ 
\cE^{(n)}_i=   \bra{E_i} V ( [E_i-H_0]^{-1}V)^{n-1}   \ket{E_i} -\text{subtraction terms}
   $. For instance, the first few terms read 
 \be
 \cE_i^{(n)} = \frac{V_{ik_1} V_{k_1k_2}  \cdots  V_{k_{n}i}}{E_{ik_1}E_{ik_2}\cdots E_{ik_{n-1}}}  - \underbrace{ \cE_i^{(2)} \,    \frac{V_{ik_1} V_{k_1k_2}  \cdots  V_{k_{n-2}i}}{E_{ik_1}E_{ik_2}\cdots E_{ik_{n-3}}}\sum_{s=1}^{n-3}\frac{1}{E_{ik_s}}    -  \cE_i^{(n-2)}
\, V_{ik} E_{ik}^{-2}V_{ki} +  \cdots    }_{\text{subtraction terms}} \,  \label{fund1}. 
  \ee

\subsection{Cancelation of disconnected two-point bubbles}
\label{2ptfact}

Next we want to prove a claim that we made in section \ref{genebub}: 
namely that disconnected two point bubble diagrams cancel. 
The proof is based on a recursive argument. First we consider a general $n$th order diagram contributing    to the first term in \reef{fund1}. Thus, this diagram can have any external number of lines and   can be either connected or disconnected -- including possibly containing further two-point bubbles. 
Next we dress the diagram with a two point bubble in every possible way, i.e. the disconnected two-point bubble spanning from zero to any number of vertices.  Finally, we need to show that the sum of all such dressings  
imply that the two-point bubble factors out and cancels against the second and third term in \reef{fund1}.
Then, it will follow that no diagrams with disconnected two point bubbles survive after doing the sum of the subtraction terms in \reef{fund1} at any order $n$. 

The dressing of a diagram by a two-point bubble in any possible way is done as follows. 
First we can dress the diagram by inserting the bubble in between any two consecutive vertices
\be
\sum_{j=1}^{n-1} \frac{1}{x_j}\frac{1}{x_j+E}\prod_{i=1}^{n-1}\frac{1}{x_i} \, ,   \label{bub1}
\ee
where $x_0=x_n=0$ and where $E$ is the energy propagating in the bubble while $x_k$ is the energy of the other states being propagated in between the vertices. 
For instance, diagramatically a fixed $j=j^*$ contribution in \reef{bub1} looks like
\be
\begin{minipage}[h]{0.24\linewidth}
\begin{tikzpicture}
\begin{feynman}[small]
 \node [dot] (j1) at (0,.-.1);
 \node [dot] (j2) at (.6,.-.1);  
  \node [dot] (a1) at (-1.6,-.4);
 \node [dot] (a2) at (-1,-.4);  
 \node [dot] (a22) at (-.4,-.4);  
 \node [dot] (a3) at (1.,-.4);   
 \node [dot] (a6) at (1.6,-.4);  
 \node [dot] (a7) at (2.2,-.4);  
  \node  (b1) at (-.3-1.6,-.4);  
    \node  (b2) at (-.3-1.6,-.1); 
    \node  (b3) at (-.3-1.6,-.7);  
  \node  (c1) at (.3+2.2,-.4);  
    \node  (c2) at (.3+2.2,-.1); 
    \node  (c3) at (.3+2.2,-.7); 
    \vertex (X1) at (-1.3,.05);
    \vertex (Y1) at (-1.3,-1.1) {\roig \scriptsize $x_1$};
    \vertex (X2) at (.3,.6)  {\roig \scriptsize $x_{j^*}+E$};
    \vertex (Y2) at (.3,-.8);
    \vertex (X3) at (1.9,.05);
    \vertex (Y3) at (1.9,-1.1)  {\roig \scriptsize $x_{n-1}$};
    \vertex (X4) at (-.2,.05);
    \vertex (Y4) at (-.2,-1.1)  {\roig \scriptsize $x_{j^*}$};
    \vertex (X5) at (.8,.05);
    \vertex (Y5) at (.8,-1.1)  {\roig \scriptsize $x_{j^*}$};
\diagram*{
   (j1) -- [ quarter right, looseness=.2,  thick] (j2)  -- [ quarter right, looseness=.2 ,  thick] (j1) ,
     (j1) -- [ half right, looseness=.6,  thick] (j2)  -- [ half right, looseness=.6 ,  thick] (j1),
        (a1) -- [line width=.7mm, red] (a2)-- [line width=.7mm, red, scalar] (a22)-- [line width=.7mm, red] (a3)  -- [line width=.7mm, red, scalar]  (a6)-- [line width=.7mm,red] (a7) ,
     (b1) -- [thick, red] (a1),
    (b2) -- [thick, red] (a1),
        (b3) -- [thick, red] (a1) ,
     (c1) -- [thick, red] (a7),
    (c2) -- [thick, red] (a7),
        (c3) -- [thick, red] (a7),
        (X1) -- [thick, scalar, red] (Y1),
        (X2) -- [thick, scalar, red] (Y2),
        (X3) -- [thick, scalar, red] (Y3),
        (X4) -- [thick, scalar, red] (Y4),
        (X5) -- [thick, scalar, red] (Y5)
};
  \end{feynman}
\end{tikzpicture}
  \end{minipage} \quad . 
\ee
Next the two-point buble can span $p$ vertices in all the following possible ways
\be
f(p)=\sum_{j=0}^{n-p}        \prod_{s=0}^p\frac{1}{x_{j+s}+E}   \prod_{i=1}^{j} \frac{1}{x_i} \prod_{i=j+p}^{n} \frac{1}{x_i} \, . 
 \label{bubs2}
\ee
Finally, upon adding  $\reef{bub1}+ \sum_{p=1}^n f(p)$ we are led to
\be
\frac{1}{E} \prod_{j=1}^{n-1} \frac{1}{x_j} \sum_{i=1}^{n-1} \frac{1}{x_i}+  \prod_{j=1}^{n-1} \frac{1}{x_j}  \frac{1}{E^2}   \, . 
\ee
The former equation has precisely the form of the subtraction terms in \reef{fund1}.
Indeed, the first term of the former equation should be interpreted as the first term of the \emph{subtraction terms} in \reef{fund1} -- with $1/E$ signifying the two-point bubble contribution to $\cE^{(2)}_i$
While the second term of the former equation is identified with the second term of  the \emph{subtraction terms}. 
Thus we have  established that diagrams containing disconnected  two point bubbles do not contribute to $\cE_i^{(n)}$.~\footnote{
Clearly, \reef{bub1} and \reef{bubs2}  should be multiplied by  kinematical factors to get the proper diagrams. But such factors are common to    \reef{bub1} and \reef{bubs2} and thus they are irrelevant for our main point. } This is of course expected from a covariant Lagrangian calculation, but it is fun to see  the intricate way in which it is realised in  non-covariant  Hamiltonian perturbation theory. 
Following a similar logic, our proof could be extended to show that all disconnected vacuum diagrams cancel.

\section{Calculation of counter-terms}
\label{acts}

In this appendix we provide details on the calculation of the counter-terms in \reef{vac1} and \reef{mass1}.
The $O(g^2)$ vacuum diagram is given by
 \be
 \begin{minipage}[h]{0.058\linewidth}
\begin{tikzpicture}
\begin{feynman}[small]
 \node [dot] (i1) at (0,0);
 \node [dot] (i2) at (.8,0);  
\diagram*{
   (i1) -- [ quarter right, looseness=.8,  thick] (i2)  -- [ quarter right, looseness=.8 ,  thick] (i1) ,
     (i1) -- [ half right, looseness=1.3,  thick] (i2)  -- [ half right, looseness=1.3 ,  thick] (i1)
};
  \end{feynman}
\end{tikzpicture}
  \end{minipage}  =  \frac{g^2  L^2}{24} \sum   \frac{L^2 \, \delta_{  k_1+k_2+k_3+k_4,0 }  }{ -(\om_{k_1}+\om_{k_2}+\om_{k_3}+\om_{k_4})}    \, , \label{vac12}
 \ee
 were we are using the notation introduced immediately after \reef{exa4}. Namely, all the sums include relativistic normalisation, they are performed over all the momenta and are restricted such that the propagating states have energies smaller or equal than $E_T$.~\footnote{
 In the case at hand the notation means   $\sum=\sum_{k_i\text{ s.t. }E_\text{in}\leq E_T}1/\prod_j 2L^2\om_{ k_j}$ where $E_\text{in}=\om_{k_1}+\om_{k_2}+\om_{k_3}+\om_{k_4}$.} Since the frequencies are restricted to $\om_{k_1}+\om_{k_2}+\om_{k_3}+\om_{k_4}\leq E_T$, we can introduce  the factor $1= \int_m^{E_T} dE \delta(E - \Sigma_ i \om_{k_i}) $ in the integrand. Then, after switching the order of integration,  we are led to 
$
   \reef{vac12}   =   - \frac{ g^2 L^2}{24} \int_{4m}^{E_T} \frac{dE}{2\pi}\frac{\Phi_4(E)}{E} 
 $, 
where $\Phi_4$ is the four-particle  phase space, given in \reef{ps1}.  
Note that the vacuum energy \reef{vac12} is UV divergent in the limit $E_T\rightarrow \infty$. The UV divergence  of the vacuum energy density can be evaluated at infinite volume.  
The   infinite volume   phase space is given by \reef{vacu22}. Therefore, we get
 \be
 \reef{vac12}= - (gL)^2 \frac{E_T+8 m \log\left(m/E_T \right)}{96 (4    \pi)^3 }+ O(E_T^0) \, .
\ee
The $d_0$ counter-term calculation proceeds as follows. 
\be
\begin{minipage}[h]{0.088\linewidth}
\begin{tikzpicture}
\begin{feynman}[small]
 \node [dot] (i1) at (0,0);
 \node [dot] (i2) at (.7,0);  
 \node [dot] (i3) at (1.4,0);  
\diagram*{
   (i1) -- [ quarter right, looseness=.8,  thick] (i2)  -- [ quarter right, looseness=.8 ,  thick] (i1) ,
     (i2) -- [ quarter right, looseness=.8,  thick] (i3)  -- [ quarter right, looseness=.8 ,  thick] (i2) ,
     (i1) -- [ half right, looseness=.8,  thick] (i3)  -- [ half right, looseness=.8 ,  thick] (i1)
};
  \end{feynman}
\end{tikzpicture}
  \end{minipage}  =\frac{g^3 L^2}{8}  \sum \frac{L^2 \delta_{k_1+k_2+q_1+q_2,0}}{\om_{k_1}+\om_{k_2}+\om_{q_1}+\om_{q_2}}  \frac{L^2 \delta_{k_1+k_2+p_1+p_2,0}}{\om_{k_1}+\om_{k_2}+\om_{p_1}+\om_{p_2}}   \, .  \label{vaac3T}
   \ee
 where the symmetry factor is is given by  the  $m=n=p=2$ of 
$
 s^4_{mnp}/4!^3=1/[(4-m-n)!(4-m-p)!(4-n-p)!\, m!n!p! ]  
$.
 Eq.~\reef{vaac3T} leads to a logarithmic UV divergence as $E_T\rightarrow \infty$.
  Different regularisation schemes like momentum cutoff, short distance cutoff or $E_T$ cutoff differ by $O(E_T^0)$.
The coefficient of the log divergent piece of \reef{vaac3T} can be computed at infinite volume [up to the overall extensive $L^2$ factor]  by introducing the phase-space functions $\Phi_n$.  
In covariant perturbation theory the $O(g^3)$ correction to the  vacuum is given by
\be \frac{g^3}{3!\cdot 8} \int_{ -\infty}^\infty d\tau_1\, d\tau_2\,    \int_{-\infty}^\infty  d^2y_1d^2y_2 \, \Delta^{2}(\vec y_1, \tau_1)\Delta^{2}(\vec y_2, \tau_2)\Delta^{2} (\vec y_1+\vec y_2,\tau_1+\tau_2)  \ .  \label{in1T}
\ee
The former expression can be written as
 $
\reef{in1T}=\frac{g^3}{8} \int_{-\infty}^\infty \frac{d^3k}{(2\pi)^3} \,  \rho^3(k)$  where we have defined $ \rho(k)=\frac{1}{3!} \int_{-\infty}^\infty d\tau \int_{-\infty}^\infty   \, \Delta^{2}(z)e^{i\bk\cdot \vec z}  d^2z   
 $. Given
$
\Delta(z)= e^{-m |z|}/(4\pi |z|)$, we get 
$
\rho(k)=   \frac{1}{3!} \frac{1}{8k} + \cdots
 $.
Therefore,~\footnote{This result agrees with dimensional regularisation or momentum cutoff regularisation, which can be found elsewhere in the literature, see for instance Eq.~(C7) of ref.~\cite{Braaten:1995cm}.
}
\be
  \reef{vaac3T} =  \frac{1}{3!}\frac{g^3 L^2}{2^{9} \, (4\pi)^2} \log(E_T/m) + O(E_T^0)   \ . \label{in22T}
\ee
Finally, the  mass gap $\cE_1^{(2)}-\cE_0^{(2)}$    at $O(g^2)$ is given by  
 \be
\begin{minipage}[h]{0.085\linewidth}
\begin{tikzpicture}
\begin{feynman}[small]
 \vertex (i0) at (-.3,0);
 \node [dot] (i1) at (0,0);
 \node [dot] (i2) at (.8,0);  
 \vertex (i3) at (1.1,0);
\diagram*{
   (i0) -- [thick]   (i3) ,
     (i1) -- [ half right, looseness=1.,  thick] (i2)  -- [ half right, looseness=1. ,  thick] (i1)
};
  \end{feynman}
\end{tikzpicture}
  \end{minipage}+ \begin{minipage}[h]{0.085\linewidth}
\begin{tikzpicture}
\begin{feynman}[small]
 \vertex (im2) at (1.1,0);
  \vertex (im1) at (1,0);
 \vertex (i0) at (.9,-.1);
 \node [dot] (i1) at (0,0);
 \node [dot] (i2) at (.8,0);  
 \vertex (i3) at (-.1,.1);
 \vertex (i4) at (-.2,0);
  \vertex (i5) at (-.3,0);
\diagram*{
(im2)--[thick](im1)--[quarter right, thick]   (i0) -- [half left, looseness=1.3, thick] (i1) --[thick]   (i2) 
   --  [thick, half right, looseness= 1.3] (i3)     -- [thick, quarter left , looseness = 1] (i4) --[thick] (i5)  ,
     (i1) -- [ half right, looseness=1.,  thick] (i2)  -- [ half right, looseness=1. ,  thick] (i1)
};
  \end{feynman}
\end{tikzpicture}
  \end{minipage} = \frac{g^2}{12m} \sum  \frac{L^2 \delta_{k_1+k_2+k_3,0}}{1-(\om_{k_1}+\om_{k_2}+\om_{k_3})}+ \frac{L^2 \delta_{k_1+k_2+k_3,0}}{1-(\om_{k_1}+\om_{k_2}+\om_{k_3}+2)}    \label{ttct} \, ,
       \ee 
       where, as drawn in the former two diagrams,  we need to take into account that there are two possible vertex orderings. 
     Again, we can evaluate the UV divergence of the former expression by introducing the phase space $\Phi_3$ at infinite volume.  Then, in the limit $m\ll E_T$ we get
       \be
        \reef{ttct}= \Big(- \frac{g^2}{96 \pi^2 }  \log(E_T/m)  + O(E_T^0) \Big) \cdot  \langle m| V_2 |m\rangle \, .
       \ee

\section{Perturbative mass}
\label{HPTapp}

In this appendix we  give details on the computation of  the mass gap and vacuum energy in Hamiltonian perturbation theory with the $E_T$ regulator. We do the calculation up to $O(g^3)$.
Note that $
V_{nn}\times V_{nk}(E_n-E_k)^{-2}V_{kn} = O(g^4)
$, since the only non-vanishing terms are
$\
\langle n | C |n \rangle  = O(g^2)$.
Thus we are left to compute  the zero, one, two and three point functions. 
The vacuum and first excited states are given by
\bea
\cE_0(E_T)&=&  
 \Va +
   \VVa       + 
   \VVVa
   + O(g^4)  \, ,    \\
\cE_1(E_T) &=&  \Vb
   +
 \Vc
   +  \Vd
   + \VVb
   +\VVc
   +   \VVd
 +
  \VVVc
   +
  \VVVd \nonumber \\
 &  +&\Big( 
  \VVe+
  \VVVe
   + 
\VVVf
+
\VVVg
+h.c.\Big)   
 +
 \VVVh
+ O(g^4)   \,  .   \quad \quad  \quad
\eea 
Where the   diagrams are given by the following expressions:
\bea
\VVa
   &=&   -    \frac{g^2}{24}  \sum \frac{1}{\prod_{j=1}^4 2L^2\om_{\vec k_j}} \frac{  L^2 \delta_{  k_1+k_2+k_3+k_4,0 }  }{\om_{k_1}+\om_{k_2}+\om_{k_3}+\om_{k_4}} \cdot L^2    \label{canci1}  \, ,  \\
  \VVVa &=& \frac{g^3}{8}  \sum    \frac{L^2\delta_{k_1+k_2+q_1+q_2,0}}{ \om_{k_1}+\om_{k_2}+\om_{q_1}+\om_{q_2} }   \frac{L^2\delta_{k_1+k_2+p_1+p_2,0}}{\om_{k_1}+\om_{k_2}+\om_{p_1}+\om_{p_2}}\cdot L^2 \, , \label{canci2} \\
\Vb  &=& m \, ,   \\
 \Vc  &=&  -c_2 \cdot  \langle m| V_2 | m \rangle  \, , \\
   \Vd   &=&  - (c_0+d_0)\cdot L^2   \, ,
\\
   \VVb &=&  -   \frac{ g^2}{6}     \sum   \frac{   L^2\delta_{k_1+k_2+k_3,0}  }{-m+\om_{k_1}+\om_{k_2}+\om_{k_3}}  \cdot \langle m| V_2 | m \rangle  \, , \\ 
  \VVc  &=&    -   \frac{ g^2}{6}     \sum   \frac{   L^2\delta_{k_1+k_2+k_3,0}  }{m+\om_{k_1}+\om_{k_2}+\om_{k_3}}  \cdot \langle m| V_2 | m \rangle  \, ,  \\ 
   \VVd  &=&    -   \frac{g^2}{24}  \sum   \frac{  L^2 \delta_{  k_1+k_2+k_3+k_4,0 }   }{\om_{k_1}+\om_{k_2}+\om_{ k_3}+\om_{k_4}}              \cdot  L^2     \, ,     \\
  \VVe    &=&   \frac{g c_2}{4}  \sum  \frac{L^2\delta_{k_1+k_2,0}}{\om_{k_1}+\om_{k_2}}  \cdot  2 \langle m| V_2 | m \rangle  \label{tocanc2}   \, , \\
    \VVVb &=&   \frac{g^3}{8}  \sum \frac{L^2\delta_{k_1+k_2+q_1+q_2,0}}{\om_{k_1}+\om_{k_2}+\om_{q_1}+\om_{q_2}} \frac{L^2\delta_{k_1+k_2+p_1+p_2,0}}{\om_{k_1}+\om_{k_2}+\om_{p_1}+\om_{p_2}} \cdot L^2  
    \, ,   \\
 \VVVc    &=&  \frac{g^3}{4}  \sum   \frac{L^2\delta_{\vec p_1+\vec p_2+\vec k,0}}{-m+ \om_{p_1}+\om_{p_2}+\om_{k}}    \frac{L^2\delta_{\vec q_1+\vec q_2+\vec k,0}}{-m+ \om_{q_1}+\om_{q_2}+\om_{k}}  \cdot   \langle m| V_2 |m \rangle \label{mmggfirst}   \, ,   \label{disco1} 
     \eea
   \bea
 \VVVd &=&    \frac{g^3}{4} \sum  \frac{L^2\delta_{\vec p_1+\vec p_2+\vec k,0}}{m+ \om_{p_1}+\om_{p_2}+\om_{k}}    \frac{L^2\delta_{\vec q_1+\vec q_2+\vec k,0}}{m+ \om_{q_1}+\om_{q_2}+\om_{k}} \cdot    \langle m| V_2 |m \rangle   \, ,
 \\
 \VVVe
 &=&  \frac{g^3}{4} \sum  \frac{L^2\delta_{p_1+p_2+p_3,0}}{m+ \om_{p_1}+\om_{p_2}+\om_{p_3}}    \frac{L^2\delta_{q_1+q_2+p_1+p_2,0}}{\om_{q_1}+\om_{q_2}+ \om_{p_1}+\om_{p_2}}  \cdot   \langle m| V_2 |m \rangle     \, ,\\
 \VVVf &=&  \frac{g^3}{4}  \sum  \frac{L^2\delta_{p_1+p_2+p_3,0}}{-m+ \om_{p_1}+\om_{p_2}+\om_{p_3}}    \frac{L^2\delta_{q_1+q_2+p_1+p_2,0}}{\om_{q_1}+\om_{q_2}+ \om_{p_1}+\om_{p_2}}    \cdot \langle m| V_2 |m \rangle     \, ,\\
\VVVg
 &=&  \frac{g^3}{12}    \sum  \frac{L^2\delta_{p_1+p_2+p_3+p_4,0}}{\om_{p_1}+\om_{p_2}+\om_{p_3}+\om_{p_4}}    \frac{L^2\delta_{q+p_1,0}}{ \om_{p_1}+\om_{q}} \cdot 2 \langle m| V_2 |m \rangle \label{tocanc1} \, , \\
 \VVVh  &=&  \frac{g^3}{12}    \sum   \frac{L^2\delta_{k_1+k_2+k_3+p,0}}{\om_{k_1}+\om_{k_2}+\om_{k_3}+\om_{p}}   \cdot   \frac{L^2\delta_{k_1+k_2+k_3+q,0}}{ \om_{k_1}+\om_{k_2}+\om_{k_3}+\om_{q}}  \cdot 2 \langle m| V_2 |m \rangle \label{mmgglast} \, .
      \eea
The sums in the previous expressions are over all momenta and involve relativistic measures, namely
 \be
\sum \equiv \sum_{\substack{\text{all } k_i \\  \text{s.t. } E_\text{in} \leq E_T}} \frac{1}{2L^2\om_{ k_i}} \,  ,
\ee where $E_\text{in}$ is the energy of the sate propagating between any to given consecutive vertices.  
Many of this expression can be computed analytically while other can be easily evaluated numerically with your favourite package. We used \texttt{vegas}, a  \texttt{python} package that implements Monte Carlo integration. 
Combined, the results lead to equation \eqref{massgappert} in the main text. 

Note that the   diagrams in  \reef{tocanc1} are  logarithmically UV divergent. 
Indeed, in the limit $E_T\sim k_i\gg q, p\sim m$, they combine into
\be
 \reef{tocanc1} = \frac{g^3}{6}  \sum \frac{L^2\delta_{q+p,0}}{ \om_{p}+\om_{q}}    \sum  \frac{L^2\delta_{k_1+k_2+k_3+p,0}}{\om_{k_1}+\om_{k_2}+\om_{k_3}+\om_{p}}     + O(1/E_T)  = -  \frac{g c_2}{4}  \sum \frac{L^2\delta_{q+p,0}}{ \om_{p}+\om_{q}}    + O(1/E_T) \, , \nonumber
\ee
and, unsurprisingly the UV divergence cancels against diagram \reef{tocanc2}.

\section{Algorithmic implementation}
\label{codeapp}

Our implementation of the truncation calculation largely follows that described in \cite{Elias-Miro:2017tup}, with some additional complications due to having one extra dimension and from the counter-terms needed to patch up for missing states.

We first construct a basis of all states with momenta purely in the first quadrant. A zero momentum state is then represented in terms of these states rotated into each of the quadrants [along with zero momentum quanta]. When computing the Hamiltonian matrix, the vast majority of the overlaps between states vanish, i.e. the matrix is extremely sparse. The key algorithmic requirement is that excessive time is not spent computing matrix elements that turn out to be zero. To avoid this we initially compute the states that can be produced from each basis element when $n=1,2,3,4$ creation or annihilation operators applied. The results of this are stored as maps [or dictionaries] with the momentum change as the keys. Many of the numerical quantities involved in the matrix elements can also be computed in advance and stored in maps. Used together these allow only non-zero matrix elements to be computed, and such computations amount to simply multiplying a handful of previously saved numbers.

The counter-terms that correct for missing states can be straightforwardly calculated in the code. We replace integrals with sums over states so that finite volume corrections are captured. Then the counterterm fixing a single two-point bubble with vertices consecutive in time, \eqref{patchvac}, is given by
\be
\delta V_{nn}^I = \sum_{4} \frac{V_{04} V_{40}}{E_{i(n+4)}} ~,
\ee
where $4$ represents a 4-particle state, and the sum is restricted to states such that $E_T- E_n < E_4 \leq E_T$. 
The counterm corresponding to towers of overlapping two-point bubbles can be calculated via matrix products. For example, the part of \eqref{itect} that corresponds to two iterated bubbles is given by
\be
\delta V_{nn}^{II} \supset \sum_{4,4'} \frac{V_{04} V_{4 (4+4')}V_{(4+4')4'}V_{4'0}}{E_{i(n+4)} E_{i(n+4+4')} E_{i(n+4')}} ~, \label{patchii_sum}
\ee
where $4$ and $4'$ represent the four particle states that make up to the two bubbles. The sum is restricted to $4$ and $4'$ such that at least one of $E_4 >E_T-E_n$ and $E_{4'} >E_T-E_n$ is satisfied.\footnote{The counterterms $\delta V_{nn}^{I}$ and $\delta V_{nn}^{II}$ are computationally expensive to implement since they must be calculated for each energy $E_n$ individually. In practice we compute the counter-terms once for state energies $E_n = 0,m/10,2m/10, \dots, E_T$ and store the results. Then when inserting the counterterms into the matrix, for a state $E_n$ we simply use the nearest saved value. We have checked this makes no difference to the numerical results.}

In our numerical calculations we also include a counterterm $\delta V_{nn}^{III}$ that accounts for the missing states in the three-point vacuum bubble diagram. This is not needed for a UV finite theory, and in practice only leads to extremely minor change to the numerical results, but it does slightly simplify the Monte Carlo integration in section~\ref{aMonteCarlo}. We have checked and find that adding further corrections to account for other classes of missing states [scaling either as $E_T^0$ or as $1/E_T^n$ with $n>1$] also does not significantly change the numerical results that we show in section~\ref{sNI} and \ref{sCM}. 

Efficient algorithms for finding the smallest few eigenvalues of sparse matrices are well known, and we simply use the library Eigen. In our implementation the majority of the computational time is spent obtaining the eigenvalues themselves from the Hamiltonian matrix, although calculating the state dependent counter-term corresponding to the towers of vacuum bubbles is also expensive. On a personal computer we reach basis sizes of order $10^6$, and using a single node of a high performance cluster we are limited to basis sizes $\lesssim 10^7$. At the maximum basis size the lowest eigenvalues can be computed in approximately an hour. We are unable to access larger basis sizes due to memory limitations on the size of the Hamiltonian matrix itself [despite  being stored in sparse form].


\section{Finite mismatch pieces} \label{app:finite}

In the main text we focused on, and corrected for, new UV divergences due to an $E_T$ regulator. In this appendix, we consider effects introduced by such a regulator that remain finite as $E_T\rightarrow \infty$. First in appendix~\ref{aaFiniteSource} we identify the types of diagram that lead to such corrections. 
Then, in appendix~\ref{aaFiniteChang} we argue that the weak/strong duality tests how important such finite corrections are for our numerical results from HT.

\subsection{Sources of finite regulator dependence}\label{aaFiniteSource}

The sources of finite corrections can be identified by continuing the analysis in section~\ref{gencase5}. Only diagrams with at least one UV divergent sub-diagram can be sensitive to the regulator, so the possibilities are limited. 

All classes of diagrams  analysed in \ref{ssPatch} that give extra  UV divergent pieces also give finite corrections. Our patches in section~\ref{ssPatch} account for the missing states in the divergent parts of such diagrams, for example the upper bubble of \eqref{detdiag1}. So there is no finite correction from the $E_T$ regulator on this loop. 

However, there can be finite corrections from the other, unpatched, loops, e.g. the lower bubble in \eqref{detdiag1}. This can be seen by considering a simplified version of the diagram with just an state of energy $x_s$ below
\be
 \begin{minipage}[h]{0.13\linewidth}
\begin{tikzpicture}
\begin{feynman}[small]
 \node [dot] (i1) at (0,0);
 \node [dot] (i2) at (1.2,0);  
 \node [dot] (j1) at (.3,.43);
 \node [dot] (j2) at (.9,.43);  
  \vertex (a1) at (-.5,-.4);
  \vertex (a2) at (1.7,-.4);  
    \vertex (b1) at (-.5,-.5);
  \vertex  (b2) at (1.7,-.5);  
  \vertex  (c1) at (-.5,-.7);
  \vertex  (c2) at (1.7,-.7);  
  \vertex  (x) at (.6,-.62){$\cdots$};  
\diagram*{
   (i1) -- [ quarter right, looseness=.1,  thick] (i2)  -- [ quarter right, looseness=.1 ,  thick] (i1) ,
     (i1) -- [ half right, looseness=.4,  thick] (i2)  -- [ half right, looseness=.4 ,  thick] (i1) ,
   (j1) -- [ quarter right, looseness=.2,  thick] (j2)  -- [ quarter right, looseness=.2 ,  thick] (j1) ,
     (j1) -- [ half right, looseness=.6,  thick] (j2)  -- [ half right, looseness=.6 ,  thick] (j1),
     (a1) --[thick] (a2),
     (b1) --[thick] (b2),
     (c1) --[thick] (c2),
};
  \end{feynman}
\end{tikzpicture}
  \end{minipage} 
  + \, 
  \begin{minipage}[h]{0.13\linewidth}
\begin{tikzpicture}
\begin{feynman}[small]
 \node [dot] (i1) at (0,0);
 \node [dot] (i2) at (1.2,0);  
 \vertex (j1) at (.3,.62);
 \vertex  (j2) at (.3,.3);  
  \vertex (a1) at (-.5,-.4);
  \vertex (a2) at (1.7,-.4);  
    \vertex (b1) at (-.5,-.5);
  \vertex  (b2) at (1.7,-.5);  
  \vertex  (c1) at (-.5,-.7);
  \vertex  (c2) at (1.7,-.7);  
  \vertex  (x) at (.6,-.62){$\cdots$};  
  \node [crossed dot, rotate=45] (yx) at (.6,.4);
\diagram*{
     (j1) --[white] (j2);
   (i1) -- [ quarter right, looseness=.1,  thick] (i2)  -- [ quarter right, looseness=.1 ,  thick] (i1) ,
     (i1) -- [ half right, looseness=.4,  thick] (i2)  -- [ half right, looseness=.4 ,  thick] (i1) ,
     (a1) --[thick] (a2),
     (b1) --[thick] (b2),
     (c1) --[thick] (c2),
};
  \end{feynman}
\end{tikzpicture}
  \end{minipage}
    \vspace{.018cm}  \, , \label{dia2bubblefinite} 
  \ee 
where the second diagram corresponds to the counter-term adding back in states to the upper bubble. Regulating the lower loop by $E_T - X_s$
\be
\eqref{dia2bubblefinite} = - \text{sym}\int_4^{E_T -X_s} dx_1 \frac{\Phi_4(x_1)}{x_1^2} \int_4^{E_T} dx_2 \frac{\Phi_4(x_1)}{x_1+x_2} ~,
\ee 
where sym is the symmetry factor, and $x_1$ and $x_2$ are the energies in the lower and upper bubbles, and the $x_2$ integral extends to $E_T$ due to the patch up. Carrying out the integrals and expanding at large $E_T$ we find a finite sensitivity to $X_s$. On the other hand, there is no finite piece from the un-patched sunset bubble in  \reef{cdfgs}, due to the faster convergence of the lower loop in this case.

There are also diagrams that do not result in UV divergent contributions but do give finite pieces. For example, this happens when a two-point vacuum bubble with its vertices consecutive in time is inserted above a convergent diagram. Indeed, we have already seen this in \eqref{tuc1}. Such diagrams are automatically fixed by \eqref{patchvac}. Meanwhile in \eqref{tsh} the sunset diagram and the three-point vacuum bubble do not give such finite pieces.

Finite pieces also arise when the two-point vacuum bubble crosses one of the vertices of a logarithmically divergent diagram. For example,
\be
 \begin{minipage}[h]{0.07\linewidth}
\begin{tikzpicture}
\begin{feynman}[small]
 \vertex (i0) at (-.3,0);
 \node [dot] (i1) at (0,0);
 \node [dot] (i2) at (.6,0);  
 \vertex (i3) at (.9,0);
 \node [dot] (j1) at (.3,.43);
 \node [dot] (j2) at (.9,.43);  
\diagram*{
   (i0) -- [thick]   (i3) ,
     (i1) -- [ half right, looseness=1.,  thick] (i2)  -- [ half right, looseness=1. ,  thick] (i1),
   (j1) -- [ quarter right, looseness=.2,  thick] (j2)  -- [ quarter right, looseness=.2 ,  thick] (j1) ,
     (j1) -- [ half right, looseness=.6,  thick] (j2)  -- [ half right, looseness=.6 ,  thick] (j1),
};
  \end{feynman}
\end{tikzpicture}
  \end{minipage}  \label{sunsetcrossed}
\ee
gives $g_{E_i}\left(x_s\right) \sim x_s^0$ in \eqref{crossvertex}  [where $x_s$ is the energy in the sunset loop]. Extracting the part of \eqref{sunsetcrossed} that depends on $X_s$, and setting $X_s= x_s$, the finite correction is evident. Further, a tower of two point bubbles crossing a vertex of the sunset diagram 
gives a similar correction.  The same happens when a two-point bubble, or a tower of these, crosses one of the vertices of the vacuum three-point bubble [calculating the correction analogously to \eqref{toreg22}, we see that this could either be one of the outer vertices or the central one]. Such effects are not corrected by our patches. 

Finally, finite pieces are produced when a sunset diagram has both vertices spanned by a two-point vacuum bubble, 
\be
 \begin{minipage}[h]{0.07\linewidth}
\begin{tikzpicture}
\begin{feynman}[small]
 \vertex (i0) at (-.6,0);
 \node [dot] (i1) at (0,0);
 \node [dot] (i2) at (.6,0);  
 \vertex (i3) at (1.2,0);
 \node [dot] (j1) at (-.2,.5);
 \node [dot] (j2) at (.8,.5);  
\diagram*{
   (i0) -- [thick]   (i3) ,
     (i1) -- [ half right, looseness=1.,  thick] (i2)  -- [ half right, looseness=1. ,  thick] (i1),
   (j1) -- [ quarter right, looseness=.1,  thick] (j2)  -- [ quarter right, looseness=.1 ,  thick] (j1) ,
     (j1) -- [ half right, looseness=.4,  thick] (j2)  -- [ half right, looseness=.4 ,  thick] (j1),
};
  \end{feynman}
\end{tikzpicture}
  \end{minipage} \label{nae}
\ee
given by
\be
\reef{nae}=-\text{sym} \int_4^{E_T} dx_4 \frac{\Phi_4(x_4)}{x_4^2} \int_3^{E_T-X_4} dx_3 \frac{\Phi_3(x_3)}{(x_3+x_4)} ~,
\ee
where $x_3$ and $x_4$ are the energies in the sunset and two-point bubble respectively. Then the dependence on the regulator can be extracted from the leading term proportional to $X_4$. Dropping order 1 factors for clarity, this scales as
\be
\sim \int_4^{E_T} dx_4 \frac{\Phi_4(x_4)}{x_4^2} \left( \frac{X_4}{E_T}\right)~,
\ee
so after setting $x_4 = X_4$ a finite piece remains.\footnote{An analogous calculation shows that a two-point vacuum bubble spanning two, or all three, of the vertices of a three-point vacuum bubble does not give a finite contribution.}

Carrying out a similar analysis for the subtraction terms shows that these also generate extra finite pieces $O(E_T^0)$ that are sensitive to the $E_T$ cutoff regulator.

Although finite corrections in themselves do not prevent the theory being well defined as $E_T \rightarrow \infty$, it might be wondered what happens once they are embedded in more complex diagrams. 
We have checked these effects in diagrams of up to $O(g^8)$ and we have not identified any such structures in the perturbation theory that are dangerous. 
An all orders proof and general understanding of this would be very desirable in the future.

\subsection{Finite pieces and the weak/strong self-duality }\label{aaFiniteChang}

As mentioned in section~\ref{sCM}, as well as testing the numerical power of HT,  the weak strong duality plays a second important role: it tests how large the effect of finite corrections due to the $E_T$ regulator is. The duality does so because the regulator dependent corrections to the mass gap of a theory depends on the particle's unperturbed mass. Consequently, the values of the finite corrections will differ between the original and dual theories.

For example, consider the diagram \eqref{dia2bubblefinite} but with the state below the bubbles simply a single particle state. Then the lower bubble is cut by $E_T- X_s $ where due to the $E_T$ regulator $X_s$. Once the corresponding integrals are evaluated and the $X_s$ dependence is extracted, it can be seen that the finite correction is proportional to the particle's  unperturbed mass.

Since we find that the original and dual theories give results that agree well over an extended range of couplings, we conclude that finite pieces are not playing a major role. Consequently, at least up to $g \simeq 100$, we are working in a scheme that could be reproduced in a calculation with a covariant regulator to reasonable accuracy.

\section{Monte Carlo integration of low order diagrams} \label{aMonteCarlo}

In this appendix we evaluate diagrams at low order in perturbation theory in $\phi^4$ in $2+1$ dimensions using Monte Carlo integration. By doing so we confirm that HT results at relatively small $E_T$ can be accurately extrapolated to the $E_T \rightarrow \infty$ limit. We also find that form \eqref{massgappert} leads to a more precise extrapolation than possible alternative choices. 

In the Monte Carlo integrals we implement exactly the same $E_T$ cutoff as occurs in a truncation calculation, and we study the dependence of the diagrams on $E_T$. Unlike truncation calculations, Monte Carlo integration can reach large values of $E_T$, although each diagram must be calculated individually.\footnote{We use the python package Vegas for our Monte Carlo results.}  The counter-terms to remove the primitive divergences, and to add back in missing states in these integrations, are taken to be identical to those that we use in truncation calculations.

The diagrams that contribute to the mass gap at order $g^2$ and $g^3$ are given in equations \eqref{deltae0g2},\eqref{deltae1g2}, \eqref{deltae0g3} and \eqref{deltae1g3}. Although not necessary to obtain a finite UV theory, we add back in the contribution from the missing states to the disconnected 3-point bubble. As a result such a diagram cancels between $\cE_1^{(3)}$ and $\cE_0^{(3)}$ in the calculation of the mass gap at finite $E_T$ as well as in the large $E_T$ limit. 

\begin{figure}[t]
\begin{center}
\includegraphics[scale=0.45]{{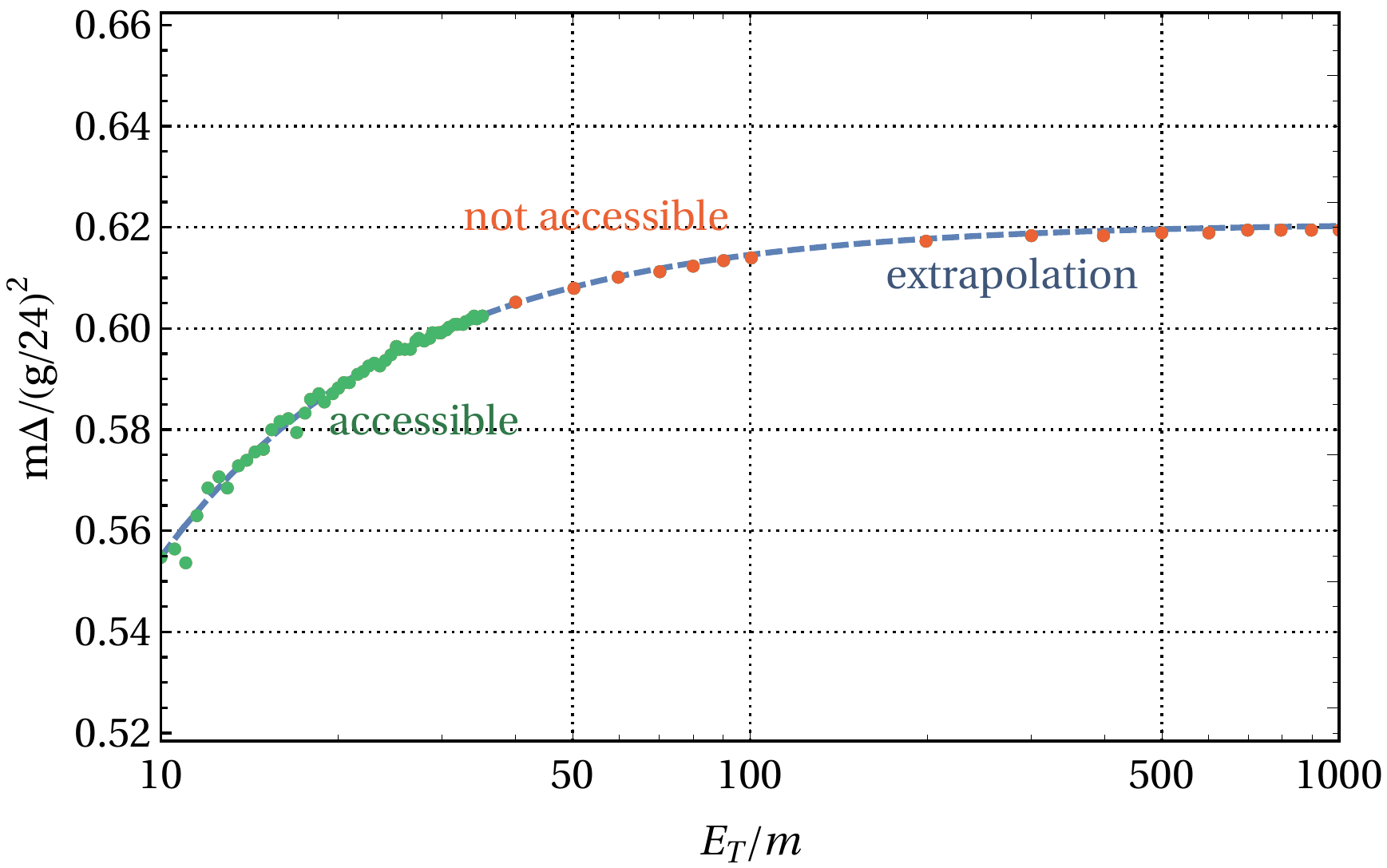}}\qquad
\includegraphics[scale=0.453]{{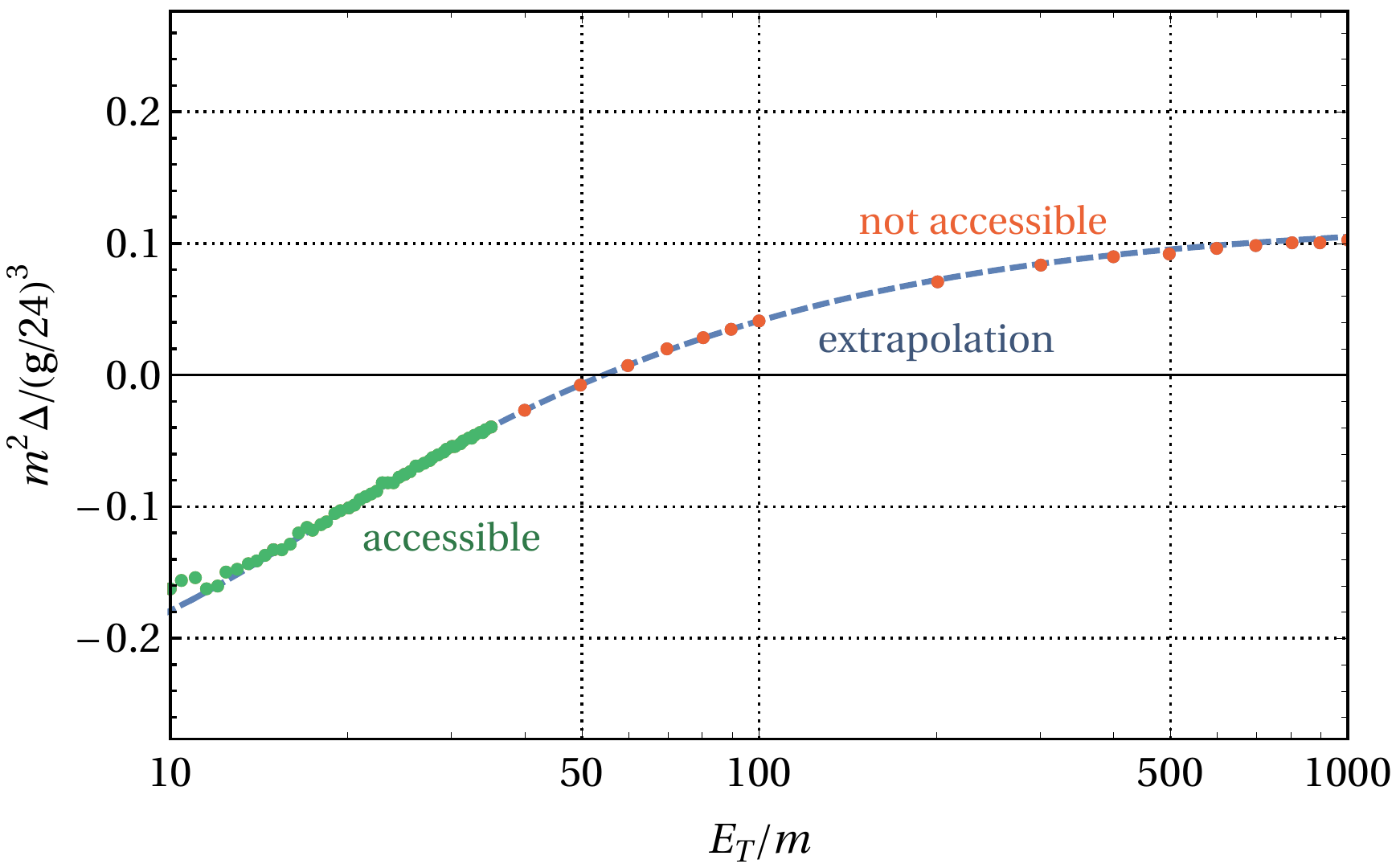}}
\end{center}
\caption{Data from Monte Carlo integration of perturbation theory diagrams showing the convergence of the $g^2$ (left) and $g^3$ (right) corrections to the mass gap $\Delta=\cE_1 - \cE_0$ as a function of the energy cutoff $E_T$ in $\phi^4$ theory in $2+1$ dimensions for a box size $L=4/m$. We indicate values of $E_T$ that are accessible and not accessible in HT calculations, and the extrapolation of the former.}
\label{fig:extrap}
\end{figure}

We calculate all of the preceding diagrams with Monte Carlo integration for a box length $L=4/m$. The results are shown as a function of $E_T$ in Figure~\ref{fig:extrap}. The prediction for the mass gap in the limit $E_T \rightarrow \infty$, \eqref{massgappert}, is obtained by extrapolating the data including all values of $E_T$. The Monte Carlo results obtained are relatively smooth once $E_T$ is not too small, which is compatible with the HT results shown in Figure~\ref{fig:ct}.

To test the potential power of the truncation method, we select only those Monte Carlo data points that can be reached in our truncation code. For $L=4/m$ this corresponds to those with $E_T < 34 m$. We also exclude points with $E_T < 17m$ since these have large fluctuations. We fit such data points with a function of the form of \eqref{fitform}. The results of these fits and their extrapolation to large $E_T$ are shown in Figure~\ref{fig:extrap}.

At the values of $E_T$ that are accessible in truncation calculations the $g^2$ correction reaches within $5\%$ of its asymptotic value. The extrapolation of this to $E_T \rightarrow \infty$ gives a prediction that coincides with the full result to better than $0.1\%$ accuracy. In contrast, at the accessible $E_T$ the $g^3$ correction is not only far from its asymptotic value but even of opposite sign. Despite this, the extrapolation of the $g^3$ correction from the small $E_T$ data matches that of the full data set to within $3\%$. 
The results obtained for the vacuum energy are qualitatively and quantitatively similar. For both the mass gap and the vacuum energy calculations, we find that the functional form \eqref{fitform} gives a more accurate extrapolation than if the $\alpha_2$ part is excluded, or if this is replaced by a piece $\alpha_2 / E_T^2$.

As discussed in section~\ref{sNI}, we estimate the coupling $g^*$ above which the theory is strongly coupled by demanding that $\cE_0^{(3)} = \cE_0^{(2)}$ at this value, which gives $g^* = 8.3$. The $g^3$ correction to the mass gap is suppressed relative to the $g^2$ correction at this value. However, most of this suppression is due to an accidental (one in four) cancellation between convergent subsets of diagrams. Taking each convergent subset individually, our estimate of strong coupling based on the mass gap would be $g^*\sim 30$.

Unlike Monte Carlo integration of diagrams, a truncation calculation is not perturbative  in $g$. Nevertheless, the Monte Carlo data is a positive indication that precise numerical results  can be obtained from truncation calculations.


\small

\bibliography{phi4_3-Biblio}
\bibliographystyle{utphys}

\end{document}